\documentclass[]{jfm}

\usepackage{graphicx}
\usepackage{newtxtext}
\usepackage{newtxmath}
\usepackage{subcaption}
\usepackage{natbib}
\usepackage{hyperref}

\usepackage[section]{placeins}
\hypersetup{
    colorlinks = true,
    urlcolor   = blue,
    citecolor  = black,
}

\newcommand{\RomanNumeralCaps}[1]
\linenumbers

\title[Evolution of internal cnoidal waves with local defects in a two-layer fluid with rotation]{Evolution of  internal cnoidal waves with local defects in a two-layer fluid with rotation}
\author{Korsarun Nirunwiroj \aff{1},
Dmitri Tseluiko \aff{1},
Karima Khusnutdinova \aff{1}
  \corresp{\email{K.Khusnutdinova@lboro.ac.uk}}}

\affiliation{\aff{1}Department of Mathematical Sciences, Loughborough University, Loughborough LE11 3TU, UK}

\begin{document}
\maketitle

\begin{abstract}
Internal waves in a two-layer fluid with rotation are considered within the framework of Helfrich's f-plane extension of the {\color{black} Miyata--Maltseva--Choi--Camassa (MMCC)} model. Within the scope of this model, we develop an asymptotic procedure which allows us to obtain a description of a large class of uni-directional waves leading to the Ostrovsky equation and allowing for the presence of shear inertial oscillations and barotropic transport. Importantly, unlike the conventional derivations leading to the Ostrovsky equation, the constructed solutions do not impose the zero-mean constraint on the initial conditions for any variable in the problem formulation. Using the constructed solutions, we model the evolution of quasi-periodic initial conditions close to the cnoidal wave solutions of the Korteweg--de Vries (KdV) equation but having a local amplitude and/or periodicity defect, and show that such initial conditions can lead to the emergence of bursts of large internal waves and shear currents.  {\color{black} As a by-product of our study, we show  that cnoidal waves with expansion defects discussed in this work are generalised travelling waves of the KdV equation: they satisfy all conservation laws of the KdV equation (appropriately understood), as well as the Weirstrass-Erdmann conditions for broken extremals of the associated variational problem and a natural weak formulation. Being smoothed in numerical simulations, they  behave, in the absence of rotation, as long-lived states with no visible evolution, while rotation changes this behaviour and leads to the emergence of strong bursts. }
%with the associated bursts in the presence of rotation being solely due to the effect of rotation. 
%We also discuss localised weak solutions of the KdV equation  representing solitons with defects.
 \end{abstract}
 
\begin{keywords}
Internal waves, Ostrovsky equation, {\color{black} generalised (shock-like) travelling waves} of the KdV equation, rogue waves
\end{keywords}

\section{Introduction}
\label{sec: 1}

The Korteweg--de Vries- and Ostrovsky-type family of models plays an important role in understanding the behaviour of long nonlinear internal waves commonly observed in coastal seas, narrow straits and river-sea interaction areas   \citep[see][and references therein]{GOSS1998, L2005, HM2006, BLS2008, OPSS2015, KZ2016, 
%HKG2021, TABK2023, 
S2022, OPSS2024}.  The Ostrovsky equation \citep{O1978}
\begin{equation}
(A_t + \nu A A_\xi + \lambda A_{\xi \xi \xi})_\xi = \gamma A
\label{I1}
\end{equation}
is a rotationally modified extension of the integrable Korteweg-de Vries (KdV) equation \citep{B1871,KdV1895,GGKM1967}, accounting for the leading order balance of weak nonlinear, dispersive and rotational effects. In the general setting of a density stratified fluid described by the Euler equations with boundary conditions appropriate for oceanic applications, equation (\ref{I1}) is written for the amplitude $A(\xi, T)$ of a single plane internal mode $\phi (z)$ in a reference frame moving with the linear long wave speed $c_0$. In physical variables where the bottom is at $z = - h$ and unperturbed surface is at $z=0$, the modal equations have the form
\begin{eqnarray}
& &\displaystyle (\rho_0 W^2 \phi_{z} )_z + \rho_0 N^2 \phi = 0 \,, \label{modal0} \\
& &\displaystyle \phi = 0  \quad \text{at} \quad z =-h\,, \quad \hbox{and} \quad
W^2 \phi_{z }= g\phi \quad \text{at} \quad z =0 \,. \label{modal0bc}
\end{eqnarray}
Here $\rho_{0} (z)$ is the stable background density profile, 
$N^2 = -g\rho_{0z} / \rho_0 $, $W=c_0-u_{0}$, where $u_{0}(z)$ is the background shear flow supported by a body force, and it is assumed that there are no critical levels, that is $W\ne 0$ for any $z$ in the flow domain. The nonlinear, dispersive and rotational coefficients $\nu, \lambda$ and $\gamma$, respectively, are given by
\begin{eqnarray}
 I\nu =  3\, \int_{-h}^0\rho_0 W^2 \phi_z^3\, dz  \,, \quad
 I\lambda =  \int_{-h}^0\rho_0 W^2 \phi^2\, dz \,,  \quad 
 I\gamma =    f^2 \int^{0}_{-h} \, \rho_{0} \Phi \phi_{z}  \, dz \,, \label{coeff} 
 \end{eqnarray}
 where
 \begin{eqnarray}
 I =  2\, \int_{-h}^0 \rho_0 W \phi_z^2 \, dz \,, \quad 
\rho_0 W\Phi  = \rho_0 W\phi_{z}  -(\rho_{0}u_{0})_z \phi  \,, \label{Phi} 
\end{eqnarray} 
and $f$ is the Coriolis parameter (a {\color{black} single mode} reduction of the bi-modal system derived in \citealt{AGK2013}). Note that when there is no shear flow, that is $u_0 (z) \equiv 0$, then $\Phi \equiv \phi_z $ and 
$\gamma = f^2/2c $;  in this case $\lambda \gamma >0$, but sufficiently strong shear near a pycnocline may lead to a situation where $\beta \gamma < 0$ \citep{AGK2014}.

\begin{figure}
    \centering
    \includegraphics[width=0.8 \linewidth]{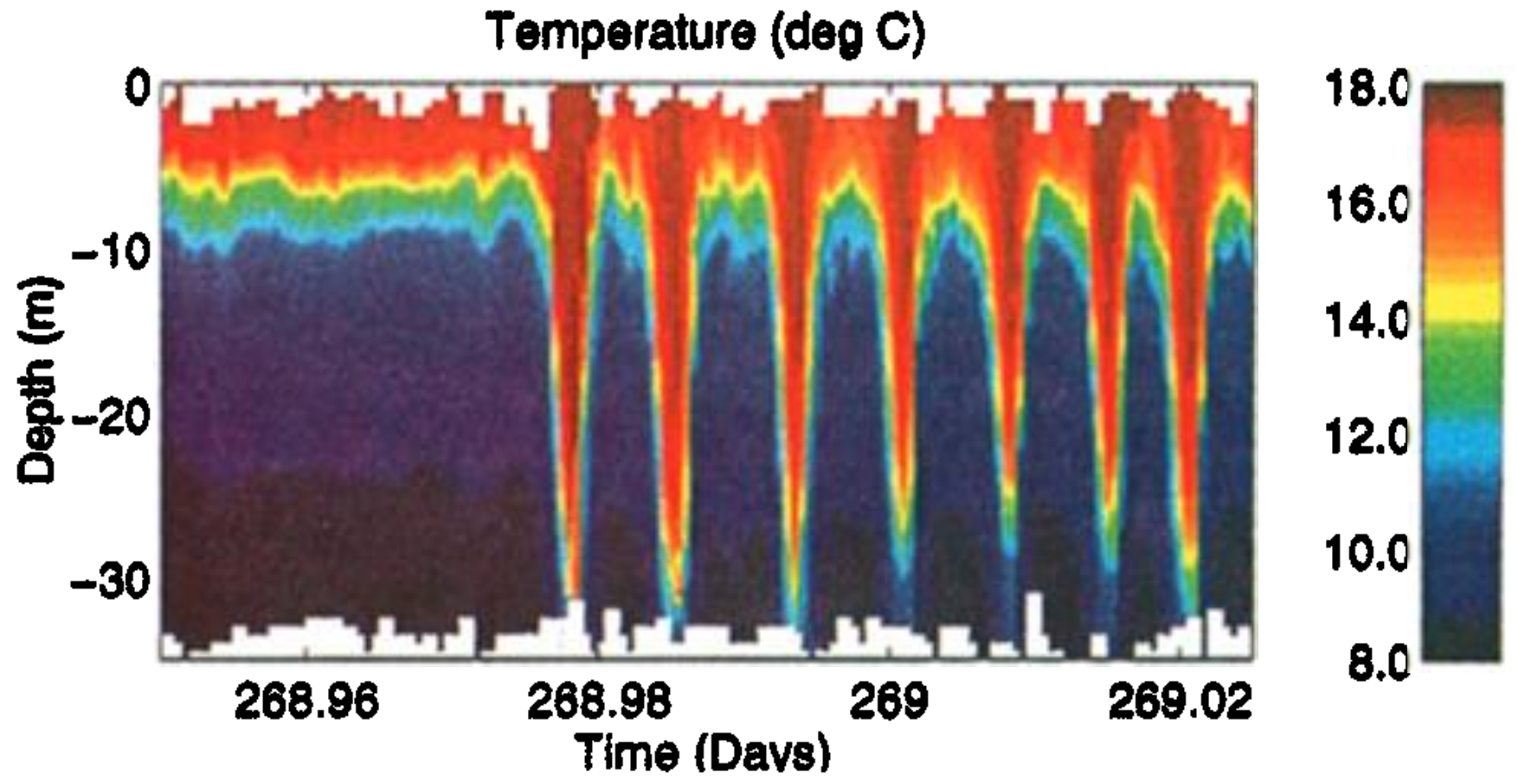}
    \caption{The first $1.7$ hours of the colour contour time series of temperature profiles off Northern Oregon from the surface to 35m depth. The figure is adapted from \citet{SO1998}.}
    \label{fig:1}
\end{figure}

The Ostrovsky equation became a paradigm forming model for studying the effects of rotation on the evolution of internal waves with the natural initial conditions in the form of KdV solitons (see \citealt{GHJ2013, S2020, OPSS2024} and references therein) and cnoidal waves, with an emphasis on modulational instability (see \citet{WJ2017, JOP2025} and references therein), as well as the related qualitative analysis of the long-time asymptotics of strongly-interacting internal modes described by solutions of coupled Ostrovsky equations \citep{AGK2014}. One of the aims of our current study is to extend the modelling to situations when the initial conditions are close to a cnoidal wave (approximately a chain of KdV solitons), but they are not perfectly periodic, and rather have some defects within the computational domain (hence, they can be viewed as being quasi-periodic). The motivation for that stems from the observational data of the type shown in Figure \ref{fig:1}, where we can see formation of a wavetrain of internal solitary waves close to an imperfect cnoidal wave. Indeed, given that the waves generally propagate in variable environment, quasi-periodic initial conditions seem to be a more natural choice than a single (pure or only slowly modulated) cnoidal wave. Recently, there appeared renewed interest in the possibility of generating rogue waves and breathers by various localised perturbations of cnoidal waves both in integrable and non-integrable settings   \citep[see][and references therein]{KM1975,ORBMA2013,CHOGDA2013,KAA2014,MH2016,CSHGDKDA2021,HWTCH2022,BJT2023,GMGJM2023,HMP2023, MCXH2023, CMH2024}, among other possible mechanisms \citep[e.g.][and references therein]{KPS2009,Z2009,PS2016,BAA2019,CBKK2022,SS2023,AGRS2024,CERT2024,FPD2024,S2024}. 
{\color{black} Moreover, internal rogue waves registered in the oceans have been linked to the KdV solitons \citep{O2010}.
Our study extends the line of research related to localised perturbations by adding the effect of rotation.}

Unlike the KdV equation, the Ostrovsky equation has a constraint on the mean value of its regular solutions. For example, for periodic solutions on the interval $[-L, L]$, any regular solution should have zero mean:
\begin{equation}
\int_{-L}^L A\  d\xi = 0.
\end{equation}
Additional constraints appear when solutions are considered in the class of functions vanishing at infinity \citep{B1992}.  The existing derivations of the Ostrovsky equation from the Euler equations do not allow one to consider the arbitrary initial conditions for the field variables of the parent system, but only those which agree with this constraint. This restriction on the choice of initial conditions of the Cauchy problem for the parent system became known as the ``zero-mean contradiction''. Hence, another aim of our study is to generalise the construction of weakly-nonlinear solutions leading to the Ostrovsky equation in such a way that the zero-mean contradiction is avoided. Indeed, this has been previously done within the scope of the derivation of Ostrovsky-type models in the simpler settings of the Boussinesq-Klein-Gordon and coupled Boussinesq equations \citep{KMP2014, KT2019, KT2022}. In contrast to the previous work, in the fluids context the mean-field equations appear to be coupled to equations for deviations from the mean values, presenting a new challenge, which is  addressed in our present paper.

Hence, the aim of our study is twofold, and the rest of the paper is organised as follows.  In Section~\ref{sec: 2} we introduce Helfrich's rotation-modified   two-layer {\color{black} Miyata--Maltseva--Choi--Camassa (MMCC-f)} model \citep{H2007}  and obtain its simpler weakly-nonlinear reduction which is used to develop the subsequent (weakly-nonlinear) derivations of the reduced model  {\color{black}  (for the original MMCC model see \citealt{M1988,M1989,CC1996, CC1999}).}
% but also \citealt{M1989}, where the same system has been derived and studied). 
While this setting is simpler than the full Euler equations, our derivations reveal that it retains the key complexity: the mean-field equations are coupled with the equations for the deviations. Therefore, this model provides an appropriate framework for our developments. Next, in Section~\ref{sec: 3} we refine the derivation of the uni-directional model by considering the simultaneous evolution of mean fields and their deviations. This approach introduces the fast characteristic variable $\xi = x - c_0 t$ and two slow-time variables $\tau = \sqrt{\alpha}t$ and $T = \alpha t$, rather than just one, where $\alpha$ is the small amplitude parameter (we assume the maximum balance conditions for the weak nonlinearity, dispersion and rotation).   We find a way to by-pass the zero-mean contradiction by developing simultaneous asymptotic expansions of both the mean-field variables and  deviations. This allows us to construct a more general class of solutions allowing the initial conditions for all fluid variables to have arbitrary (and generally time-dependent) mean values, while the emerging Ostrovsky equations have zero mean by construction. In Section~\ref{sec: 4}  we use the constructed weakly-nonlinear solution combined with numerical modelling using the Ostrovsky equation in order to investigate the effect of rotation on the evolution of cnoidal waves of the KdV equation close to their solitonic limit and having local amplitude and/or periodicity defects.  We begin the section by considering the effect of rotation on simple soliton and cnoidal wave solutions, as well as dark and bright breathers of the KdV equation and {\color{black} expansion and contraction  periodicity defects in order to set up the phenomenological framework for the discussion of our main numerical results concerning generic localised perturbations. The expansion   / contraction  defects are introduced by cutting the cnoidal wave at the trough and symmetrically inserting  a  piece of a straight line / extracting a small symmetric piece around the trough and gluing the remaining parts together, respectively, see the first two rows of Figure \ref{fig:WS} in Appendix A.} In all cases we take a sufficiently large computational domain and impose the periodic boundary conditions to model the resulting quasi-periodic solutions.  We show that, combined with the effect of rotation,  initial conditions in the form of cnoidal waves with local defects can lead to bursts of large amplitude internal waves and shear currents.  For the generic local defects introduced by adding a localised perturbation, this can be attributed to the formation of a pair of bright and dark breathers and contraction and expansion periodicity defects, with the subsequent effect of rotation. For the pure expansion and contraction defects  the bursts do not take place without rotation. In fact, in numerical runs the smoothed counterparts of such initial conditions evolve almost like travelling wave solutions of the KdV equation,  for a very long time. We {\color{black} argue} that this happens because these functions (and some other functions constructed from the known KdV solutions by similar procedures) satisfy all (infinitly many) conservation laws of the KdV equation {\color{black} (see Appendix A for the details of this and related discussions).} Moreover, a cnoidal wave with an expansion defect has a continuous first derivative, satisfying the Weirstrass-Erdmann corner condition for broken (non-smooth) extremals (e.g. \citealt{CF1954}) {\color{black} and it is a generalised (`shock-like') travelling wave, using the terminology introduced by \citet{GS2022}. } {\color{black} We also make remarks about a possible weak formulation for  the generalised travelling waves of the KdV equation. The Cauchy problem for the KdV equation with periodic boundary conditions  in $L^2$ is globally well-posed,  including uniqueness and continuous dependence with respect to the initial data \citep{B1993}.}
% (the global existence of the weak $L^2$ solutions on the infinite line was proven by \citet{K1983,KF1984}).   } 
The bursts observed in our modelling with pure periodicity defects are then attributed solely to the effect of rotation. Moreover, we show that the effects discussed in the paper are structurally stable with respect to the natural (compatible with the period of the background cnoidal wave) variations in the size of the computational domain. We finish with a discussion in Section \ref{sec: 5}, where we also show an example where a local perturbation of a cnoidal wave of the type considered in our paper has led to the generation of a rogue wave.  A pseudospectral scheme used to solve the Ostrovsky equation is discussed in Appendix B.

\section{MMCC-f model}
\label{sec: 2}

\begin{figure}
  \centerline{\includegraphics[width=0.8 \linewidth]{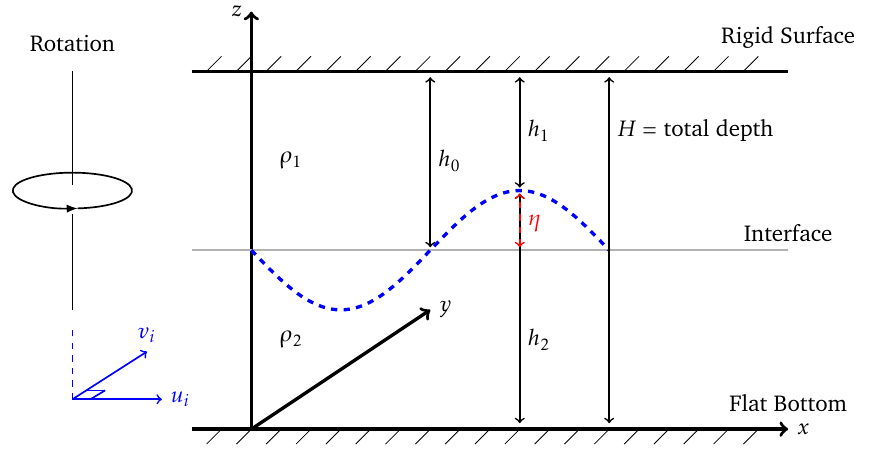}} 
  \caption{Schematic of a two-layer fluid with rotation in the rigid-lid approximation.}
\label{fig: 2}
\end{figure}

\noindent We consider the f-plane extension of the MMCC model for an inviscid, incompressible two-layer fluid with the rigid lid introduced by \citet{H2007}:
\begin{align}
&    h_{it} + (h_i \overline{u}_i)_x =0, \label{eq:1}\\
&    \overline{u}_{it} + \overline{u}_i \overline{u}_{ix} -f\overline{v}_i=-g\eta_x+\dfrac{1}{\rho_i}P_x + D_i, \quad \text{where }\label{eq:2}\\
     \hspace{0.5cm} & D_i = h_i^{-1}\Big\{ \dfrac{h_i^3}{3}\big[ \overline{u}_{ixt} + \overline{u}_i \overline{u}_{ixx} -(\overline{u}_{ix})^2 \big]\Big\}_x , \label{eq:3}\\
  &  (h_i \overline{u}_i)_t+(h_i \overline{u}_i \overline{v}_i)_x + fh_i \overline{u}_i = 0, \label{eq:4}\\
   & \overline{v}_{it} + \overline{u}_i \overline{v}_{ix} + f\overline{u}_{i} = 0. \label{eq:5}
\end{align}
Here, $h_i$, $\rho_i$ are the layer depths and densities, $\overline{u}_i, \overline{v}_i$ denote the depth-averaged over each layer horizontal velocities in the $x$ and $y$ directions with $i=1$ and $2$ referring to the upper and lower layers, respectively; $f$ is the Coriolis parameter, $g$ is gravity, $P$ is the pressure at the interface. The subscripts $t$ and $x$ denote partial derivatives.  In the absence of motion $h_1=h_0$ and $h_1+h_2=H$ (total depth) (see Figure \ref{fig: 2}).

In the Boussinesq approximation $(\frac{\rho_2-\rho_1}{\rho_1} \ll 1)$, the equations can be simplified by eliminating $P_x$. Using $\sqrt{g'H},\, H,\, l$ and ${l}/{\sqrt{g'H}}$ to non-dimensionalise $(u_i,\,v_i),\, h_i,\, x$ and $t$, respectively, 
and changing the variables to $s=\overline{u}_2-\overline{u}_1$, $v=\overline{v}_2-\overline{v}_1$, $h_2=1-h_1, U = h_1\overline{u}_1 +h_2\overline{u}_2$ and $V = \overline{v}_1h_1 + \overline{v}_2h_2$ with the barotropic transport in $x$ direction $U = F(t) \neq 0$ in general, we obtain
\begin{align}
    &\eta_{t} + (c_0^2 s +\sigma s\eta - s\eta^2 + F\eta)_x = 0, \label{1}\\
    &s_t + (\dfrac{\sigma}{2}s^2 -s^2\eta +\eta +Fs)_x - \tilde \gamma v = \tilde \beta (D_2-D_1), \label{2}\\
    &v_t  {\color{black} -}  sv\eta_{x} + \sigma sv_x -2 s\eta v_x + Fv_x + sV_x + \tilde \gamma s = 0, \label{3}\\
    &V_t + ( c_0^2 sv + \sigma sv\eta - sv\eta^2 + FV )_x +\tilde \gamma F = 0, \label{4}
\end{align}
where {\color{black} $\sigma = 2 h_0 - 1, g' = g \Delta \rho/\rho_1,  L_R = \sqrt{g'H}/f,  \tilde \gamma = {l}/{L_R},  \tilde \beta = \left ({H}/{l} \right )^2$.}

Next, we consider the Ostrovsky equation regime and assume that $\tilde \beta=\mathcal{O}(\alpha)$ and $\tilde \gamma=\mathcal{O}(\alpha^{1/2})$, where $\alpha$ is the small amplitude parameter and scale $\tilde \beta = \alpha \beta,\, \tilde \gamma = \sqrt{\alpha} \gamma$, where $\beta, \gamma = \mathcal{O}(1)$. Then, considering the asymptotic expansions 
\begin{align}
     (\eta, s, v, V) &= \alpha  (\eta, s, v, V)^{(1)} + \alpha^2  (\eta, s, v, V)^{(2)} + O(\alpha^3), \label{5}\\
  %  s &= \alpha s^{(1)} + \alpha^2 s^{(2)} +...,\\
  %  v &= \alpha v^{(1)} + \alpha^{2} v^{(2)} +...,\\
  %  V &= \alpha V^{(1)} + \alpha^{2} V^{(2)} +...,\\
    F &= \alpha F^{(1)} + \alpha^{2} F^{(2)}  + O(\alpha^3), \label{6}
\end{align}
and dropping the terms of $O(\alpha^3)$ (not used in our subsequent derivations), gives the simpler weakly-nonlinear equations 
\begin{align}
   & \eta_{t} + c_0^2 s_x = - \alpha (\sigma s\eta + F\eta)_x, \label{7}\\
   &s_t + \eta_x = \sqrt{\alpha} \gamma v - \alpha (\sigma s s_x + Fs_x  - \dfrac{ \beta c_0^2}{3} s_{xxt}), \label{8}\\
   &v_t = - \sqrt{\alpha} \gamma s - \alpha (\sigma sv_x + Fv_x + sV_x), \label{9}\\
   &V_t =  - \sqrt{\alpha} \gamma F - \alpha (c_0^2 sv + FV )_x. \label{10}
\end{align}
For $F = 0$, the equations reduce to the model derived by \citet{H2007, G1996}.

We consider the periodic solutions on the interval $[-L, L]$ (a typical setting for numerical runs using pseudospectral methods). Alongside with the equations (\ref{7})--(\ref{10}) we will consider the equations for the evolving mean fields by averaging the system with respect to $x$ over this interval and denoting the mean values by hats:
\begin{align}
    &\hat{\eta}_t = 0, \label{11} \\
    &\hat{s}_t = \sqrt{\alpha} \gamma \hat{v},  \\
    &{\color{black} \hat{v}_t = - \sqrt{\alpha} \gamma \hat{s}  - \alpha \frac{1}{2L} \int_{-L}^{L} (\sigma s v_x 
    %+Fv_x 
    +sV_x) dx, }  \label{12} \\
    &\hat{V}_t = -\sqrt{\alpha} \gamma F.  \label{12a}
\end{align}
Here, the mean fields are generally time-dependent.
Hence, the equations for the evolving mean values and deviations from the mean values are generally coupled because of the integral term in the equation (\ref{12}).  This differs from our previous derivations of Ostrovsky-type equations free from zero-mean contradiction within the scope of Boussinesq-Klein-Gordon equation \citep{KMP2014, KT2019} and coupled Boussinesq equations \citep{KT2022}, where the mean fields were described by independent equations.

\section{Weakly-nonlinear uni-directional solution free from zero-mean contradiction}
\label{sec: 3}

In this section, we construct a large class of solutions of the system (\ref{7})--(\ref{12a}) describing uni-directionally propagating waves on top of evolving mean fields by presenting each variable in the form of the sum of its mean value (generally, time-dependent) and deviation from this mean value:
\begin{equation}
\eta = \hat \eta + \tilde \eta, \ s = \hat s + \tilde s, \ v = \hat v + \tilde v, \ V = \hat V + \tilde V, 
\label{AS1}
\end{equation}
where the unknowns $\hat \eta, \hat s, \hat v, \hat V$ and the given function $F$ are assumed to be functions of $t, \tau = \sqrt \alpha t, T = \alpha t$, while the deviations $\tilde \eta, \tilde s, \tilde v, \tilde V$ are assumed to depend on the fast characteristic variable $\xi = x - c_0 t$, where $c_0$ is the linear long-wave speed, and two slow-time variables $\tau$ and $T$. Both the averages and deviations from the averages are sought in the form of asymptotic multiple-scale expansions in powers of $\sqrt{\alpha}$:
\begin{eqnarray}
(\hat \eta, \hat s, \hat v, \hat V)  &=& (\hat \eta, \hat s, \hat v, \hat V)^{(0)} + \sqrt{\alpha}  (\hat \eta, \hat s, \hat v, \hat V)^{(1)}+ \alpha (\hat \eta, \hat s, \hat v, \hat V)^{(2)} + O(\alpha^{3/2}), \label{AS2} \\
(\tilde \eta, \tilde s, \tilde v, \tilde V) &=& (\tilde \eta, \tilde s, \tilde v, \tilde V)^{(0)} + \sqrt{\alpha}  (\tilde \eta, \tilde s, \tilde v, \tilde V)^{(1)} + \alpha (\tilde \eta, \tilde s, \tilde v, \tilde V)^{(2)} + O(\alpha^{3/2}). \label{AS3}
\end{eqnarray}
The prescribed function $F(t)$ defining the barotropic transport in the $x$ direction is also assumed to be given in the form
\begin{equation}
F = F^{(0)} + \sqrt{\alpha} F^{(1)} + \alpha F^{(2)} + O(\alpha^{3/2}). \label{AS4}
\end{equation}
The average interfacial displacement $\hat \eta$ is a constant, and if $\hat \eta^{(0)} \ne 0$, then without loss of generality we assume that $\hat \eta = \hat \eta^{(0)}$. If $\hat \eta^{(0)} = 0$, but  $\hat \eta^{(1)} \ne 0$ then without loss of generality $\hat \eta = \sqrt{\alpha} \hat \eta^{(1)}$, etc. Similarly, if $F^{(0)} \ne 0$, then we can assume $F = F^{(0)}$, etc.

At $O(1)$ we obtain the equations
\begin{eqnarray}
&-c_0 \tilde \eta^{(0)}_{\xi} + c_0^2 \tilde s^{(0)}_{\xi} =0, \quad \hat s^{(0)}_t - c_0 \tilde s^{(0)}_{\xi}  + \tilde \eta^{(0)}_{\xi} =0,& \label{AS5} \\ 
&\hat v^{(0)}_t - c_0 \tilde v^{(0)}_\xi = 0, \quad
  \hat V^{(0)}_t - c_0 \tilde V^{(0)}_\xi = 0, \quad
   \hat s^{(0)}_t =0,\quad  \hat v^{(0)}_t=0, \quad  \hat V^{(0)}_t = 0, &
   \label{AS6}
\end{eqnarray}
implying
\begin{eqnarray}
& \hat s^{(0)} = A^{(0)}(\tau, T), \quad  \hat v^{(0)} = B^{(0)}(\tau, T), \quad  \hat V^{(0)} = C^{(0)}(\tau, T),& \label{AS7} \\
 & \displaystyle \tilde s^{(0)} = \frac{1}{c_0} \tilde \eta^{(0)}, \quad   \tilde v^{(0)} = 0, \quad   \tilde V^{(0)} = 0.&
  \label{AS8}
\end{eqnarray}
Here, the functions $(A, B, C)^{(0)}$ are arbitrary functions of their variables, and we have used that all deviations should have zero mean, by construction of the solution. 

Next, collecting the terms of order $O(\sqrt{\alpha})$, using relations (\ref{AS7}), (\ref{AS8}) and simplifying the resulting system by virtue of its averaged members,  we obtain the system
\begin{eqnarray}
&\displaystyle -c_0 \tilde \eta^{(1)}_{\xi} + c_0^2 \tilde s^{(1)}_{\xi} = - \tilde \eta^{(0)}_\tau, \quad - c_0 \tilde s^{(1)}_{\xi}  + \tilde \eta^{(1)}_{\xi} = - \frac{1}{c_0} \tilde \eta^{(0)}_{\tau}, \label{AS9}& \\
&\displaystyle \tilde v^{(1)}_\xi =  \frac{\gamma}{c_0^2} \tilde \eta^{(0)}, \quad
  \tilde V^{(1)}_\xi = 0,  \label{AS10}&\\
&\displaystyle   \hat s^{(1)}_t = \gamma B^{(0)} - A^{(0)}_{\tau}, \quad  \hat v^{(1)}_t= - \gamma A^{(0)} - B^{(0)}_{\tau}, \quad  \hat V^{(1)}_t = - \gamma F^{(0)} - C^{(0)}_\tau.
   \label{AS11}&
\end{eqnarray}

Equations (\ref{AS9}) imply 
\begin{equation}
\tilde \eta^{(0)}_{\tau} = 0, \quad \tilde s^{(1)} = \frac{1}{c_0} \tilde \eta^{(1)}, 
\label{AS12}
\end{equation}
again using that all deviations have zero mean.

To avoid secular growth in $\hat s^{(1)}$ and $\hat v^{(1)}$, we require that 
\begin{equation}
A^{(0)}_{\tau} = \gamma B^{(0)}, \quad B^{(0)}_{\tau} = - \gamma A^{(0)},
\label{AS13}
\end{equation}
which implies 
%\begin{equation}
$A^{(0)}_{\tau \tau} = - \gamma^2 A^{(0)}, $
%\label{AS14}
%\end{equation}
yielding
\begin{equation}
A^{(0)} = A(T) \cos \gamma [\tau + \phi(T)], \quad B^{(0)} = -A(T) \sin \gamma [\tau + \phi(T)],
\label{AS15}
\end{equation}
where $A(T)$ and {\color{black} $\phi(T)$} are arbitrary functions of $T$.

To avoid secular growth in $\hat V^{(1)}$ one can consider two options:
\begin{enumerate}
\item $F^{(0)} = F^{(0)}(\tau, T)$, then $C^{(0)}_{\tau} = -\gamma F^{(0)}$, implying
\begin{equation}
\hat V^{(0)} = C^{(0)} = -\gamma \int_0^{\tau} F^{(0)}(\tilde \tau, T) d \tilde \tau,
\label{AS16}
\end{equation}
{\color{black} assuming that $F^{(0)}$ is such that the $C^{(0)}$ is bounded};
\item $F^{(0)} = F^{(0)}(t, \tau, T)$ such that $\hat V^{(1)} = - \int_0^t [\gamma F^{(0)} (t, \tau, T) + C^{(0)}_{\tau}(\tau, T)] d \tilde t$ is bounded, e.g.\ ${\color{black} \gamma} F^{(0)} = - C^{(0)}_{\tau} + \sin \omega t, \ \mbox{where} \  \omega = \omega(\tau, T)$. 
\end{enumerate}
{\color{black} If $F^{(0)} \ne 0$, we can assume that $F^{(1)} = 0$, etc.}

In what follows, we consider the first case. Then, summarising, we have
\begin{eqnarray}
 &\displaystyle \hat s^{(0)} = A(T) \cos \gamma [\tau + \phi(T)], \quad  \hat v^{(0)} = - A(T) \sin \gamma [\tau + \phi(T)], \label{AS17} &\\
  &\displaystyle \hat V^{(0)} =  -\gamma \int_0^{\tau} F^{(0)}(\tilde \tau, T) d \tilde \tau , \ \
  \tilde s^{(0)} = \frac{1}{c_0} \tilde \eta^{(0)}(\xi, T), \ \
  \tilde v^{(0)} = 0, \ \   \tilde V^{(0)} = 0; 
  \label{AS18} &\\  
   &\displaystyle \hat s^{(1)} = A^{(1)}(\tau, T), \quad  \hat v^{(1)} = B^{(1)}(\tau, T), \quad  \hat V^{(1)} = C^{(1)}(\tau, T), \quad   \tilde s^{(1)} = \frac{1}{c_0} \tilde \eta^{(1)}, \label{AS19} &\\
 &\displaystyle \tilde v^{(1)} = \frac{\gamma}{c_0^2} \left (\int_{-L}^\xi \tilde \eta^{(0)}(\tilde \xi, T) d \tilde \xi - \left\langle \int_{-L}^\xi \tilde \eta^{(0)}(\tilde \xi, T) d \tilde \xi \right\rangle \right ), \quad   \tilde V^{(1)} = 0.
  \label{AS20}&
 \end{eqnarray}
Here, $\left\langle\int_{-L}^\xi \tilde \eta^{(0)}(\tilde \xi, T) d \tilde \xi \right\rangle = \frac{1}{2L}  \int_{-L}^{L} \left ( \int_{-L}^\xi \tilde \eta^{(0)}(\tilde \xi, T) d \tilde \xi \right ) d\xi$ is the mean value.

Finally, collecting the terms at $O(\alpha)$, using (\ref{AS17})--(\ref{AS20}) and simplifying the resulting system by virtue of its averaged members, we obtain
\begin{eqnarray}
&&-c_0 \tilde \eta^{(2)}_{\xi} + c_0^2 \tilde s^{(2)}_{\xi} = - \frac{\sigma}{c_0} (\hat \eta^{(0)} +\tilde \eta^{(0)})  \tilde \eta^{(0)}_\xi  \nonumber \\
&&\quad-\sigma \left ( A(T) \cos \gamma [\tau + \phi(T)] + \frac{1}{c_0} \tilde \eta^{(0)} \right ) \tilde \eta^{(0)}_\xi  - 
F^{(0)} \tilde \eta^{(0)}_\xi  - \tilde \eta^{(1)}_\tau - \tilde \eta^{(0)}_T, 
\label{AS21} \\
&&- c_0 \tilde s^{(2)}_{\xi}  + \tilde \eta^{(2)}_{\xi} =  \frac{{\color{black} \gamma^2}}{c_0^2} \left (\int_{-L}^\xi \tilde \eta^{(0)}(\tilde \xi, T) d \tilde \xi - \left\langle\int_{-L}^\xi \tilde \eta^{(0)}(\tilde \xi, T) d \tilde \xi \right\rangle \right ) -  
 \frac{\beta c_0^2}{3} \tilde \eta^{(0)}_{\xi \xi \xi}  \nonumber \\
 &&\quad- {\color{black} \frac{\sigma}{c_0}} \left (A(T) \cos \gamma [\tau + \phi(T)] + \frac{1}{c_0} \tilde \eta^{(0)}\right ) \tilde \eta^{(0)}_\xi - 
 \frac{1}{c_0} (F^{(0)} \tilde \eta^{(0)}_\xi {\color{black}  +} \tilde \eta^{(1)}_\tau {\color{black} +} \tilde \eta^{(0)}_T),
\label{AS22} \\
&&{\color{black} \tilde v^{(2)}_\xi = \frac{\gamma}{c_0^2} \tilde \eta^{(1)},}
%- \frac{1}{c_0} \left ( A_T \sin \gamma [\tau + \phi(T)] + \gamma A \phi_T \cos \gamma [\tau + \phi(T)] \right ), 
\label{AS23} \\
&&{\color{black} \tilde V^{(2)}_\xi = - A(T) \sin \gamma [\tau + \phi(T)] \tilde \eta^{(0)}_\xi,}
%- \frac{\gamma}{c_0} \int_0^{\tau} F^{(0)}_T(\tilde \tau, T) d \tilde \tau, 
\label{AS24} \\
 &&  \hat s^{(2)}_t = \gamma B^{(1)} - A^{(1)}_{\tau} - A_T \cos \gamma [\tau + \phi(T)] + \gamma A \phi_T \sin \gamma [\tau + \phi(T)],
\label{AS25} \\   
 && \hat v^{(2)}_t= - \gamma A^{(1)} - B^{(1)}_{\tau} + A_T \sin \gamma [\tau + \phi(T)] + \gamma A \phi_T \cos \gamma [\tau + \phi(T)],
\label{AS26} \\    
 && \hat V^{(2)}_t = - \gamma F^{(1)} - C^{(1)}_\tau + \gamma  \int_0^{\tau} F^{(0)}_T(\tilde \tau, T) d \tilde \tau.
   \label{AS27}
\end{eqnarray}

Equations (\ref{AS21}) and (\ref{AS22}) yield the equation
\begin{eqnarray}
&&- 2 \tilde \eta^{(1)}_\tau  = 2 \tilde \eta^{(0)}_T + \left [\frac{\sigma \hat \eta^{(0)}}{c_0} + 2 \sigma A \cos \gamma (\tau + \phi) + 2 F^{(0)}(\tau, T) \right ] \tilde \eta^{(0)}_\xi + \frac{3 \sigma}{c_0} \tilde \eta^{(0)} \tilde \eta^{(0)}_\xi    \nonumber \\
&&\quad+\frac{\beta c_0^3}{3} \tilde \eta^{(0)}_{\xi \xi \xi} - \frac{\gamma^2}{c_0} \left ( \int_{-L}^\xi \tilde \eta^{(0)}(\tilde \xi, T) d \tilde \xi - <\int_{-L}^\xi \tilde \eta^{(0)}(\tilde \xi, T)d \tilde \xi> \right ).
\label{AS28}
\end{eqnarray}

{\color{black} Integrating equation $(\ref{AS28})$ with respect to $\tau$ and avoiding secular growth with $\tau$, we obtain
\begin{eqnarray}
&&2 \tilde \eta^{(0)}_T + \frac{\sigma \hat \eta^{(0)}}{c_0}  \tilde \eta^{(0)}_\xi + \frac{3 \sigma}{c_0} \tilde \eta^{(0)} \tilde \eta^{(0)}_\xi +   \frac{ \beta c_0^3}{3} \tilde \eta^{(0)}_{\xi \xi \xi}  \nonumber \\
&&\quad-\frac{\gamma^2}{c_0} \left (\int_{-L}^\xi \tilde \eta^{(0)}(\tilde \xi, T) d \tilde \xi - \left\langle\int_{-L}^\xi \tilde \eta^{(0)}(\tilde \xi, T)d \tilde \xi\right\rangle \right )  = 0
\label{AS31}
\end{eqnarray}
and 
\begin{equation}
\tilde \eta^{(1)} = - \left [\frac{\sigma A}{\gamma}  \sin \gamma (\tau + \phi) + \int_0^{\tau} F^{(0)}(\tilde \tau, T) d\tilde \tau \right] \tilde \eta_\xi^{(0)},
\label{AS3}
\end{equation}
where we did not add an arbitrary zero mean field function since this just redefines $\tilde \eta ^{(0)} (\xi, T)$.

Differentiating equation (\ref{AS31}) with respect to $\xi$ 
%leads to the traditional form of
%\begin{eqnarray}
%\left (2 \tilde \eta^{(0)}_T + \frac{\sigma \hat \eta^{(0)}}{c_0}  \tilde \eta^{(0)}_\xi + \frac{3 \sigma}{c_0} \tilde \eta^{(0)} \tilde \eta^{(0)}_\xi +   \frac{\hat \beta c_0^3}{3} \tilde \eta^{(0)}_{\xi \xi \xi} \right)_\xi =  \frac{\gamma^2}{2c_0} \tilde \eta^{(0)}.
%\label{AS32}
%\end{eqnarray}
and changing the variable $\xi$ to $\zeta = \xi - \frac{\sigma \hat \eta^{(0)}}{2c_0} T$ leads to the traditional form of the Ostrovsky equation:
\begin{eqnarray}
\left (\tilde \eta^{(0)}_T +  \alpha_1 \tilde \eta^{(0)} \tilde \eta^{(0)}_\zeta +  \beta_1  \tilde \eta^{(0)}_{\zeta \zeta \zeta} \right)_\zeta =  \gamma_1 \tilde \eta^{(0)},   \label{AS33} 
\end{eqnarray}
where
\begin{eqnarray}
\alpha_1 =  \frac{3 \sigma}{2 c_0}, \quad \beta_1 =   \frac{ \beta c_0^3}{6}, \quad \gamma_1 =  \frac{\gamma^2}{2 c_0}. \nonumber
\end{eqnarray}

Next, substituting equation (\ref{AS3}) for $\tilde \eta^{(1)}$ into equation (\ref{AS23}) for $\tilde v^{(2)}_{\xi}$ 
%and requiring that there is no secular growth in $\tilde v^{(2)}$, and that the function has zero mean, 
we conclude that
\begin{equation}
%A_T = 0, \quad \phi_T = 0, \quad 
\tilde v^{(2)} = - \frac{1}{c_0^2}  \left [\sigma A \sin \gamma (\tau + \phi) + \gamma \int_0^\tau F^{(0)}(\tilde \tau, T) d \tilde \tau \right] \tilde \eta^{(0)},
\label{AS34}
\end{equation}
using the zero-mean condition once again.
Similarly, equation (\ref{AS24}) yields 
\begin{equation}
%F^{(0)}_T = 0, \quad \mbox{and} \quad \tilde 
V^{(2)} = - A \sin \gamma (\tau + \phi) \tilde \eta^{(0)}.
\label{AS35}
\end{equation}

Finally, the equations for the mean values yield  $A_T = \phi_T = 0$ in order to avoid secular growth, and then simplify to take the form 
\begin{equation}
\hat s^{(2)}_t = \gamma B^{(1)} -  A^{(1)}_{\tau}, \quad \hat v^{(2)}_t = -\gamma A^{(1)} -  B^{(1)}_{\tau}, \quad 
\hat V^{(2)}_t = -\gamma F^{(1)} -  C^{(1)}_{\tau} + \gamma \int_0^{\tau} F_T^{(0)} (\tilde \tau, T) d \tilde \tau.
\label{AS36}
\end{equation}
Assuming that $F^{(1)} = F^{(1)}(\tau, T)$, we then have
\begin{equation}
A^{(1)}_{\tau} = \gamma B^{(1)}, \quad  B^{(1)}_{\tau} = -\gamma A^{(1)}, \quad C^{(1)}_{\tau} = - \gamma F^{(1)}+ \gamma \int_0^{\tau} F_T^{(0)} (\tilde \tau, T) d \tilde \tau.
\label{AS37}
\end{equation}
In what follows, we consider $(F, A, B)^{(1)} = 0$, while
\begin{equation}
C^{(1)} = \gamma \int_0^{\tau} \left ( \int_0^{\hat \tau} F_T^{(0)} (\tilde \tau, T) d \tilde \tau \right ) d \hat \tau,
\end{equation}
provided this function is bounded. A sufficient condition for the latter is given by $F_T^{(0)} = 0$, implying that $F^{(0)} = F^{(0)} (\tau)$, which we assume here.}

Hence, considering the equations up to $O(\alpha)$, i.e.\ up to the accuracy of the governing equations (\ref{7})--(\ref{12a}), allows us to fully define all terms at $O(1)$ and $O(\sqrt{\alpha})$. The procedure can be continued to any order, but instead of using the truncated  weakly-nonlinear formulation, we would need to use the original strongly-nonlinear equations.

To summarise, up to the accuracy of the governing equations, we obtained the following large class of uni-directional waves described by the Ostrovsky equation and propagating over the non-zero, generally evolving, mean fields:
\begin{eqnarray}
&&\eta = \hat \eta^{(0)} + \tilde \eta^{(0)} - \sqrt{\alpha} \left [\frac{\sigma A}{\gamma} \sin \gamma (\tau + \phi) + \int_0^\tau F^{(0)}(\tilde \tau) d \tilde \tau \right] \tilde \eta^{(0)}_{\zeta} + O(\alpha), 
\label{AS38} \\
&&s = A \cos \gamma (\tau + \phi) + \frac{1}{c_0}  \tilde \eta^{(0)}  \nonumber \\
&& \qquad-\sqrt{\alpha}  \frac{1}{c_0} \left [\frac{\sigma A}{\gamma} \sin \gamma (\tau + \phi) +  \int_0^\tau F^{(0)}(\tilde \tau) d \tilde \tau \right ] \tilde \eta^{(0)}_{\zeta} + O(\alpha), 
\label{AS39} \\
&&v = - A \sin \gamma (\tau + \phi)  \nonumber \\
 &&\qquad+\sqrt{\alpha} \frac{\gamma}{c_0^2} \left [\int_{-L}^\zeta \tilde \eta^{(0)}(\tilde \zeta, T) d \tilde \zeta - 
\left\langle\int_{-L}^\zeta \tilde \eta^{(0)}(\tilde \zeta, T) d \tilde \zeta\right\rangle\right ] + O(\alpha),
\label{AS40}\\
&&V = -\gamma \int_0^{\tau} F^{(0)}(\tilde \tau) d \tilde \tau + O(\alpha),
\label{AS41}
\end{eqnarray}
where $\hat \eta^{(0)},\, A,\, \phi$ are arbitrary constants, $F^{(0)}(\tilde \tau)$ is a function such that $\int_0^\tau F^{(0)}(\tilde \tau) d \tilde \tau$ is bounded (e.g.\ $\sin \omega \tilde \tau$), and $\tilde \eta^{(0)} = \tilde \eta^{(0)}(\zeta, T)$ satisfies the Ostrovsky equation (\ref{AS33}), where 
\begin{equation}
\zeta = \xi - \frac{\sigma \hat \eta^{(0)}}{2 c_0} T = x - \left (c_0 + \frac{\alpha \sigma \hat \eta^{(0)}}{2 c_0}\right ) t, \quad T = \alpha t.
\label{AS42}
\end{equation}

{\color{black}  
We note that this weakly-nonlinear solution can be rewritten in a more convenient and asymptotically equivalent form:
\begin{eqnarray}
&&\eta = \hat \eta^{(0)} + \tilde \eta^{(0)}\left (T, \theta \right ) + O(\alpha), 
\label{F1} \\
&&s = A \cos \gamma (\tau + \phi) + \frac{1}{c_0}  \tilde  \eta^{(0)}\left (T, \theta \right ) + O(\alpha),   
\label{F2} \\
&&v = - A \sin \gamma (\tau + \phi)  \nonumber \\
 &&\qquad+\sqrt{\alpha} \frac{\gamma}{c_0^2} \left [\int_{-L}^\theta \tilde \eta^{(0)}(\tilde \zeta, T) d \tilde \zeta - 
\left\langle\int_{-L}^\theta \tilde \eta^{(0)}(\tilde \zeta, T) d \tilde \zeta\right\rangle\right ] + O(\alpha),
\label{F3}\\
&&V = -\gamma \int_0^{\tau} F^{(0)}(\tilde \tau) d \tilde \tau + O(\alpha),
\label{F4}
\end{eqnarray}
where
\begin{equation}
\theta = \zeta - \frac{\sqrt{\alpha} \sigma A}{\gamma} \sin \gamma (\tau + \phi) - \sqrt{\alpha} \int_0^\tau F^{(0)}(\tilde \tau) d \tilde \tau,
\label{F4}
\end{equation}
and
\begin{eqnarray}
\left (\tilde \eta^{(0)}_T +  \alpha_1 \tilde \eta^{(0)} \tilde \eta^{(0)}_\theta +  \beta_1  \tilde \eta^{(0)}_{\theta \theta \theta} \right)_\theta =  \gamma_1 \tilde \eta^{(0)},   \label{Oe} 
\end{eqnarray}
with
$
\alpha_1 =  \frac{3 \sigma}{2 c_0}, \quad \beta_1 =   \frac{ \beta c_0^3}{6}, \quad \gamma_1 =  \frac{\gamma^2}{2 c_0}. 
$
Note that we should not add any additional transport terms to the equation (\ref{Oe}) since we simply replace $\zeta$ with $\theta$ in the solution of (\ref{AS33}) in order to combine the $O(1)$ and $O(\sqrt{\alpha})$ terms in the asymptotic expansions.   This representation, where the effective phase $\theta$ now depends on $\zeta$ and $\tau$, gives a clear description of the main effects related to nonzero mass: generally there appear $O(1)$ inertial oscillations in both shear variables, as well as $O(\sqrt{\alpha})$ oscillations with the same frequency in the phase of the interfacial displacement and shear in the propagation direction. Nonzero barotropic transport  $F^{(0)}$ in the propagation direction also gives $O(\sqrt{\alpha})$ shift in the phase of the interfacial displacement and shear in the propagation direction, as well as giving $O(1)$ contribution to the transverse barotropic transport.

Thus, the solution $\tilde \eta^{(0)}$ of the Ostrovsky equation (\ref{Oe})  has zero mean by construction, $\eta$ has an arbitrary constant mean value, while the shear variables $s$ and $v$ generally have time-dependent mean values. The barotropic transport variables $F$ and $V$ are also generally time-dependent. 

We finish the section by considering the limit of the constructed weakly-nonlinear solution (\ref{F1}) - (\ref{Oe}) to the case when there is no rotation. Assuming $\phi = 0$, consider a finite value of $\tau$ and let $\gamma \to 0$. Then,
\begin{eqnarray}
&&\eta = \hat \eta^{(0)} + \tilde \eta^{(0)}\left (T, \theta \right ) + O(\alpha), 
\label{K1} \\
&&s = A  + \frac{1}{c_0}  \tilde  \eta^{(0)}\left (T, \theta \right ) + O(\alpha),   
\label{K2} \\
&&v =  O(\alpha), \quad 
V =  O(\alpha),
\label{K3}
\end{eqnarray}
where
\begin{eqnarray}
\theta &=& \zeta - \sigma A T - \sqrt{\alpha} \int_0^\tau F^{(0)}(\tilde \tau) d \tilde \tau 
  %        &=& x - \left ( c_0 + \frac{\alpha \sigma \hat \eta^{(0)}}{2 c_0} + \alpha \sigma A \right ) t - \sqrt{\alpha} \int_0^\tau F^{(0)}(\tilde \tau) d \tilde \tau
\end{eqnarray}
and
\begin{eqnarray}
\tilde \eta^{(0)}_T +  \alpha_1 \tilde \eta^{(0)} \tilde \eta^{(0)}_\theta +  \beta_1  \tilde \eta^{(0)}_{\theta \theta \theta} = 0   \label{KdV} 
\end{eqnarray}
with 
$
\alpha_1 =  \frac{3 \sigma}{2 c_0}, \quad \beta_1 =   \frac{ \beta c_0^3}{6}. 
$

In the rest of this paper we consider the case $F^{(0)} = 0$ (i.e., the barotropic transport is absent).
A particular choice $A = \frac{\hat \eta^{(0)}}{c_0}$ corresponds to the case when there is no background shear flow, and then
\begin{equation}
\theta =  \xi -  \alpha_1 \hat \eta^{(0)} T = x  - \left ( c_0 + \frac{3 \alpha  \sigma \hat \eta^{(0)}}{2 c_0} \right ) t.
\end{equation}
In general, $A = \frac{\hat \eta^{(0)}}{c_0} + s_0$, where $s_0 =  {\rm const}$ represents a constant background shear, and then we have
\begin{equation}
\theta =  \xi -  (\alpha_1 \hat \eta^{(0)} + \sigma s_0)  T = x  - \left ( c_0 + \frac{3 \alpha  \sigma \hat \eta^{(0)}}{2 c_0} + \alpha \sigma s_0 \right ) t.
\end{equation}
  Suppose one allows for the presence of a weak piecewise-constant shear flow in the two layers and assume that $s_0 > 0$. In the absence of rotation, the sufficient conditions allowing one  to avoid the appearance of the long-wave instability and critical levels are given, in non-dimensional variables used in this paper, by the conditions
\begin{equation}
s_0 <  \left (\frac{\rho_1 h_0 + \rho_2 (1 - h_0)}{ \rho_2} \right ) ^{1/2} \quad \mbox{and} \quad s_0 < c_0
\label{shear}
\end{equation}
(see \citet{O1979,O1985,B1991, BM2011,LM2015,KZ2016} for the details and the necessary conditions). 
In the following sections we consider the basic case $s_0 = 0$.

The constructed weakly-nonlinear solution can be used to describe waves of small amplitude. The conservation of energy to leading order was verified using the asymptotic approximation of \citet{H2007}. Fully nonlinear waves in the absence of rotation are significantly different from the weakly-nonlinear description provided by the KdV equation (see \citealt{CCMRS2006, C2006, CRST2010, JC2002, JC2008, BCM2020, DBM2022} and references therein). The extended   weakly-nonlinear models can be used to describe MMCC waves of greater amplitude than that described by the KdV equation \citep{STCK2025}, paving the way for similar extensions in the presence of rotation.

}

\section{The effect of rotation on the evolution of cnoidal waves with defects}
\label{sec: 4}

{\color{black} The MMCC system admits periodic nonlinear travelling wave solutions \citep{CRST2010}  (see also \citealt{STCK2025} for the weakly-nonlinear modelling of the cnoidal waves of small and moderate amplitude). } The aim of this section is to model the {\color{black} effect of rotation on the } evolution of the waves generated by initial conditions close to the cnoidal waves of the KdV equation but having local amplitude and/or periodicity defects. We use the weakly-nonlinear solution constructed in Section \ref{sec: 3} and first model the effect of rotation on the exact classical solutions of the KdV equation: solitons and cnoidal waves \citep{B1871,KdV1895}, as well as bright and dark breathers \citep{KM1975, HMP2023}. {\color{black} Then, we consider cnoidal waves with the local expansion and contraction periodicity defects introduced  by cutting at the trough and symmetrically inserting a piece of a straight line, and by symmetrically cutting away a part close to the trough between the two neighbouring peaks and gluing together the remaining parts of the solution, respectively (see the first two rows in Figure \ref{fig:WS} of the Appendix A).  This allows us to set up a framework for the discussion of the effects observed in our subsequent modelling of cnoidal waves with generic localised perturbations. As a by-product of our study of initial conditions with local periodicity defects we came across an observation that cnoidal waves with expansion defects can be viewed as generalised (`shock-like') travelling waves of the KdV equation. In Appendix \ref{sec:Appendix A}, we prove that all conservation laws of the KdV equation (inifinitely many) are identically satisfied for them, provided that we understand the conserved quantities as the natural sum of integrals, treating the points of discontinuity similarly to shocks.  Moreover, a cnoidal wave with an expansion defect has a continuous first derivative, satisfying the Weirstrass-Erdmann corner condition for the associated variational problem. We also make comments about a possible weak formulation.
}

In our numerical experiments, we use two sets of parameters. 
In the first set, which is used for the majority of our simulations (except the runs shown in Figures \ref{fig:A30}, \ref{fig:A24} (right) and \ref{fig:A24.1} (right)), the pycnocline is closer to the surface, and internal waves are the waves of depression. Hence, we show the bottom view in the majority of our figures in this section.
% of parameters of the two-layer fluid system is defined as follows: 
The depths of the upper and lower layers are $37.5$ m and $112.5$ m, respectively, and the total depth of the system, denoted by $H$, is $150.0$ m. The densities of the upper and lower layers are chosen to be
$\rho_{1} = 1000.0 ,\, \rho_{2} = 1003.1$ kg m$^{-3}$, respectively. The small parameters are defined as $\alpha = a/H $, $\tilde \beta = (H/l)^2$ and $\tilde \gamma=(lf)/\sqrt{(g'H)}$, where $a$ is a wave amplitude, $l$ is a wavelength, $f$ is the Coriolis parameter, $g' = g\Delta\rho/\rho_1$ is the reduced gravity. 
We let $\alpha = 0.005$ and $\tilde \beta = 0.030$. We assume midlatitude oceanic values for the Coriolis parameter $f = 5\times10^{-5} \text{ s}^{-1}$, and reduced gravity $g'= 0.030 \text{ m s}^{-2}$. Hence, $\tilde \gamma \approx 0.020$. The non-dimensional unperturbed upper layer depth $h_0=0.250$ gives values of the linear long wave speed $c_0 = (h_0-h_0^{2})^{1/2} \approx 0.433$ and coefficient $\sigma = 2h_0-1 = -0.500$. The scaled $\mathcal{O}(1)$ parameters are $\beta = \tilde \beta/\alpha $ and $\gamma = \tilde \gamma/\sqrt{\alpha}$. Then, the coefficients of the Ostrovsky equation (\ref{AS33}) are given by $\alpha_1 \approx -1.732, \beta_1 \approx 0.081, \gamma_1 \approx 0.096$. This regime is close to one of the regimes considered by \citet{H2007}, and internal waves are waves of depression.
% The remained numerical values are $u_1 = -10^{-3}, u_2 = 0$ and $u_3 = 1$ for the cnoidal wave initial condition \eqref{eq: cn_1}.
The second set of parameters is used to model the internal waves of elevation shown in  Figures \ref{fig:A30}, \ref{fig:A24} (right) and \ref{fig:A24.1} (right), and here the pycnocline is closer to the bottom:
%is employed in section $[ref]$, where the depths of the two layers are reversed, resulting in the upper layer being deeper than the lower layer. 
the depths of the upper and lower layers are $120.0$ and $30.0$ m, respectively. The total depth of the system is again $H=150.0$ m. The other parameters are unchanged except the non-dimensional unperturbed upper layer depth $h_0=0.800$, giving values of the linear long wave speed $c_0 = (h_0-h_0^{2})^{1/2} = 0.400$ and coefficient $\sigma = 2h_0-1 = 0.600$. 
%The scaled $\mathcal{O}(1)$ parameters are $\beta = \tilde \beta/\alpha $ and $\gamma = \tilde \gamma/\sqrt{\alpha}$. 
Then, the coefficients of the Ostrovsky equation  (\ref{AS33})  are found to be $\alpha_1 = 2.250, \beta_1 = 0.064, \gamma_1 \approx 0.104$. The Ostrovsky equation is solved using a pseudospectral method described in Appendix \ref{sec:Appendix B}. {\color{black} The constant shear is absent, $s_0=0$,  and the initial phase of the inertial oscillations is $\phi =0$.}
% and we assume that the barotropic transport is absent, i.e. $F^{(0)}=0$.

%%% Subsection starts here %%%

\subsection{The effect of rotation on solitons and cnoidal waves}

\begin{figure}  
    \centering
    \includegraphics[width=0.51\linewidth]{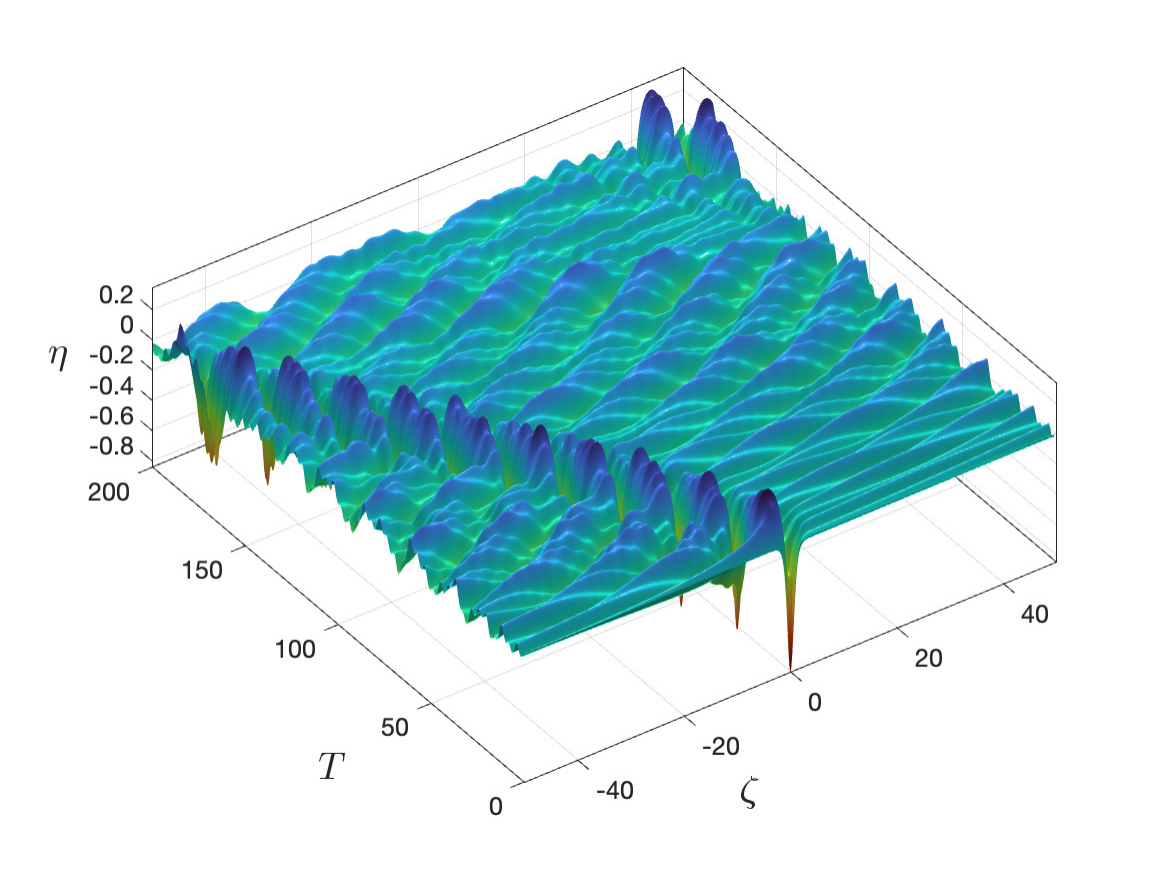}\hspace{-0.5cm}
    \includegraphics[width=0.51\linewidth]{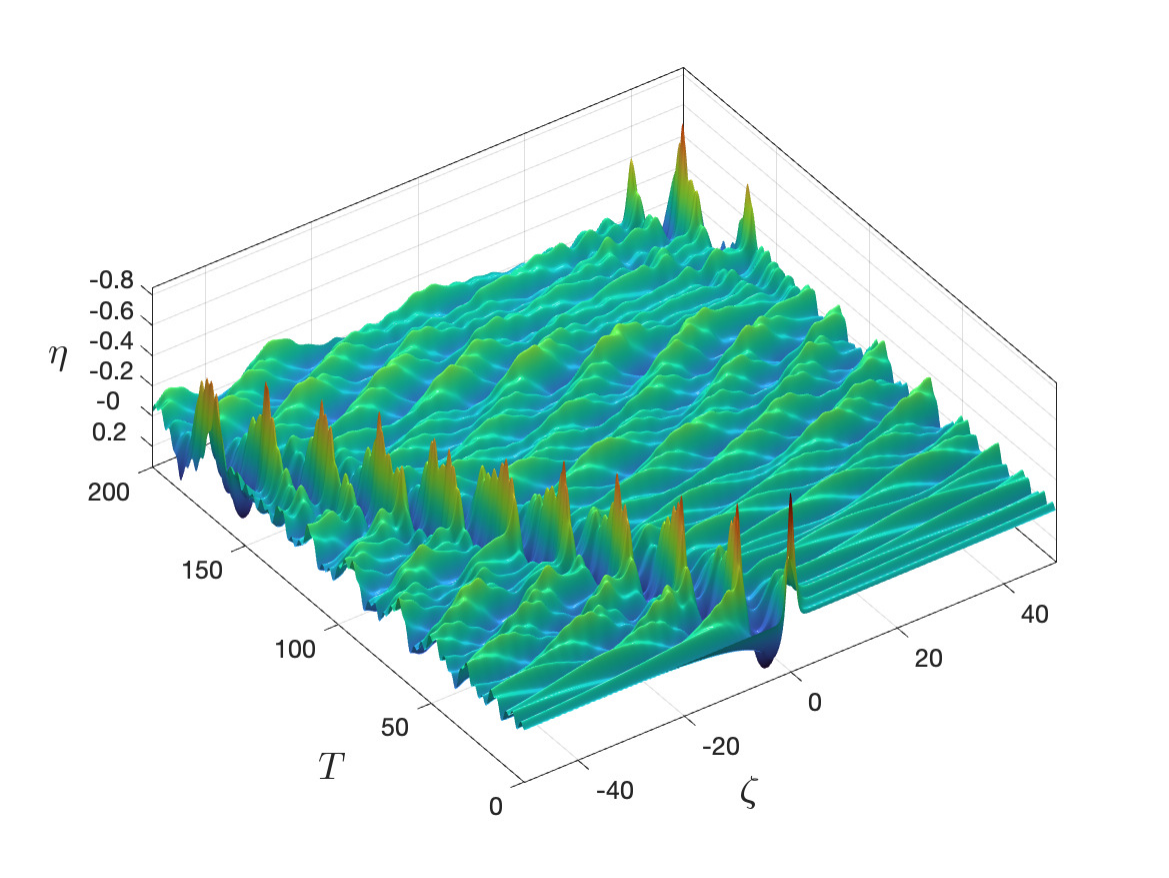} \\[-0.5cm]
        \includegraphics[width=0.49\linewidth]{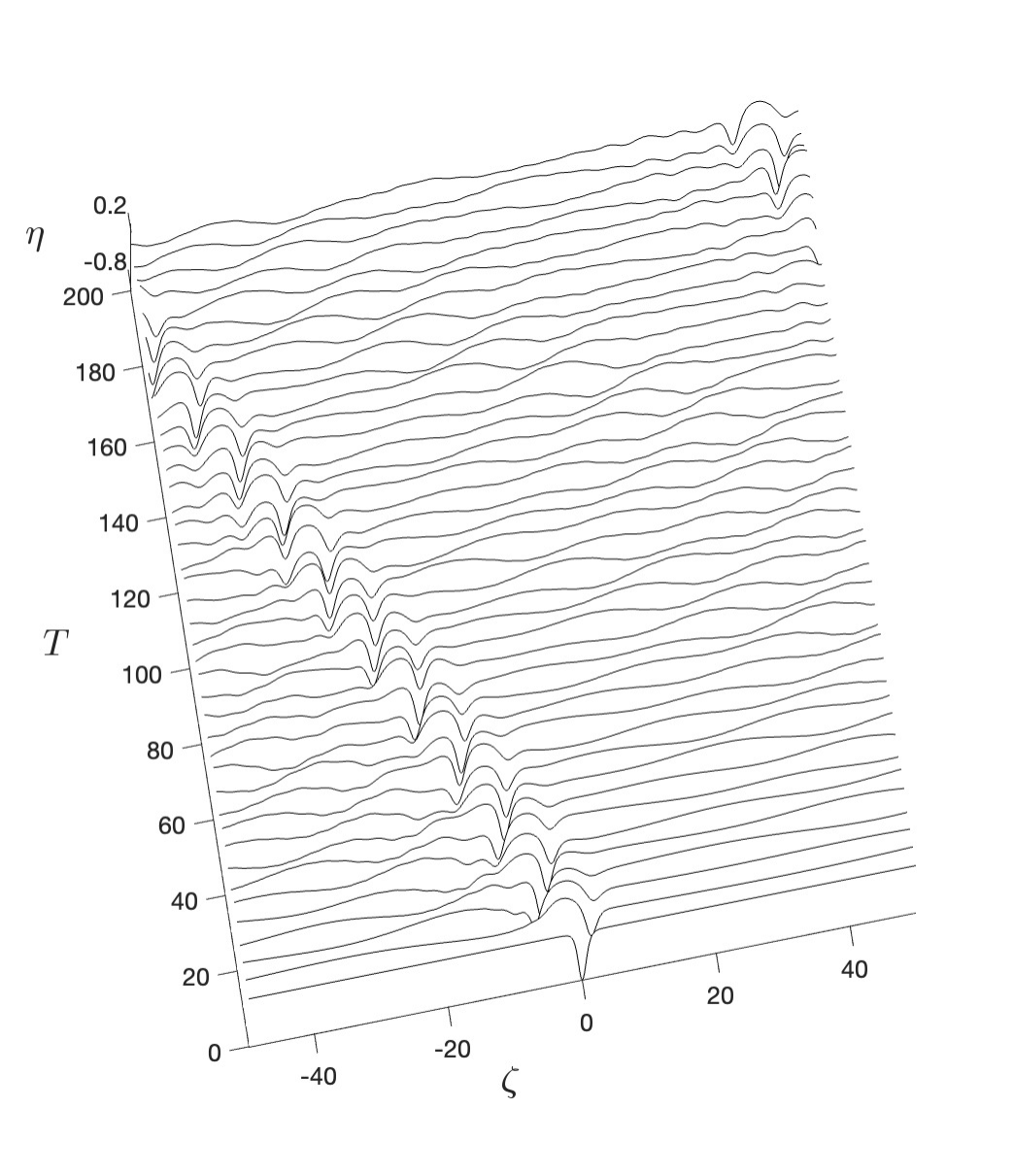}%\quad
        \includegraphics[width=0.49\linewidth]{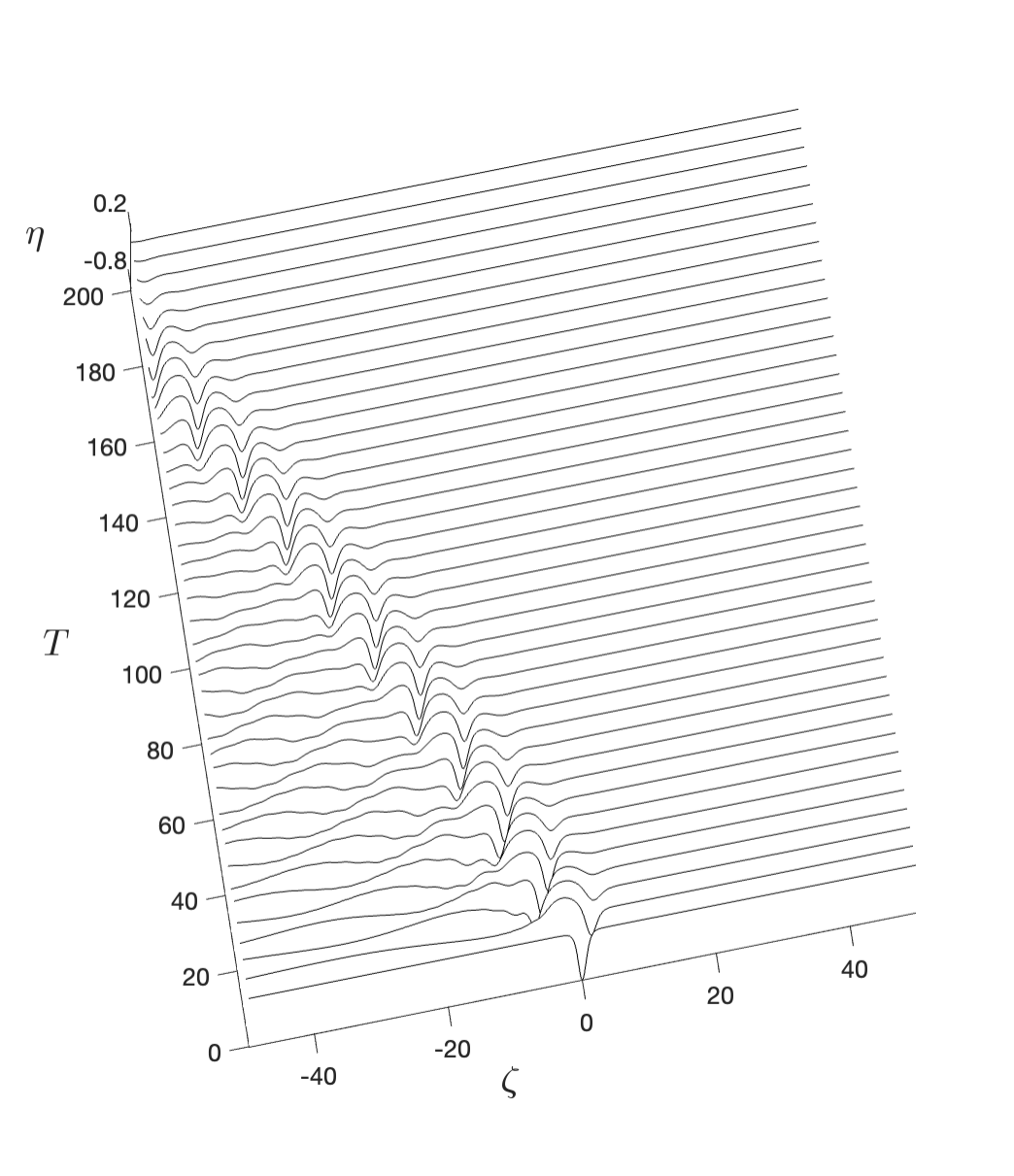}\\[2ex]
            \includegraphics[width=0.3\linewidth]{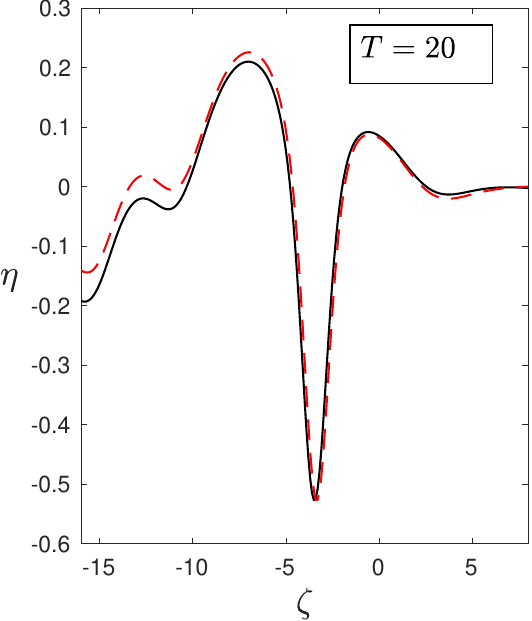} 
            \includegraphics[width=0.3\linewidth]{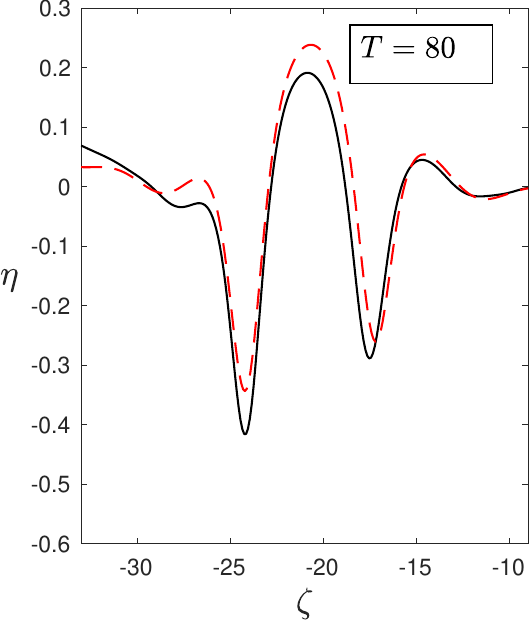}
            \includegraphics[width=0.3\linewidth]{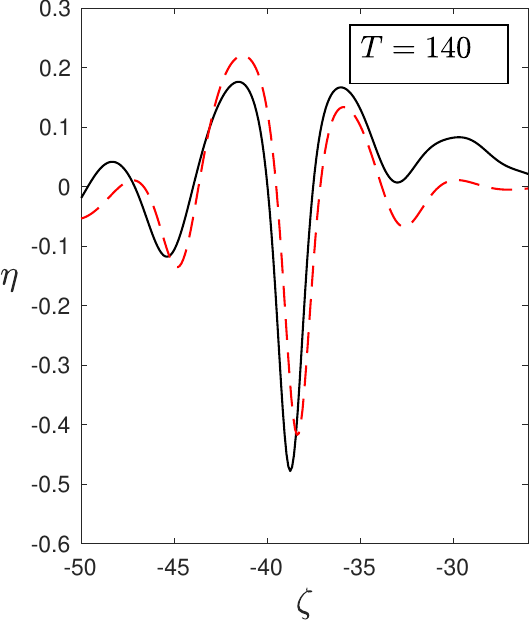}
        \caption{The effect of rotation on the KdV soliton initial condition  in simulations with periodic boundary conditions. First row: view from above (left) and view from below (right) of the interfacial displacement. Second row: interfacial displacement in simulations without the sponge layers (left) and with the sponge layers (right).
        Third row: comparison of the lead wavepacket in simulations  without the sponge layers (black) and with the sponge layers (red dashed). }
        \label{fig:A3}
\end{figure}

{\color{black} 
We begin by modelling the evolution of initial conditions in the form of solitary and cnoidal wave solutions of the KdV equation associated with the Ostrovsky equation (\ref{Oe}), which we write in the form
\begin{eqnarray}
\eta^{(0)}_T +  \alpha_1 (\eta^{(0)} - \hat \eta^{(0)})  \eta^{(0)}_\theta +  \beta_1   \eta^{(0)}_{\theta \theta \theta}  =  0,  \label{KdV} 
%\alpha_1 =  \frac{3 \sigma}{2 c_0}, \quad \beta_1 =   \frac{ \beta c_0^3}{6}. 
\end{eqnarray}
where $\eta^{(0)} = \hat \eta^{(0)} + \tilde \eta^{(0)}$.
%Hence, at $T=0$ the initial condition for the Ostrovsky equation (\ref{AS33}) corresponding to the soliton solution is given by the mean-free part of the function
We consider the initial condition for $\eta$ in the form of a soliton solution
\begin{align}
 \eta^{(0)} \Big |_{T=0}  = \dfrac{3 v_s}{\alpha_1} \text{sech}^2 \Big[ \dfrac{1}{2} \sqrt{\dfrac{v_s}{\beta_1}}(\theta + \alpha_1 \hat \eta^{(0)} T -v_s T-\zeta_0) \Big ]_{T=0}, \label{soliton} 
\end{align}
where $\zeta_0$ and $v_s>0$ are arbitrary constants, 
% while
%the initial condition for the Ostrovsky equation (\ref{AS33}) corresponding to 
and the cnoidal wave solution 
%of equation (\ref{KdV}) takes the form of the mean-free part of the solution given 
\begin{align}
    \eta^{(0)} = \dfrac{6\beta_1}{\alpha_1}\Bigg\{ u_2 + (u_3 - u_2) \text{ cn}^2\Big[ (\theta + \alpha_1 \hat \eta^{(0)} T  - v_c T - \zeta_0) \sqrt{\dfrac{u_3 -u_1}{2}} ; m \Big] \Bigg\}_{T=0}, \label{cnoidal}
\end{align}
given in terms of the Jacobi elliptic functions, where $u_1 < u_2 < u_3$ are real, $v_c = 2 \beta_1 (u_1 + u_2 + u_3)$, and
%$\alpha_1 = \frac{3\sigma}{2c_0}$, $\beta_1=\frac{ \beta c_0^3}{6}$ are the coefficients of the Ostrovsky equation \eqref{eq:ost}, 
$m ={(u_3-u_2)}/{(u_3-u_1)}$ is the elliptic modulus.  We recall that in the case under study
 \begin{equation}
 \theta = \zeta  - \frac{2 \alpha_1 \hat \eta^{(0}}{3} T
 %x - \left ( c_0 + \alpha \alpha_1 \hat \eta^{(0)} + \alpha \sigma s_0 \right ) t,
 \end{equation}
 and, accounting for the transport term in (\ref{KdV}), we get the phase
 \begin{equation}
  \theta + \alpha_1 \hat \eta^{(0)} T - v_{s/c} T - \zeta_0 = \zeta + \frac{\alpha_1 \hat \eta^{(0}}{3} T  - v_{s/c} T - \zeta_0,
% \theta + \alpha_1 \hat \eta^{(0)} T - v_{s/c} T - \zeta_0 = x - \left ( c_0 + \alpha \sigma s_0 \right ) t - v_{s/c} T - \zeta_0,
 \end{equation}
 choosing the variable $\zeta$ as a convenient variable defining the speed of the moving reference in both cases, with and without rotation. Here, $ \zeta + \frac{\alpha_1 \hat \eta^{(0}}{3} T = \xi = x - c_0 t$. }
 In the limit $m \to 1$, the cnoidal wave approaches a soliton (generally, on a non-zero pedestal).
%Hence, the cnoidal wave initial condition is written as
%\begin{align}
%    \tilde{\eta}^{(0)}|_{T=0} = \dfrac{6\beta_1}{\alpha_1}\Bigg\{ u_2 + (u_3 - u_2) \text{ cn}^2\Big[ \zeta\sqrt{\dfrac{u_3 -u_1}{2}} ; m \Big] \Bigg\}.
%    \label{eq: cn_1}
%\end{align}
The period of the cnoidal wave is given by 
\begin{align}
    \lambda = \dfrac{2K(m)}{\sqrt{(u_3 - u_1)/2}},
\end{align}
where $K(m)$ is the complete elliptic integral of the first kind (see, for example, \citealt{K2000}). {\color{black} We use the same initial conditions for $\eta^{(0)}$ when rotation is present, and obtain the corresponding weakly-nonlinear solutions using (\ref{F1}) - (\ref{F4}) and the Ostrovsky equation (\ref{Oe}). }

In Figure \ref{fig:A3}, we compare the evolution of a single soliton in a two-layer fluid system with rotation in numerical experiments with periodic boundary conditions, either with or without the sponge layers near the boundaries (see, for example, \citealt{AGK2013} for the discussion of the sponge layers). The sponge layers near the boundaries act as a filter absorbing the radiation. 
The computational domain is $2L = 100$, with the number of modes $M = 998$, the spatial step $\Delta x \approx 10^{-1}$, the total simulation time $T_{max} = 200$, and the temporal step $\Delta T = 10^{-2}$. Other constants are $v_s \approx 0.487$ and $\zeta_0 = 0$.

\begin{figure}
    \centering
    \includegraphics[width=0.51\linewidth]{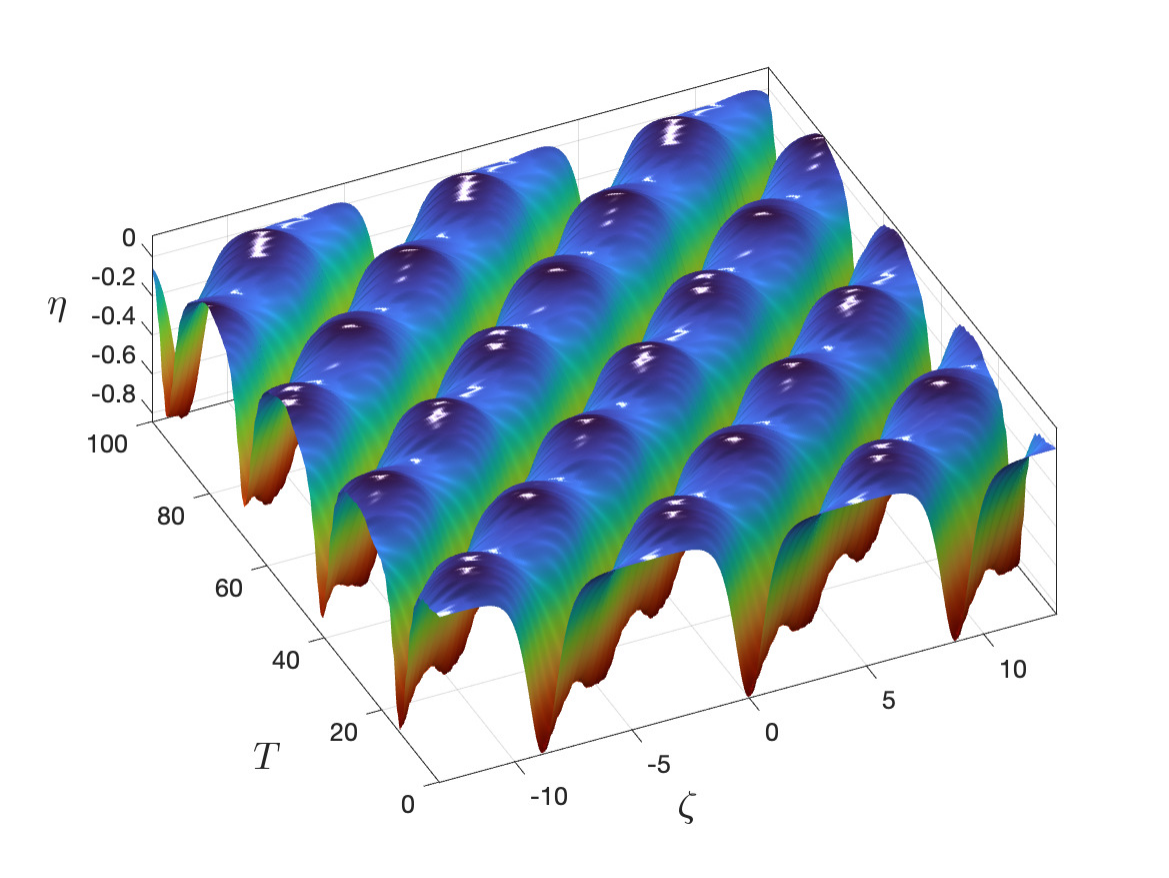}\hspace{-0.5cm}
    \includegraphics[width=0.51\linewidth]{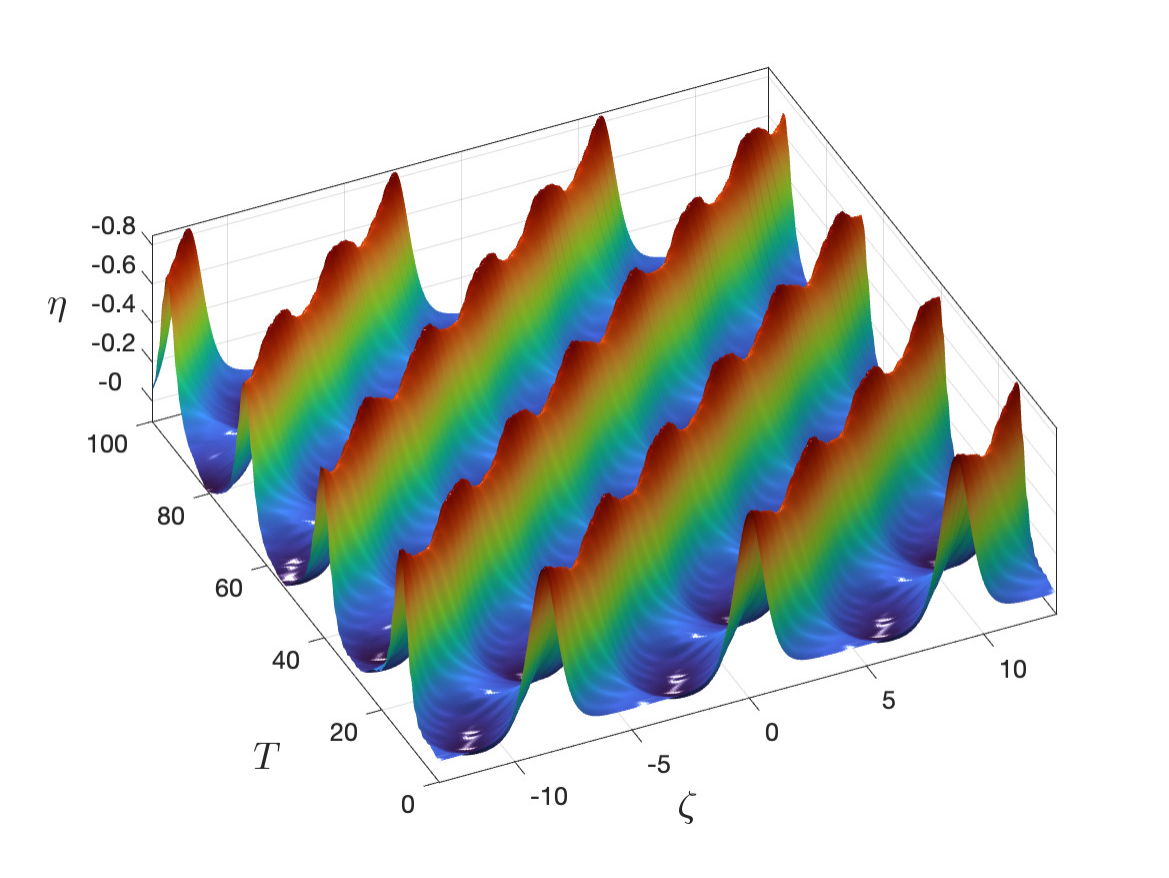}
    \caption{The effect of rotation on the KdV cnoidal wave initial condition: view from above (left) and view from below (right) of the interfacial displacement.
    }
    \label{fig:A7}       
    \end{figure}

Top row of Figure \ref{fig:A3} shows three-dimensional plots of the time evolution of the interfacial displacement for initial conditions in the form of a single soliton, without (left) and with  (right) the sponge layers near the boundaries of the periodic domain. The soliton evolves into a wavepacket as it propagates, and for a long time the radiation penetrating through the boundaries in the absence of sponge layers does not have a large effect on the shape or speed of the main wavepacket, which is shown in more detail in the middle and bottom rows of Figure \ref{fig:A3}. The overall wave pattern remains stable and consistent throughout the simulation.

       % \vspace{1cm}
       \begin{figure}
    \centering
    \includegraphics[width=0.47\linewidth]{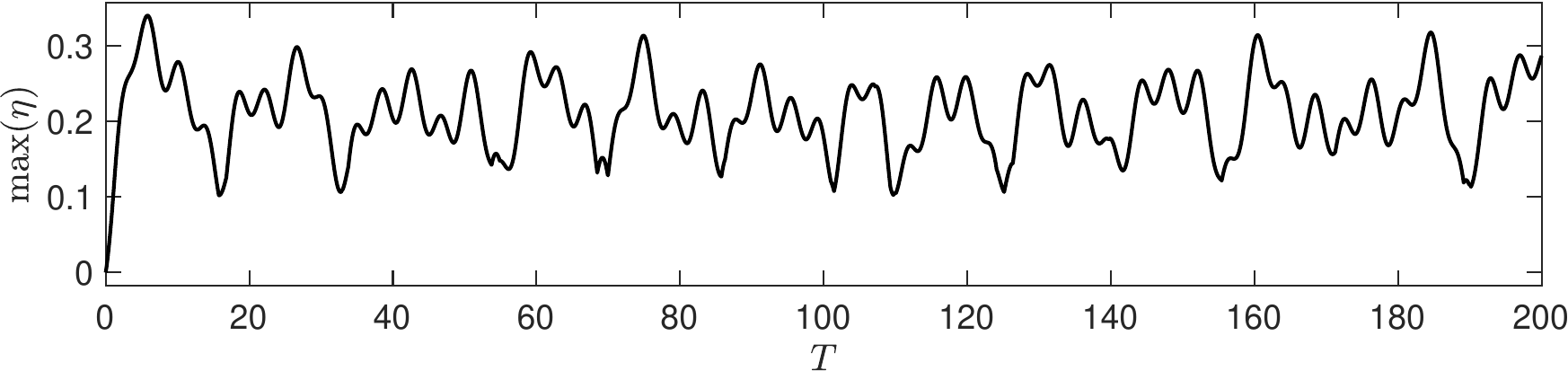}
    \includegraphics[width=0.47\linewidth]{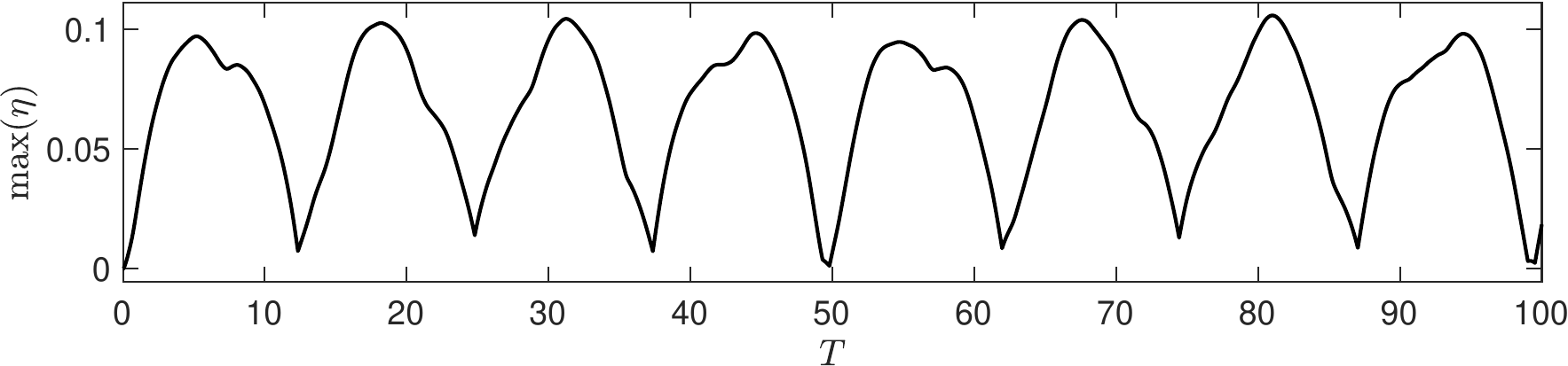}\\[0.5cm]
    \includegraphics[width=0.47\linewidth]{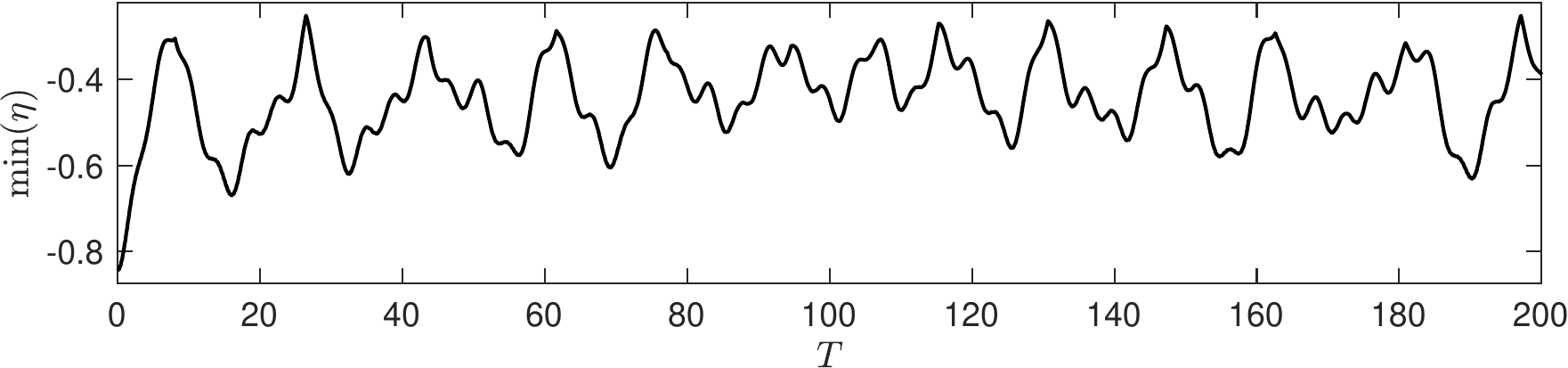}
    \includegraphics[width=0.47\linewidth]{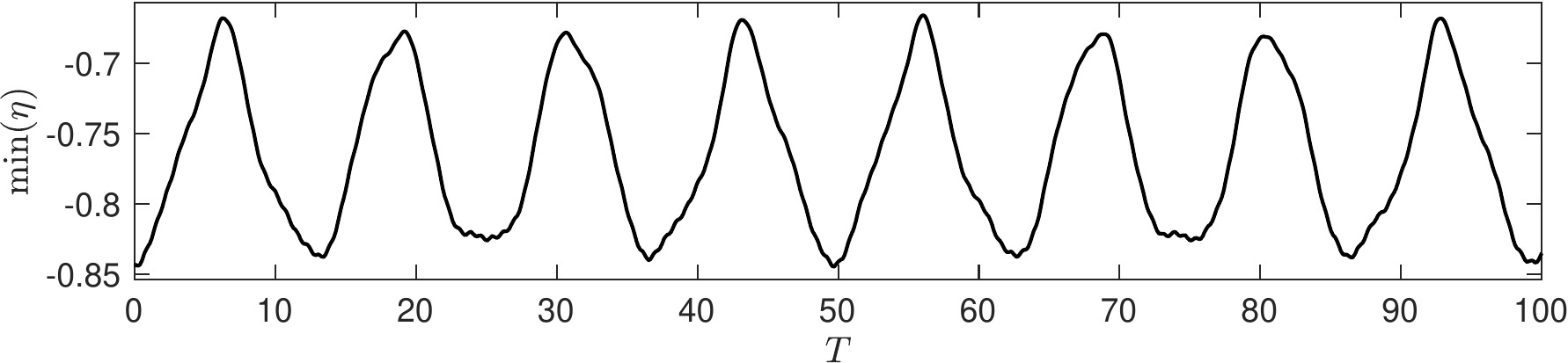}\\[0.5cm]
    \includegraphics[width=0.47\linewidth]{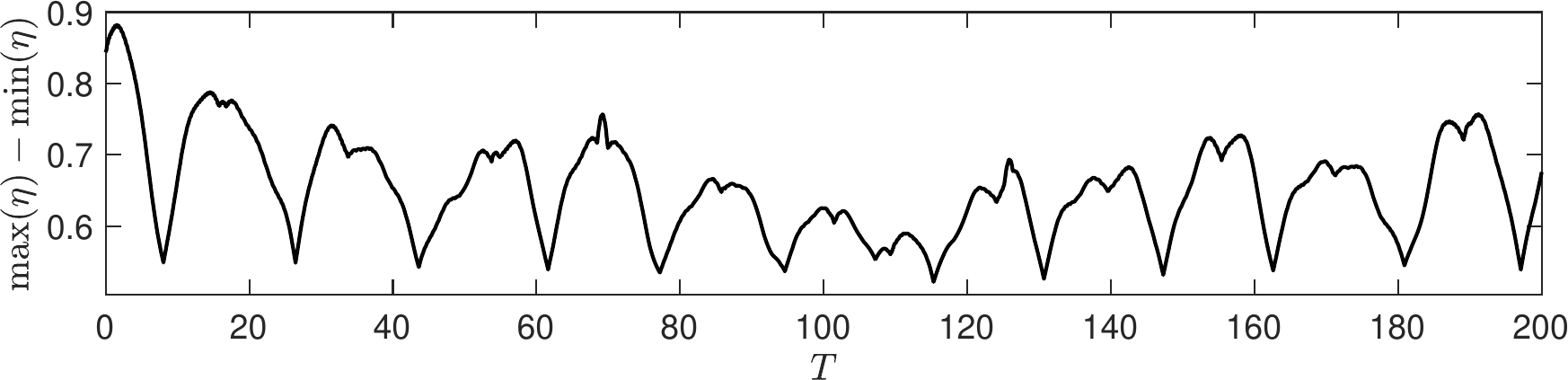}
    \includegraphics[width=0.47\linewidth]{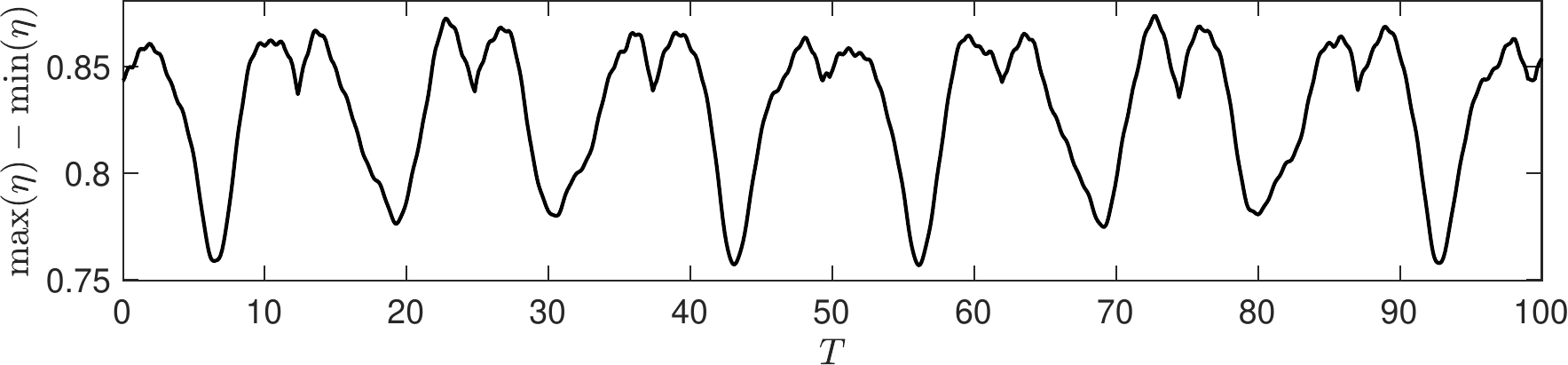}
    \caption{
    %Maximum and minimum plots of interface $\eta$ over time $T$ initiated by a cnoidal wave initial condition under the effect of rotation. Same parameters as in Fig. \ref{fig:A7}.
    Evolution of the maximum, minimum and amplitude of the interfacial displacement for a soliton (left) and cnoidal wave  (right) initial condition on a periodic domain under the effect of rotation. 
    %Same parameters as in Figures \ref{fig:A3} and  \ref{fig:A7}, respectively.
    % {\color{red} Consider adding the amplitude}
    }  
    \label{fig:A8}
\end{figure}

Similarly, in Figure \ref{fig:A7} we show the effect of rotation on the cnoidal wave initial condition.
% $\rho_{1} = 1000 ,\, \rho_{2} = 1003.1$ kg m$^{-3}$, with the depths of the upper and lower layers being $37.5$ and $112.5$ m, respectively. 
The computational domain is $2L = 26.40$, with the number of modes $M = 262$, the spatial step $\Delta x \approx 10^{-1}$, the total simulation time $T_{max} = 100$, and the temporal step $\Delta T = 10^{-2}$. The parameters characterising the initial condition are $u_1 = -10^{-3}, u_2 = 0$ and $u_3 = 3$. It is evident that rotation leads to formation of a rather regular quasi-periodic wave pattern shown in Figure \ref{fig:A7}, without the formation of bursts of large waves. {\color{black} Overall, this wave looks similar to a cnoidal wave, but it has a slowly oscillating amplitude.}

We also note that both for the soliton and cnoidal wave initial conditions the time evolution of the maximum and minimum of the free-surface elevation is quasi-periodic, with no significant bursts, as shown in Figure \ref{fig:A8}. All computational parameters are the same as in the previous figures.
%The plots showing the time evolution of the maximum and minimum of the free surface elevation are shown in Figure \ref{fig:A8}, and again there are no significant bursts. 
Hence, we conclude that rotation alone does not trigger formation of large waves in these simulations.

\subsection{The effect of rotation on breathers on a cnoidal wave background}

The exact breather on a cnoidal wave solutions of the KdV equation were studied by \citet{KM1975, HMP2023}. There are two types of breathers: bright and dark (this terminology refers to the canonical form of the KdV equation, where solitons are waves of elevation).
%%% Definitions for bright and dark breathers %%%
For a bright breather propagating as a dislocation of a cnoidal wave the exact solution  of the KdV equation (\ref{KdV})  takes the form 
\begin{align}
    \eta^{(0)} = \frac{12 \beta_1}{\alpha_1} \Big\{ k^2 -1 +\frac{E(k)}{K(k)} +\frac{\partial^2}{\partial \zeta^2} \Big[ \log \big( \tau({\color{black} \zeta + \frac{\alpha_1 \hat \eta^{(0}}{3} T}, \beta_1 T) \big) \Big] \Big\}, 
    \label{dislocation A1}
\end{align}
where $k\in(0,1)$ is the elliptic modulus (the elliptic parameter $m = k^2$),  $K(k)$ is the complete elliptic integral of the first kind, $E(k)$ is the complete elliptic integral of the second kind, and the $\tau-$function %for the first family (bright breathers)
is given by
\begin{align*}
    \tau(x,t) := \Theta(x-c_0t +\alpha_b) \exp\{\kappa_b(x-c_bt+x_0)\} + \Theta(x-c_0t -\alpha_b) \exp\{-\kappa_b(x-c_bt+x_0)\}
\end{align*}
with $\kappa_b>0,\, c_b > c_0$ and $\alpha_b\in(0,K(k))$. Here, $\Theta(x) = \theta_4 \left (\frac{\pi x}{2K(k)}\right )$, where $\theta_4$ is the Jacobi theta function of type four, given by $\theta_4(u) = 1+2 \sum\limits_{n=1}^{\infty} (-1)^n q^{n^2}\cos(2nu)$. The solution is parameterised by $\lambda \in (-\infty,-k^2)$. The parameters are defined as follows \citep{HMP2023}:
\begin{align}
    \varphi_\gamma &= \arcsin \Big( \frac{\sqrt{-\lambda-k^2}}{\sqrt{1-2k^2-\lambda}} \Big),\quad \\
    \alpha_b &= F(\varphi_\gamma,k),\quad 
    \kappa_b = \frac{\sqrt{1-\lambda-k^2}\sqrt{-\lambda-k^2}}{\sqrt{1-2k^2-\lambda}} - Z(\varphi_\gamma,k),\\
    c_0 &= 4(2k^2-1),\quad 
    c_b = c_0 + \frac{4\sqrt{1-\lambda-2k^2} \sqrt{1-\lambda-k^2} \sqrt{-\lambda -k^2}}{\kappa_b},
\end{align}
where $F(\varphi,k):= \int_0^\varphi \frac{d\theta}{\sqrt{1-k^2 \sin^2\theta}}$ is the elliptic integral of the first kind, and $Z(\varphi,k):= \int_0^\varphi \sqrt{1-k^2 \sin^2\theta}d\theta$ is the elliptic integral of the second kind.

For a dark breather propagating as a dislocation of a cnoidal wave the exact solution is again given by equation \eqref{dislocation A1}, where the solution is now parameterised by $\lambda \in (1-2k^2, 1-k^2)$, and the $\tau-$function takes the form
\begin{align*}
    \tau(x,t) := \Theta(x-c_0t +\alpha_d) \exp\{-\kappa_d(x-c_dt+x_0)\} + \Theta(x-c_0t -\alpha_d) \exp\{\kappa_d(x-c_dt+x_0)\}
\end{align*}
with $\kappa_d>0,\, c_d < c_0$ and $\alpha_d\in(0,K(k))$. The parameters are given by
\begin{align}
    \varphi_\alpha &= \arcsin \Big( \frac{\sqrt{1-k^2-\lambda}}{k} \Big),\quad \\
    \alpha_d &= F(\varphi_\alpha,k),\quad 
    \kappa_d = Z(\varphi_\alpha,k),\quad \\
    c_0 &= 4(2k^2-1),\quad  
    c_d = c_0 - \frac{4\sqrt{(k^2+\lambda) (\lambda-1+2k^2) (1-k^2 -\lambda)}}{\kappa_d}.
\end{align}
%where $F(\varphi,k):= \int_0^\varphi \frac{d\theta}{\sqrt{1-k^2 \sin^2\theta}}$ is the elliptic integral of the first kind, and $Z(\varphi,k):= \int_0^\varphi \sqrt{1-k^2 \sin^2\theta}d\theta$ is the elliptic integral of the second kind.

\begin{figure}
    \centering
    \includegraphics[width=0.51\linewidth]{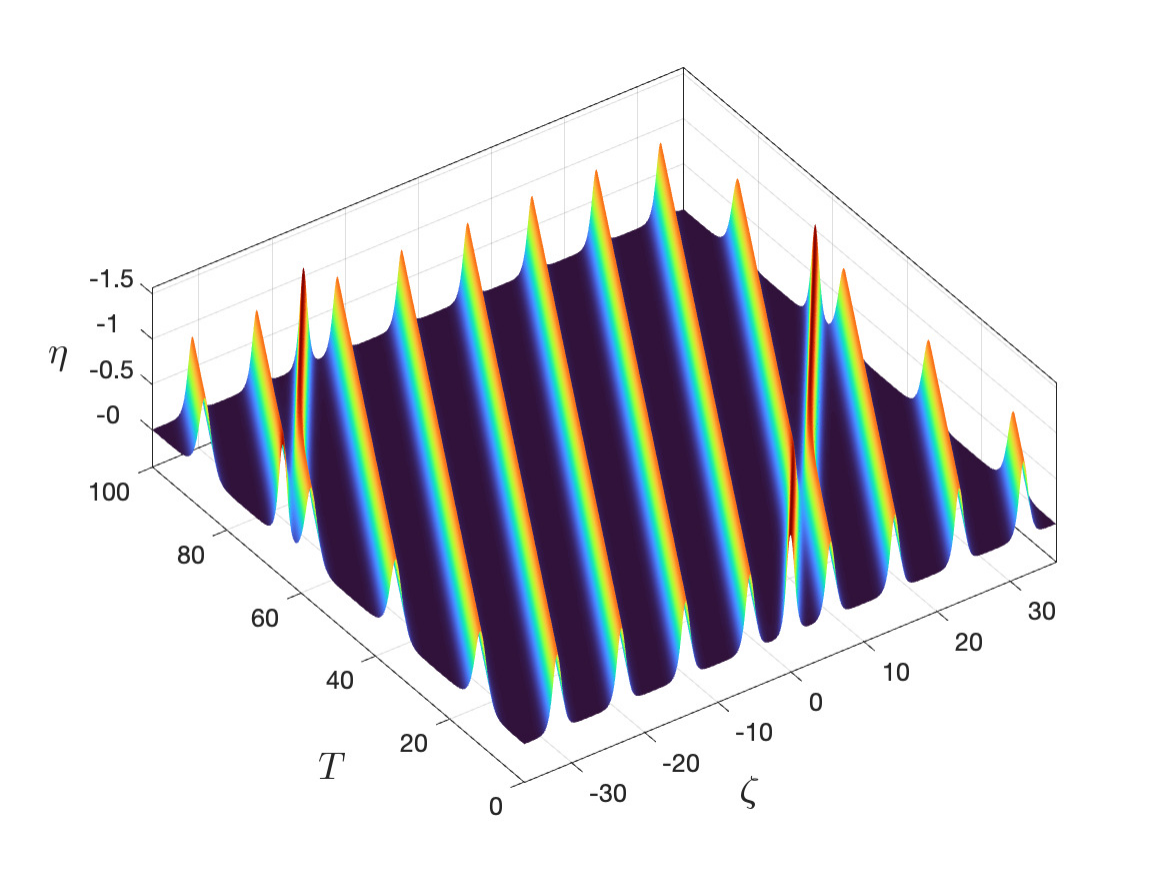}\hspace{-0.5cm}
     \includegraphics[width=0.51\linewidth]{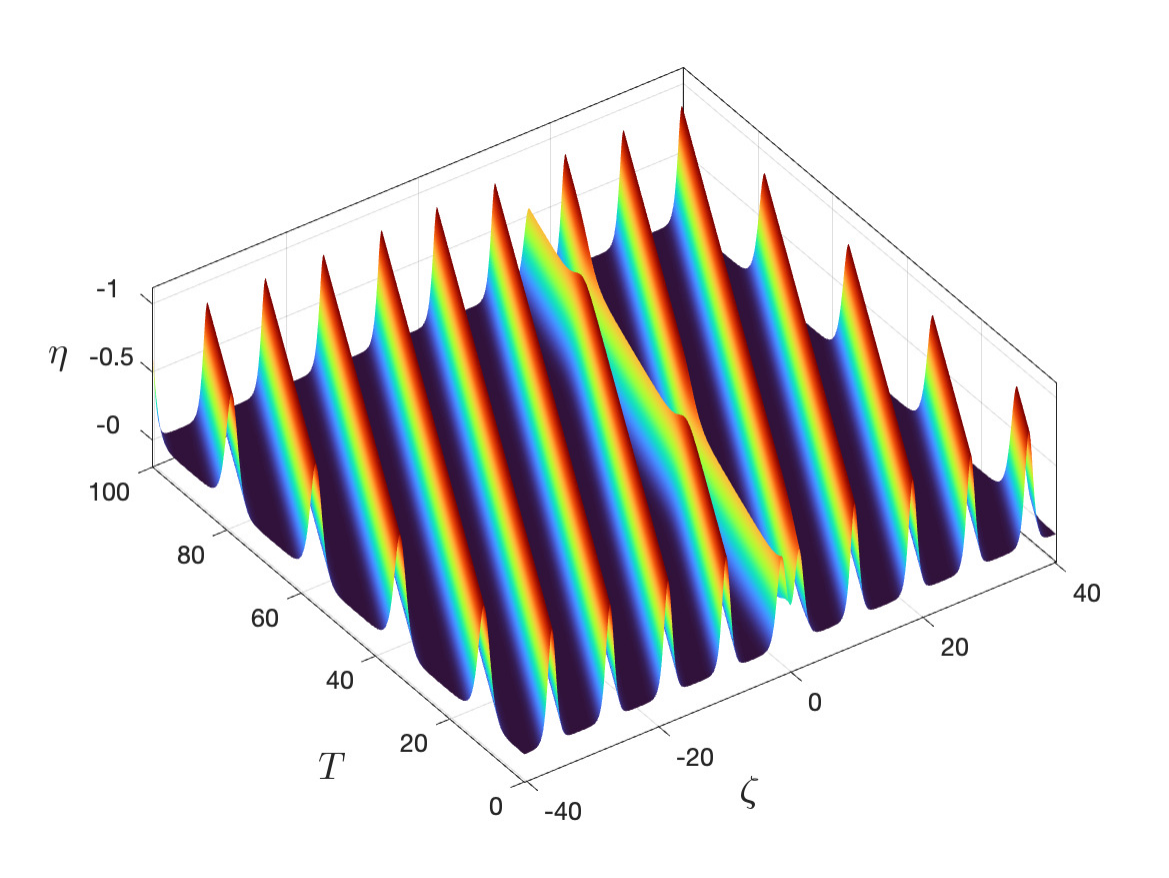}\\[-0.3cm] 
    
     \includegraphics[width=0.51\linewidth]{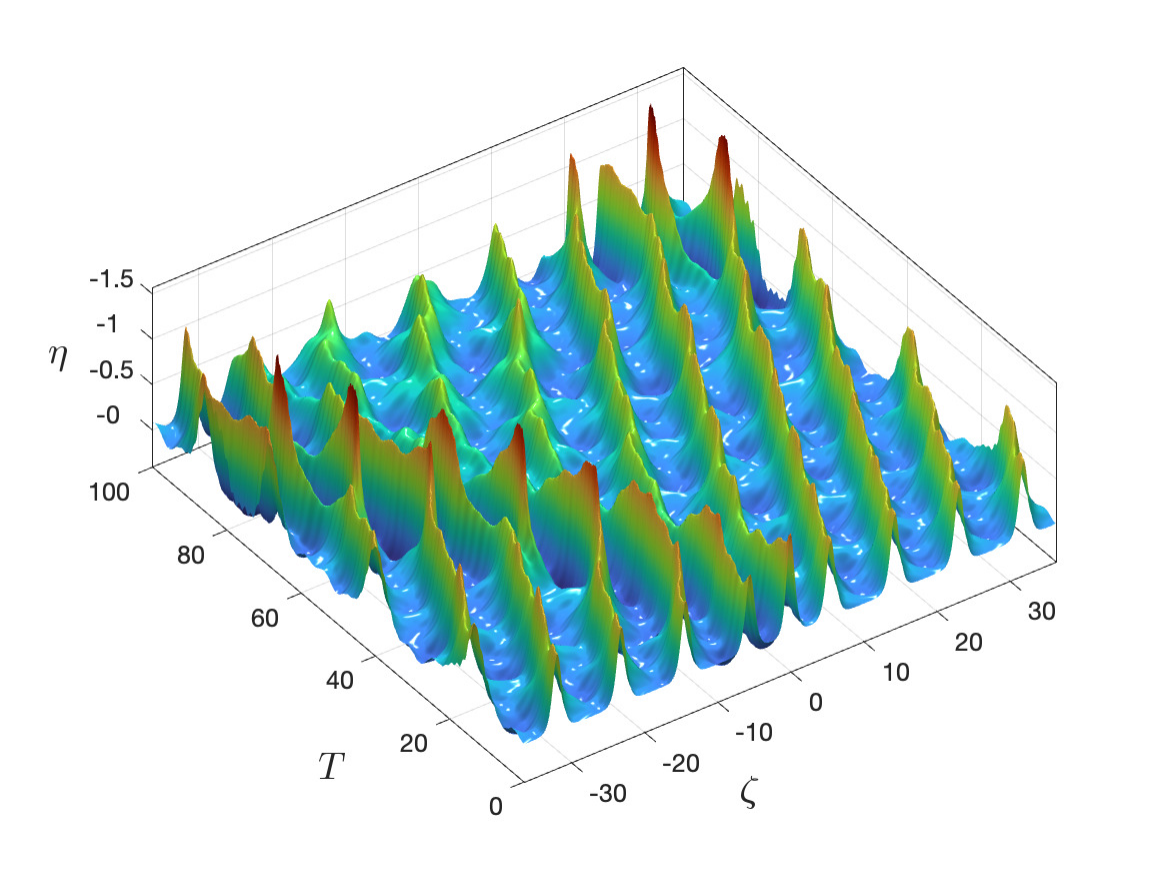}\hspace{-0.5cm}  
     \includegraphics[width=0.51\linewidth]{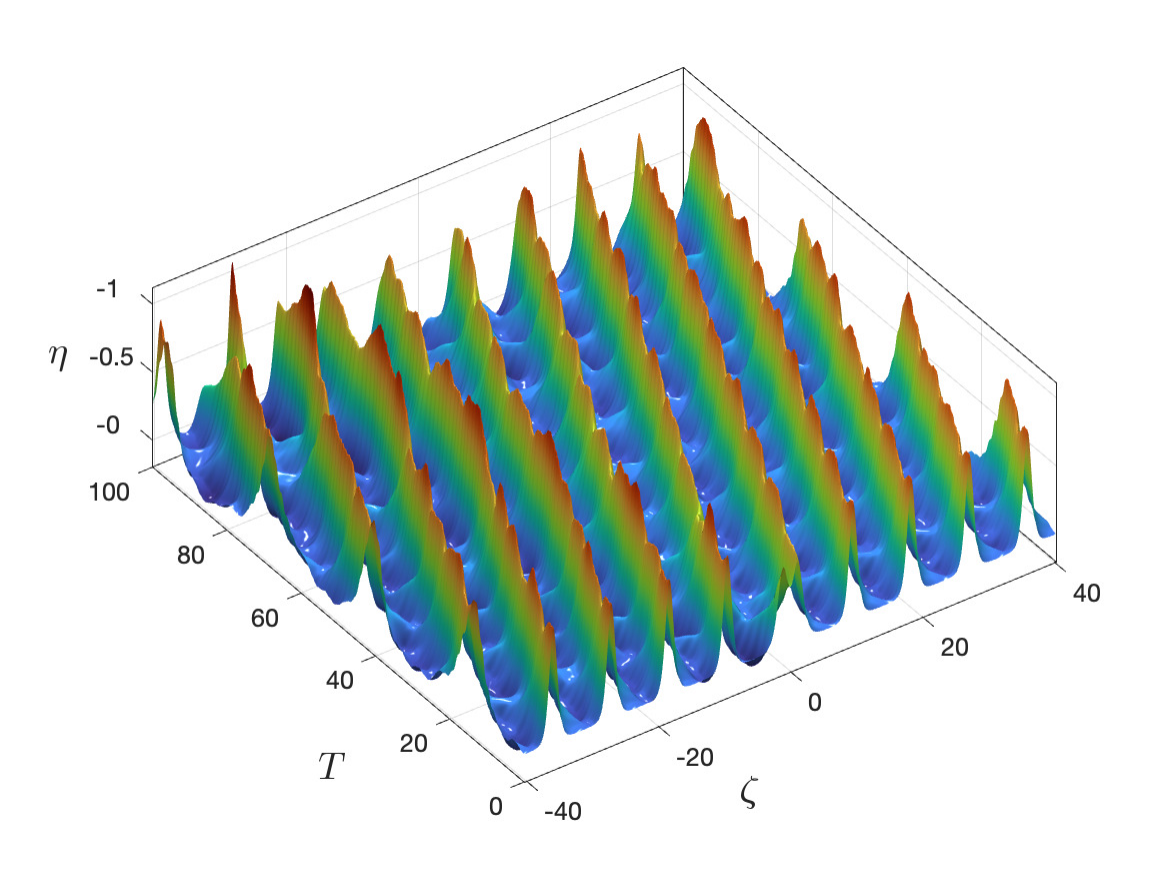}\\[-0.3cm]  
   
     \includegraphics[width=0.51\linewidth]{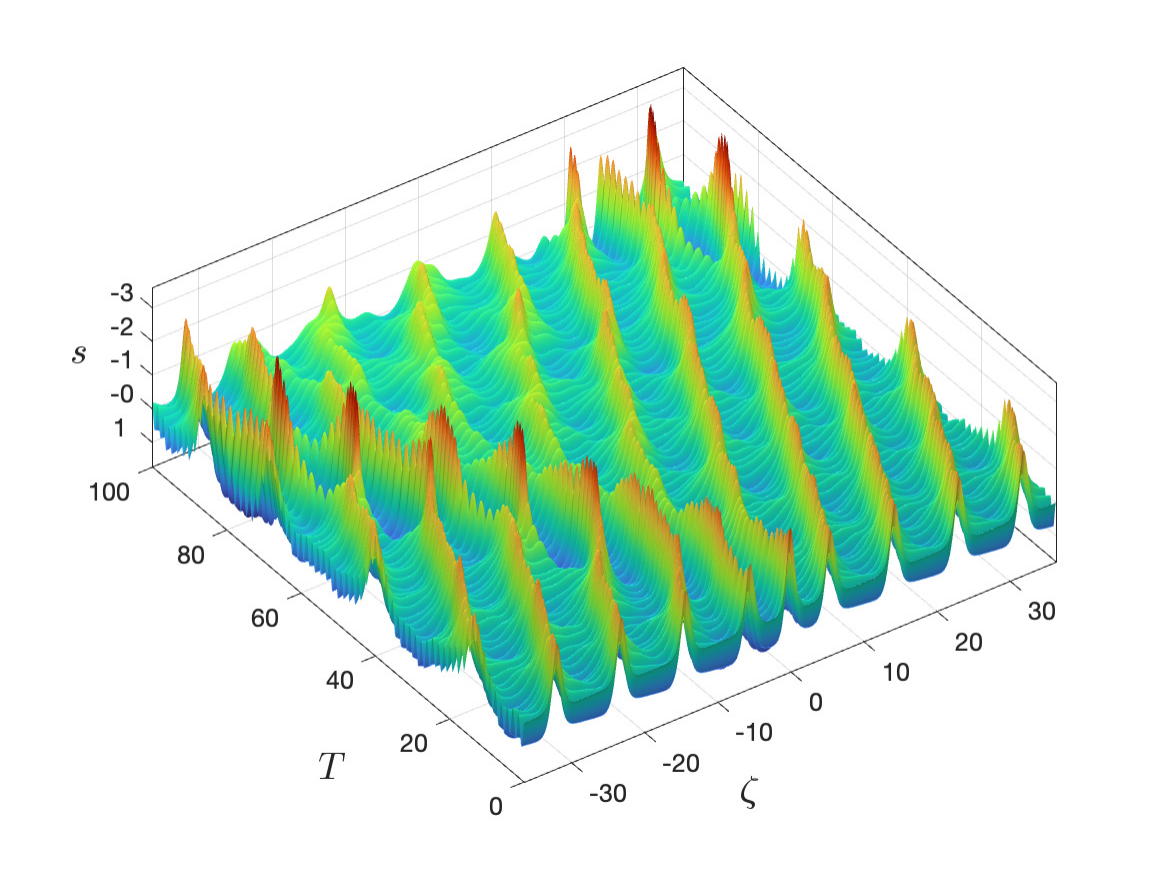}\hspace{-0.5cm}  
     \includegraphics[width=0.51\linewidth]{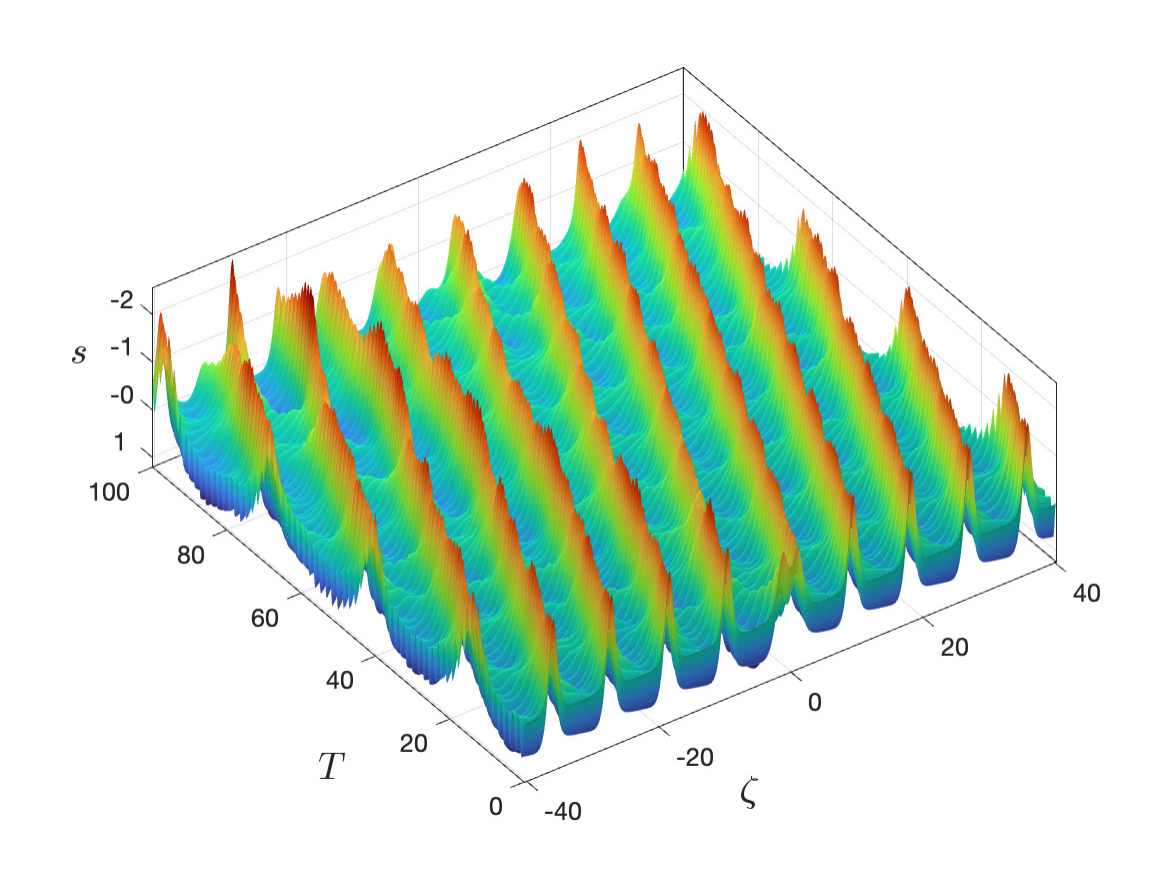}\\[-0.3cm]  
     
     \includegraphics[width=0.51\linewidth]{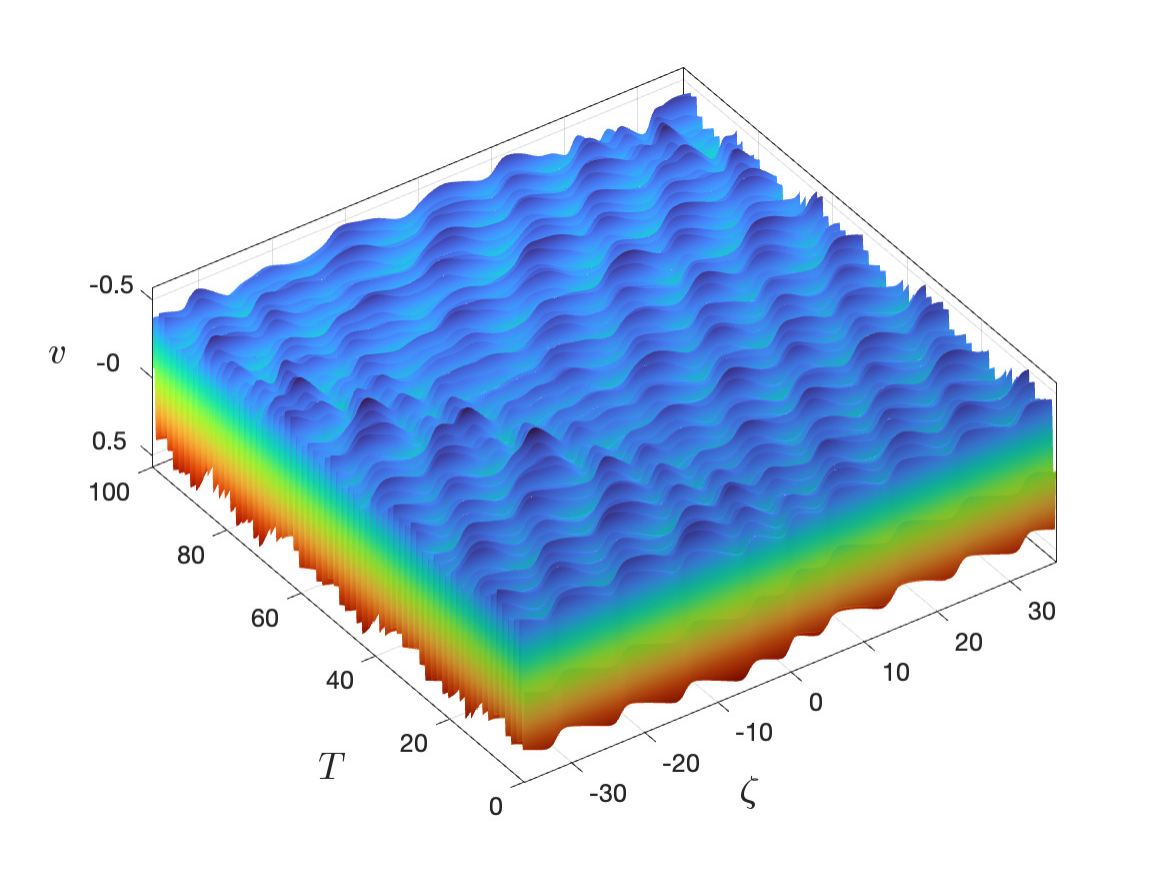}\hspace{-0.5cm}  
     \includegraphics[width=0.51\linewidth]{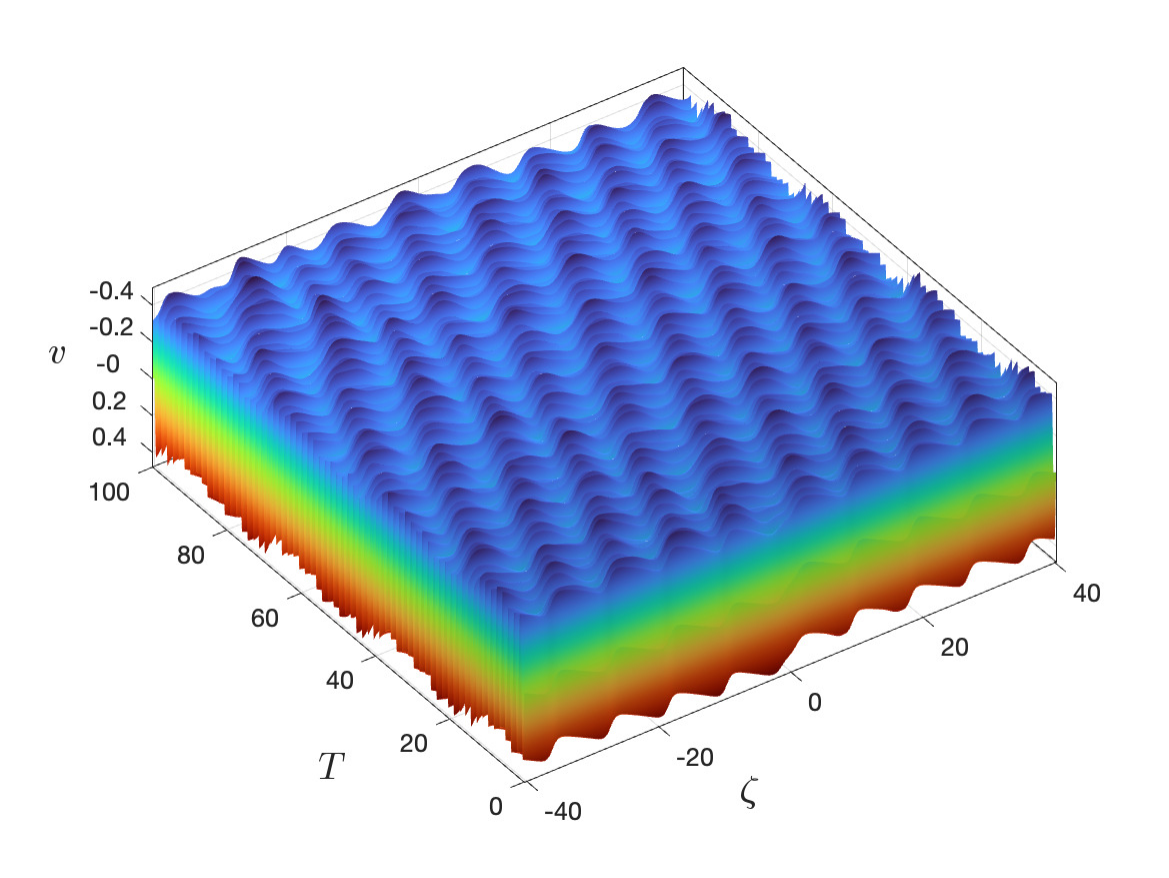} 
    \caption{Numerical solution for a bright (left) and dark (right)  breather on a cnoidal wave initial condition  (view from below). First row: interfacial displacement in the absence of rotation. Second row: interfacial displacement under the effect of rotation. Third / fourth row: shear  in the direction of wave propagation /  orthogonal direction, under the effect of rotation. }
    \label{fig:A9}
\end{figure}

\begin{figure}
    \centering
	\includegraphics[width=0.47\linewidth]{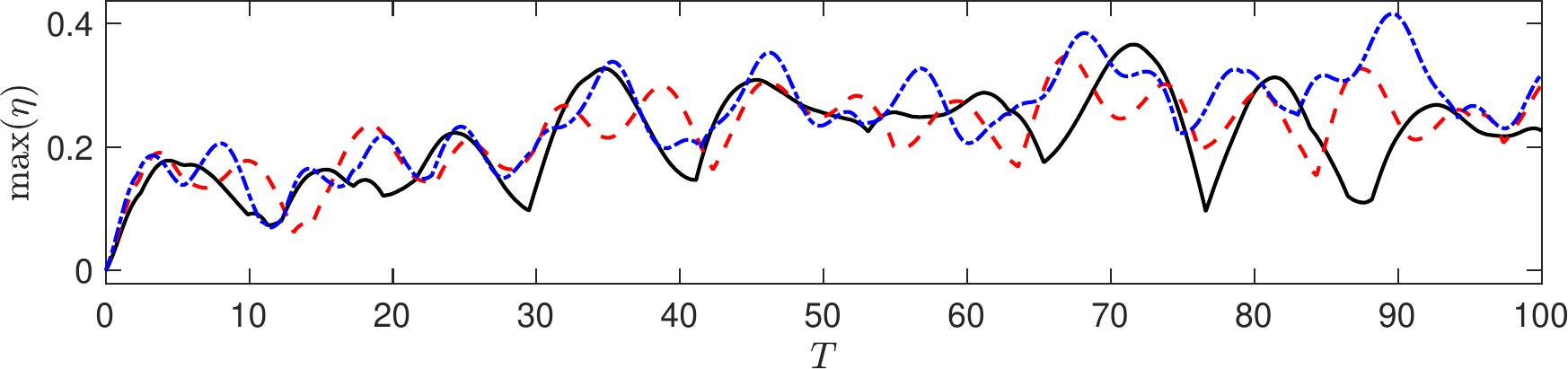}
     	\includegraphics[width=0.47\linewidth]{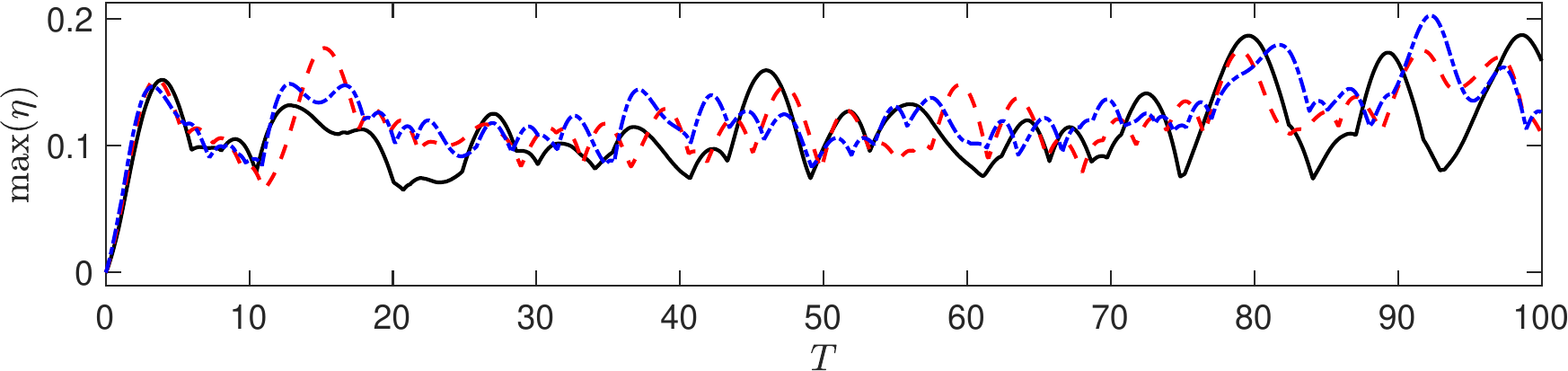}\\[0.5cm]
     	\includegraphics[width=0.47\linewidth]{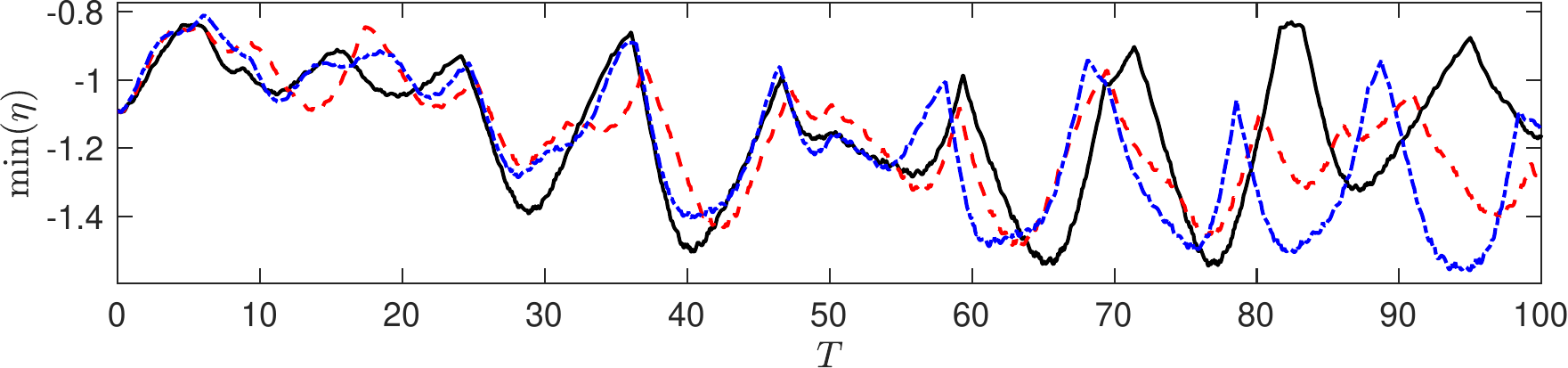}
     	\includegraphics[width=0.47\linewidth]{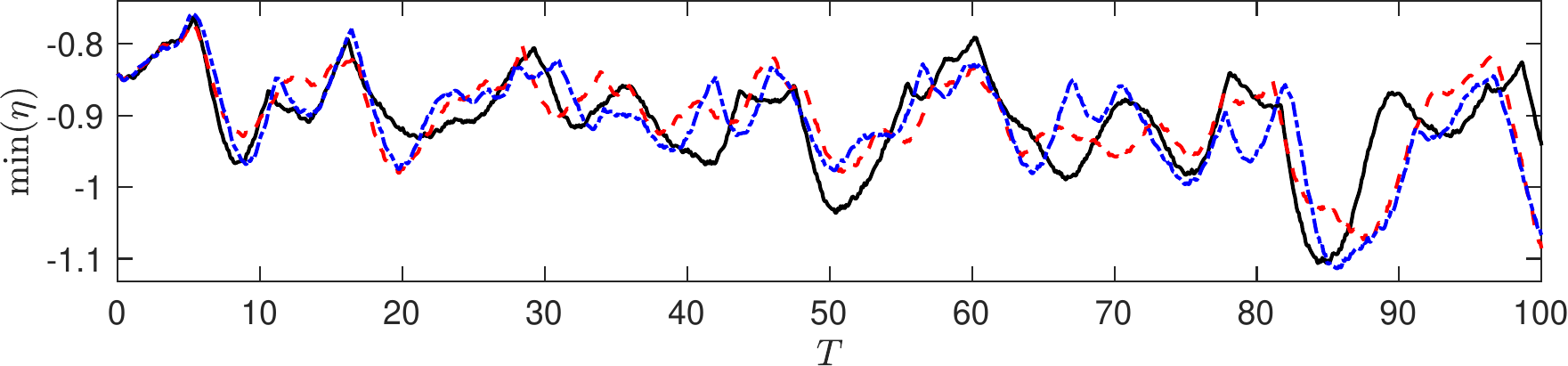}\\[0.5cm]
     	\includegraphics[width=0.47\linewidth]{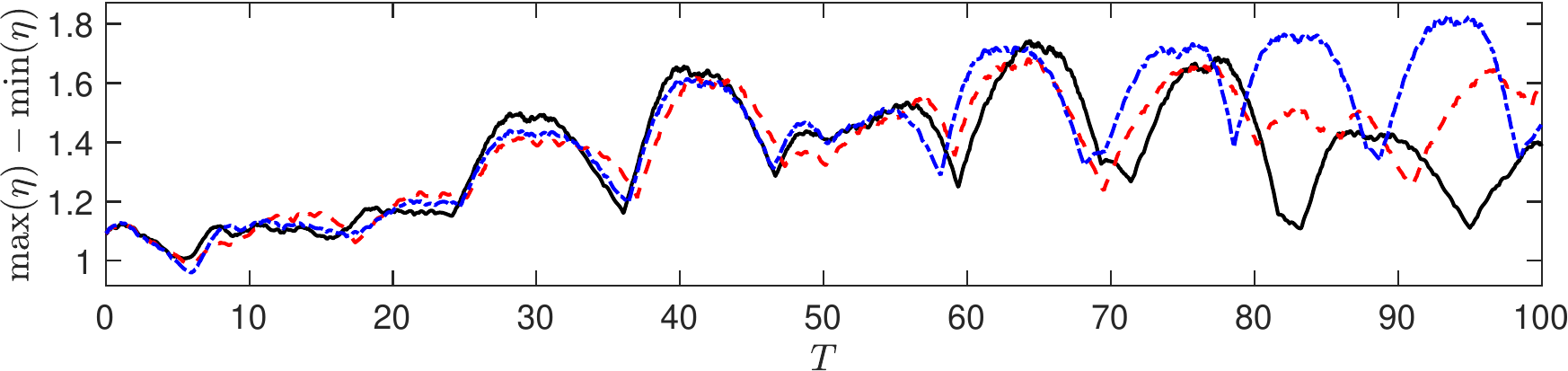}
     	\includegraphics[width=0.47\linewidth]{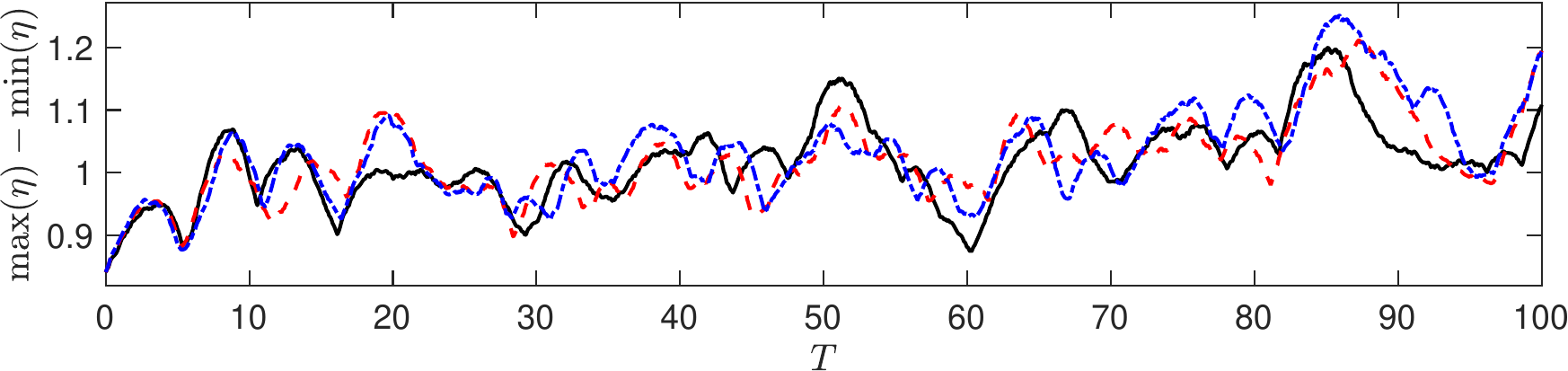}
    \caption{The effect of rotation on the evolution of the maximum, minimum and amplitude of the interfacial displacement for a bright (left) and dark (right)  breather on a cnoidal wave initial condition. Black solid, red dashed and blue dot-dashed lines correspond to 5, 7 and 9 peaks in the domain, respectively.
%    {\color{red} Please, check.  The computational domains are $2L= 37.42$ (5 peaks), $2L = 55.02$ (7 peaks), $2L = 72.86$ (9 peaks). Right column: black solid, red dashed and blue dot-dashed lines correspond to 5, 7 and 9 peaks in the domain, respectively. The computational domains are $2L= 45.44$ (6 peaks), $2L = 63.04$ (8 peaks), $2L = 80.64$ (10 peaks). The remaining physical and numerical parameters are the same as in Fig. \ref{fig:A9}. }
}
    \label{fig:A16}
\end{figure}

\begin{figure}
    \centering   
       	\includegraphics[width=0.5\linewidth]{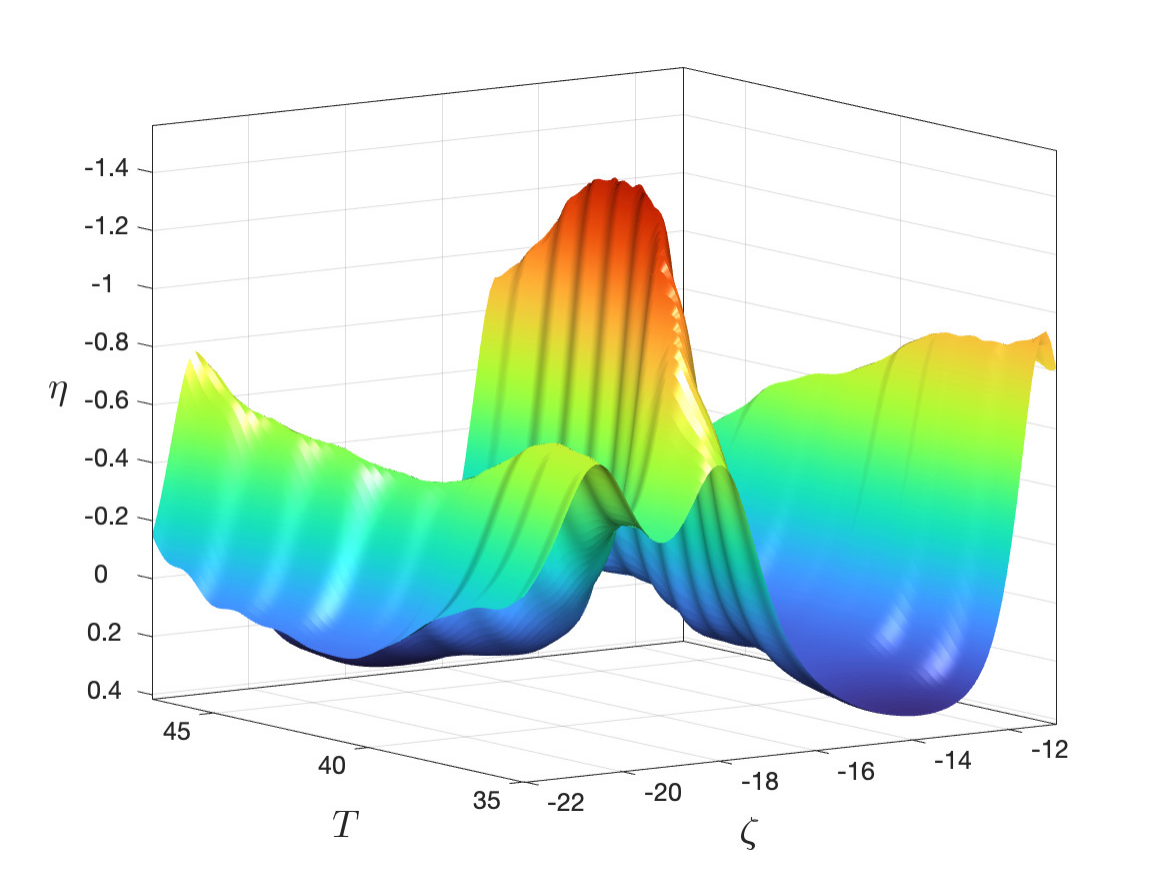}\hspace{-0.5cm}
	\includegraphics[width=0.5\linewidth]{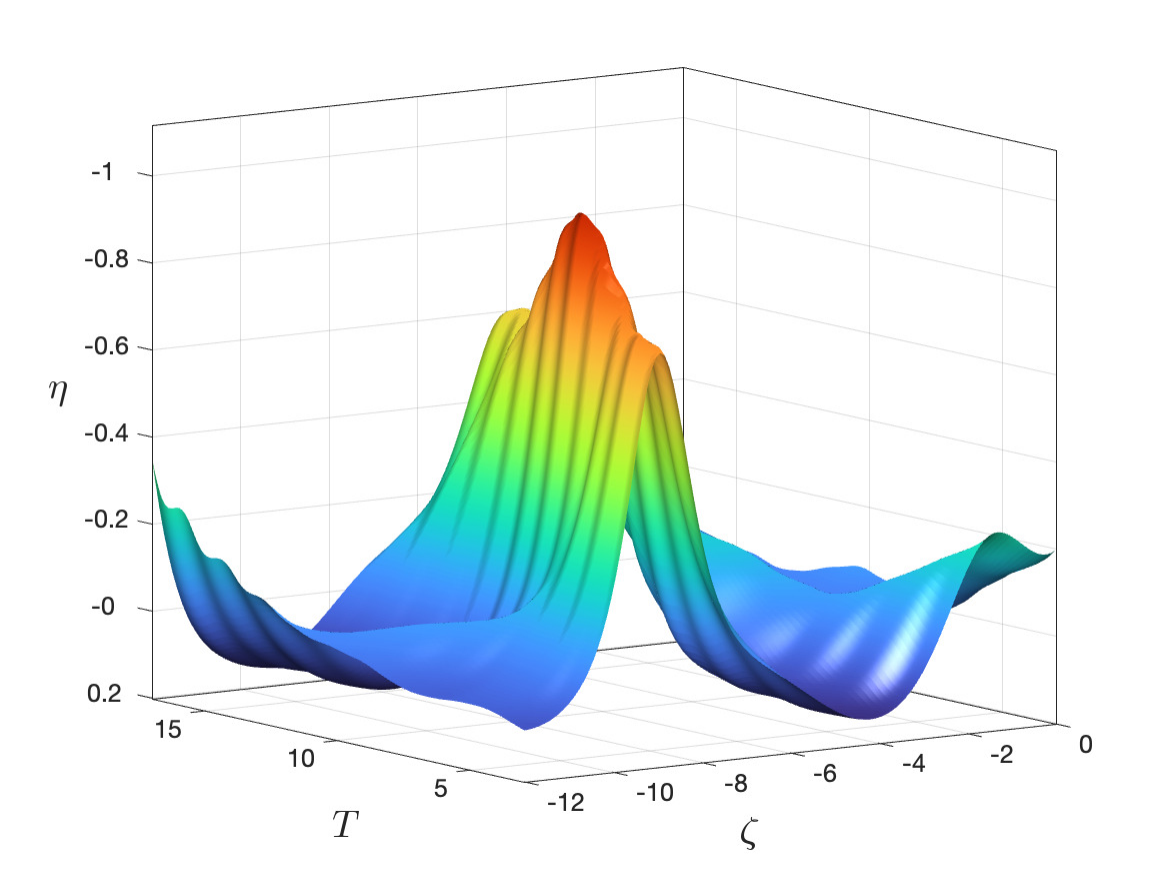}        
        \caption{Close-up view from below of the large burst in the interfacial displacement for a bright (left) and dark (right) breather on a cnoidal wave initial condition around $T = 40$ (left) and $T = 9$ (right), respectively. }
        \label{fig:A14}
\end{figure}

Numerical solutions initiated with the bright and dark breather on a cnoidal wave initial conditions at $T=0$ in the absence of rotation are shown in the first row of Figure \ref{fig:A9}. 
%$\rho_{1} = 1000 ,\, \rho_{2} = 1003.1$ kg m$^{-3}$, with the depths of the upper and lower layers being $37.5$ and $112.5$ m, respectively. 
For the bright breather simulations, shown in the left column, the computational domain length is $2L = 72.86$, with the number of modes $M=728$, the spatial step $\Delta x \approx 10^{-1}$, the total simulation time $T_{max} = 100$, and the temporal step $\Delta T = 10^{-2}$. The parameters used in the initial condition are $[k,\lambda] \approx [0.9998, -1.30]$. For the dark breather simulations, shown in the right column, the computational domain length is $2L = 80.64$, with the number of modes $M=804$, the spatial step $\Delta x \approx 10^{-1}$, the total simulation time $T_{max} = 100$, and  the temporal step $\Delta T = 10^{-2}$. The other parameters are $[k,\lambda] \approx [0.9998, -0.50]$. In the absence of rotation, the plots show stable propagation of the bright and dark breather solutions on top of the cnoidal wave, in {\color{black} good} agreement with the available analytical solution (\ref{dislocation A1}).

Next, the effect of rotation on the evolution of the bright breather on a cnoidal wave initial condition is shown  in  the subsequent rows of the same Figure \ref{fig:A9}, for the interfacial displacement (second row), shear in the direction of wave propagation (third row) and shear in the orthogonal direction (fourth row).
% for the bright breather and Figures \ref{fig:A19} - \ref{fig:A16} for the dark breather, 
%The physical parameters are $\rho_{1} = 1000 ,\, \rho_{2} = 1003.1$ kg m$^{-3}$, with the depths of the upper and lower layers being $37.5$ and $112.5$ m, respectively. The computational domains is $2L = 72.86$, with number of modes of $M=728$, a spatial step size of $\Delta x \approx 10^{-1}$, and total simulation time is $T_{max} = 100$ with a temporal step size of $\Delta T = 10^{-2}$. The other parameters are $[k,\lambda] \approx [0.9998, -1.30]$. 
We notice a striking difference with the results shown in the top row of the figure (no rotation): there emerges a rather strong left-propagating localised burst clearly visible both in the free surface elevation (second row) and the shear in the direction of wave propagation (third row). The counterpart of the wave is also present in the shear  in the orthogonal direction (fourth row), but this signal is weak.  It must be noted that the wave forms soon after the initiation of the simulation, and it continues to grow for a long time.

In these runs, the effect on a dark breather on a cnoidal wave is qualitatively similar  to that on the bright breather, though less pronounced. In both cases, the background cnoidal wave continues propagating to the right (in the moving reference frame), and there appears a significant burst both in the interfacial displacement and the shear in the direction of wave propagation, moving to the left.  
%For the dark breather, this burst moves  slower than in the bright-breather case.  
For the dark breather, the signal in the shear in the orthogonal direction  is also present, but it is barely noticeable to the naked eye.

Figure \ref{fig:A16} shows the time evolution of the maximum, minimum and the amplitude of the interfacial displacement for both cases of a bright (left column) and dark (right column) breather on a cnoidal wave initial condition. We see that the observed effects of rotation are structurally stable with regard to the size of the computational domain.   We experimented with 5, 7 and 9 major peaks in the domain.  In the bright-breather case, the computational domains are $2L= 37.42$ (5 peaks), $2L = 55.02$ (7 peaks), $2L = 72.86$ (9 peaks).  For the dark-breather case, the computational domains are $2L= 45.44$ (5 peaks), $2L = 63.04$ (7 peaks), $2L = 80.64$ (9 peaks). The remaining numerical parameters are the same as before.  It is again evident that the wave amplitude grows. The burst forms soon after the initiation of the numerical runs. The close-up views of the large waves visible in the free surface elevation are shown in Figure \ref{fig:A14} for the time around {\color{black} $T=40$ (bright breather case, left) with approximately  $29\%$ increase in the amplitude compared to the initial condition, and  around $T=9$ (dark breather case, right)  with  approximately  $15\%$ increase in the amplitude compared to the initial condition. }Hence, we conclude that under the effect of rotation the moving dislocation on top of the otherwise regular cnoidal wave can lead to the emergence of strong bursts of interfacial waves and shear currents in the direction of propagation of the cnoidal wave.

{\color{black} We also note that we preformed preliminary simulations with higher-amplitude breathers, and these suggest more complex dynamics: the burst initially propagates to the right, but part of the mass is subsequently emitted into the background cnoidal wave, and the burst subsequently decreases in amplitude and changes direction. A detailed investigation of this behaviour is  beyond the scope of the present work and is left for future work.}

\subsection{The effect of rotation on cnoidal waves with periodicity defects}

Recent research related to the wave packets described by the Schr\"odinger equation has shown that localised phase defects can lead to the emergence of rogue waves \citep{HWTCH2022}.  Here, we investigate whether periodicity defects introduced into the long cnoidal waves in the KdV-Ostrtovsky regime can also lead to the emergence of large localised bursts of energy, under the effect of rotation. Also, can it happen already in the absence of rotation, i.e.\ in the KdV regime? 

\begin{figure}
    \centering
    \includegraphics[width=0.51\linewidth]{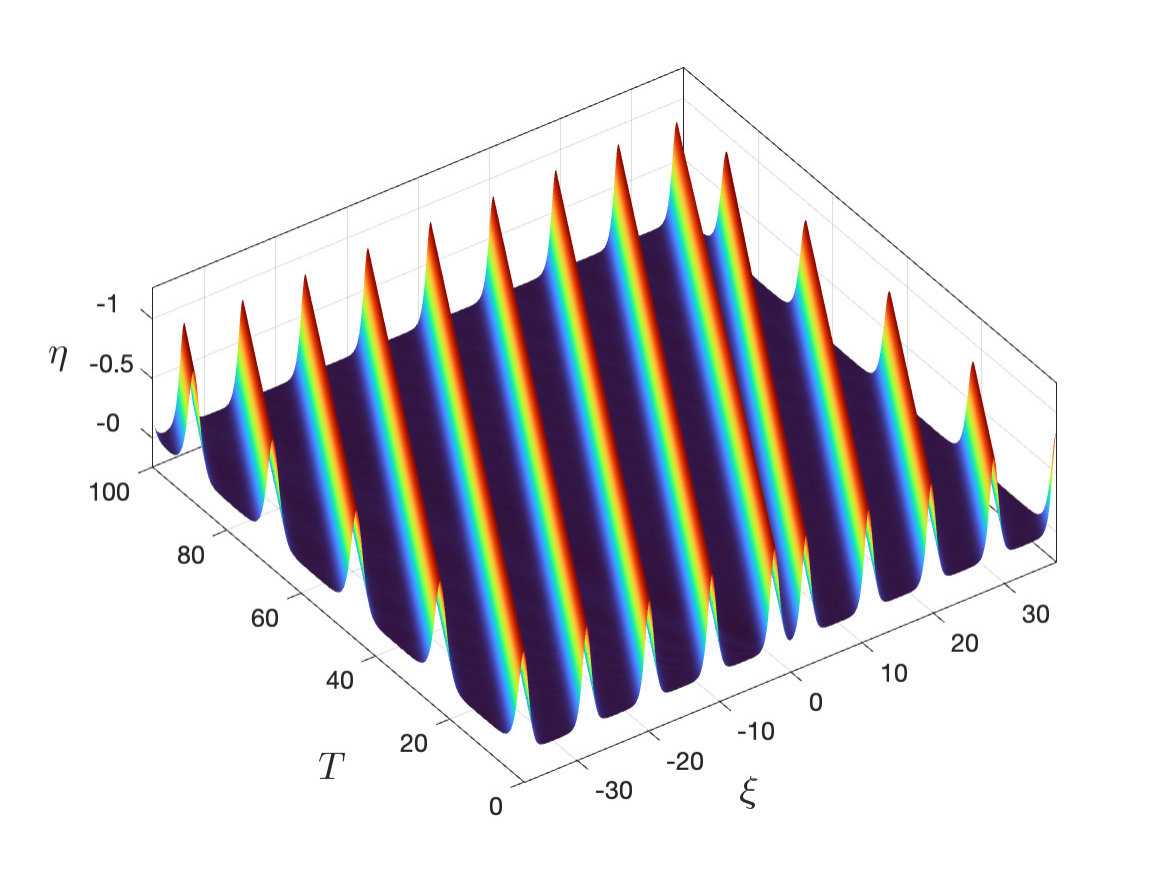}\hspace{-0.5cm}      
    \includegraphics[width=0.51\linewidth]{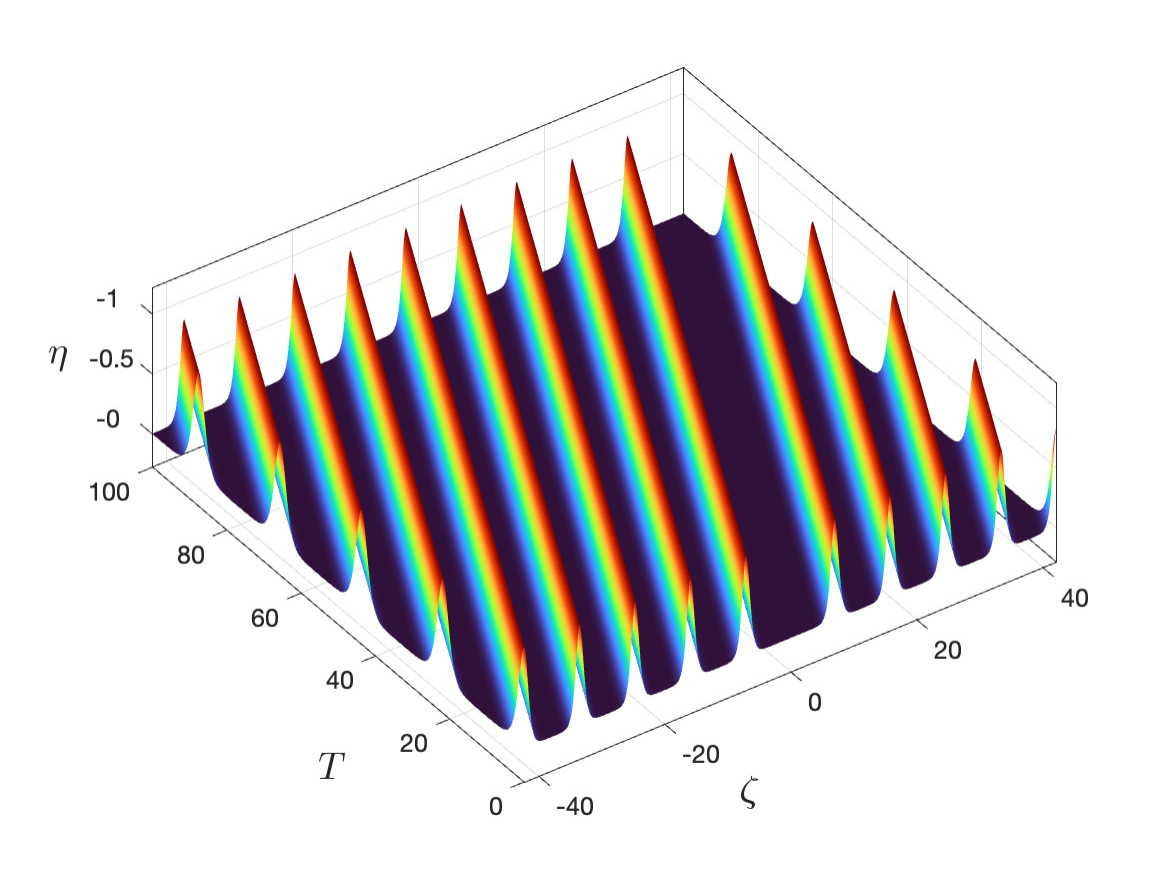}\\[-0.3cm]
    
    \includegraphics[width=0.51\linewidth]{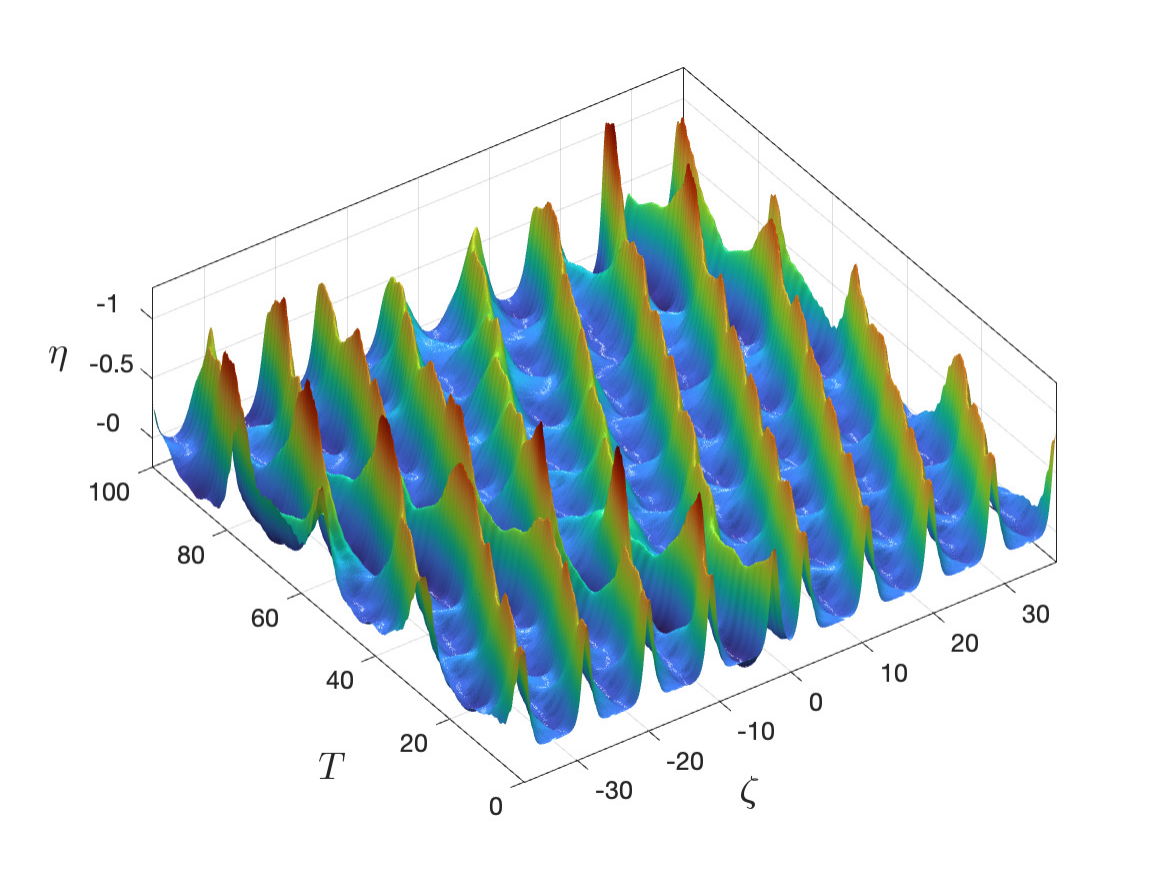}\hspace{-0.5cm}       
    \includegraphics[width=0.51\linewidth]{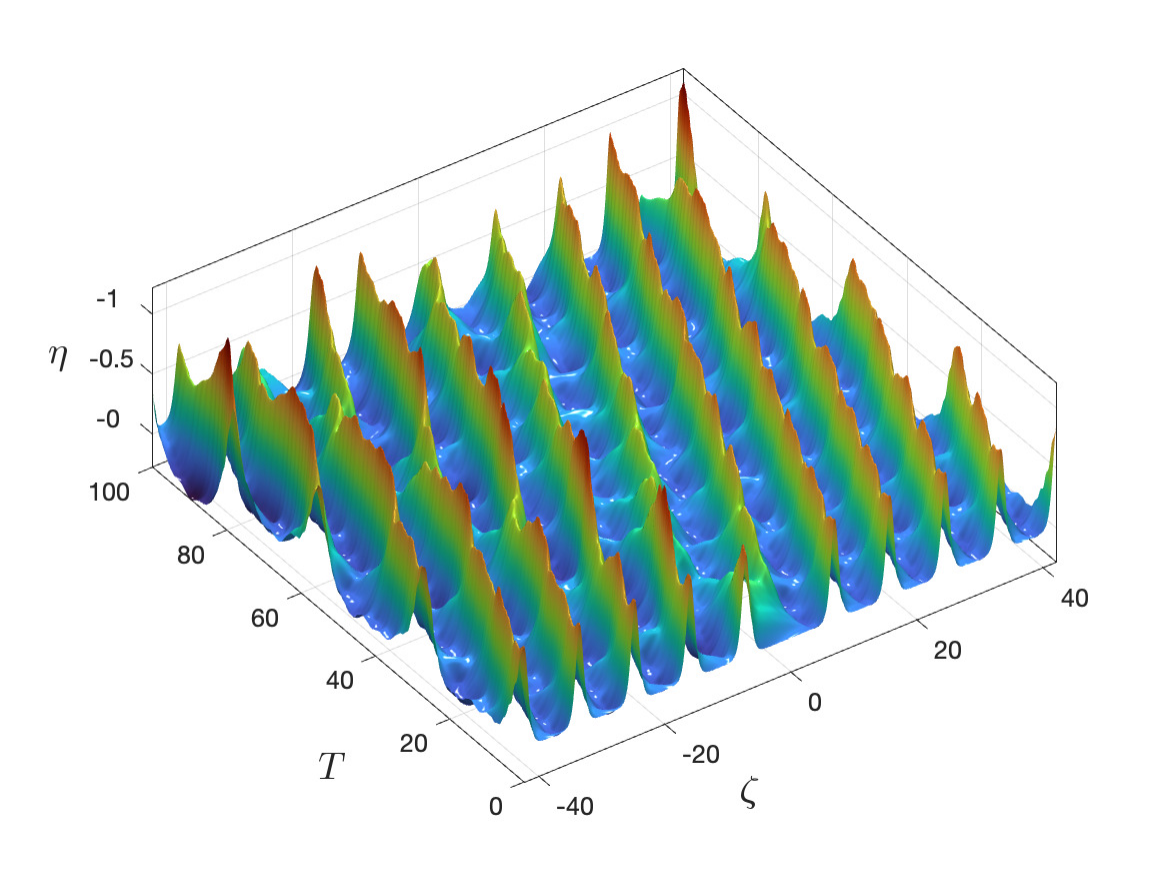}\\[-0.3cm]

       \includegraphics[width=0.51\linewidth]{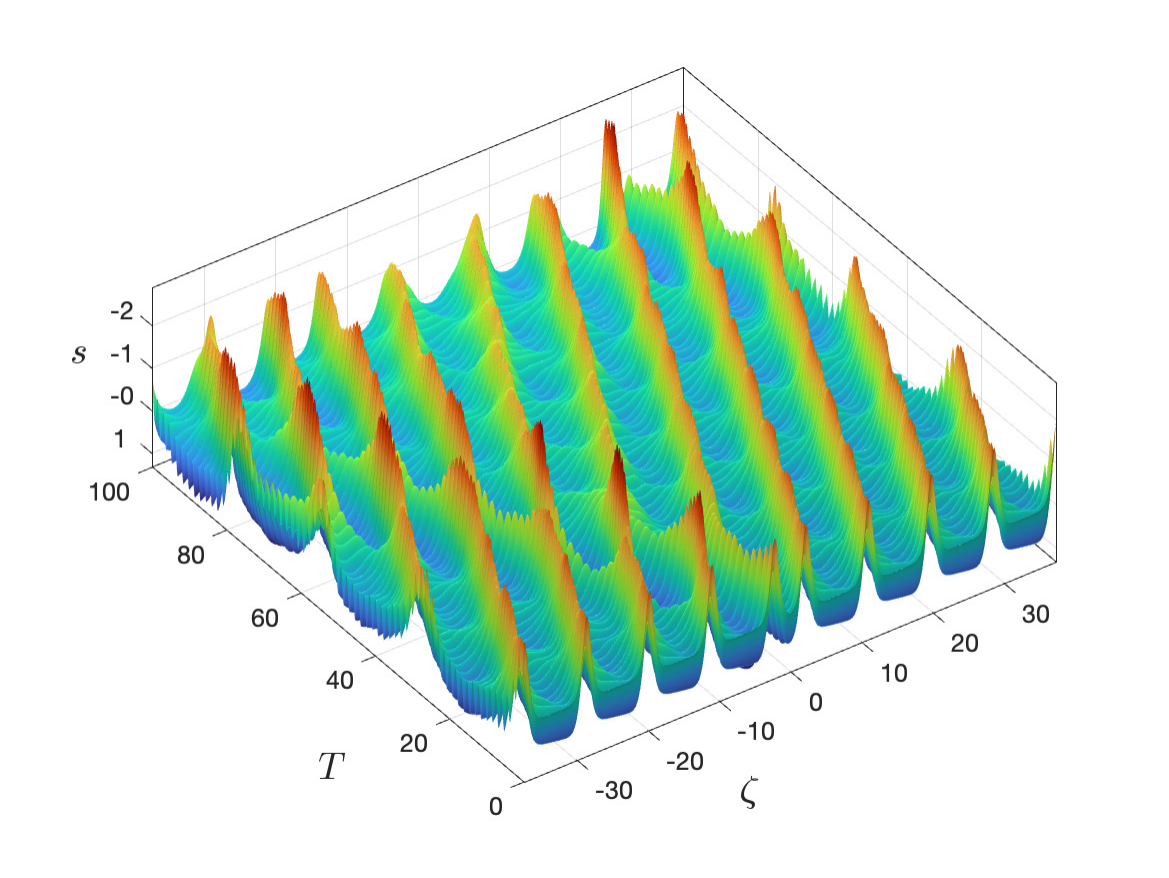}\hspace{-0.5cm}
         \includegraphics[width=0.51\linewidth]{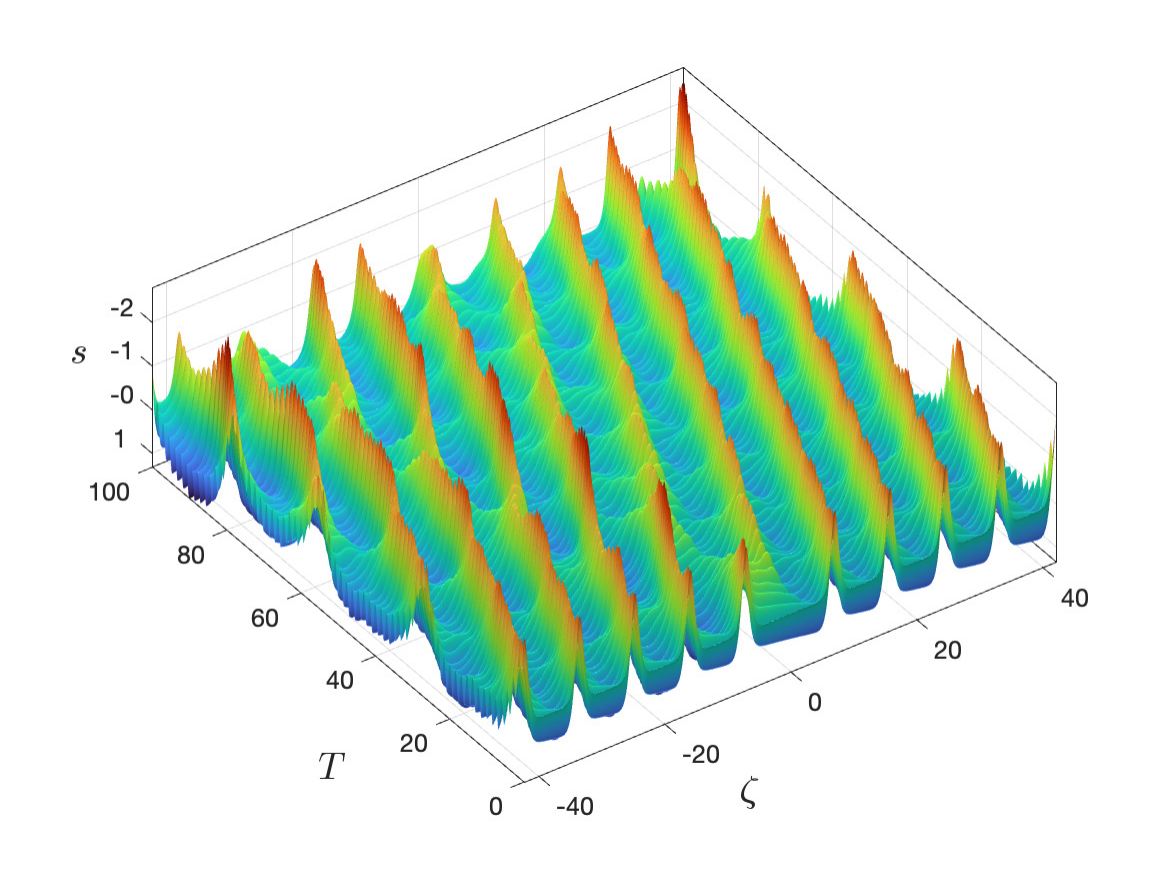}\\[-0.3cm]  
         
                \includegraphics[width=0.51\linewidth]{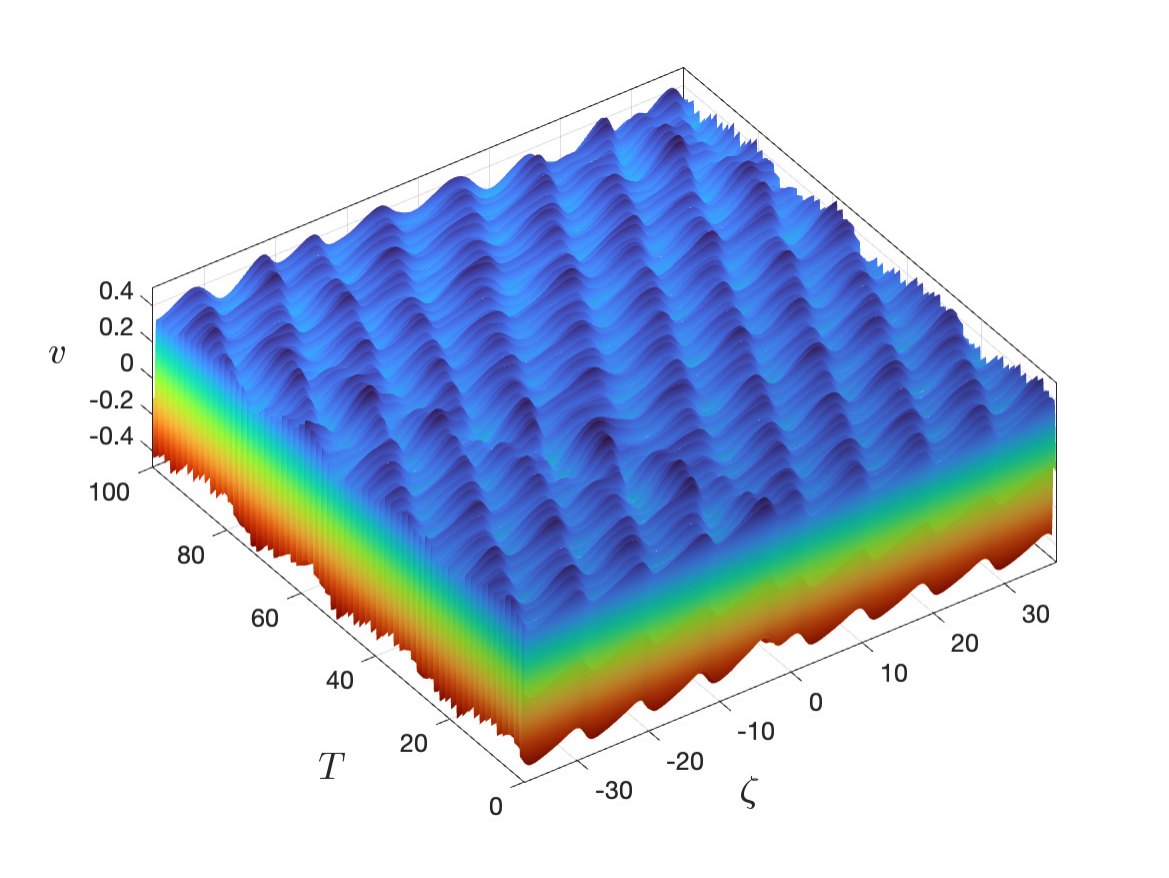}\hspace{-0.5cm}
         \includegraphics[width=0.51\linewidth]{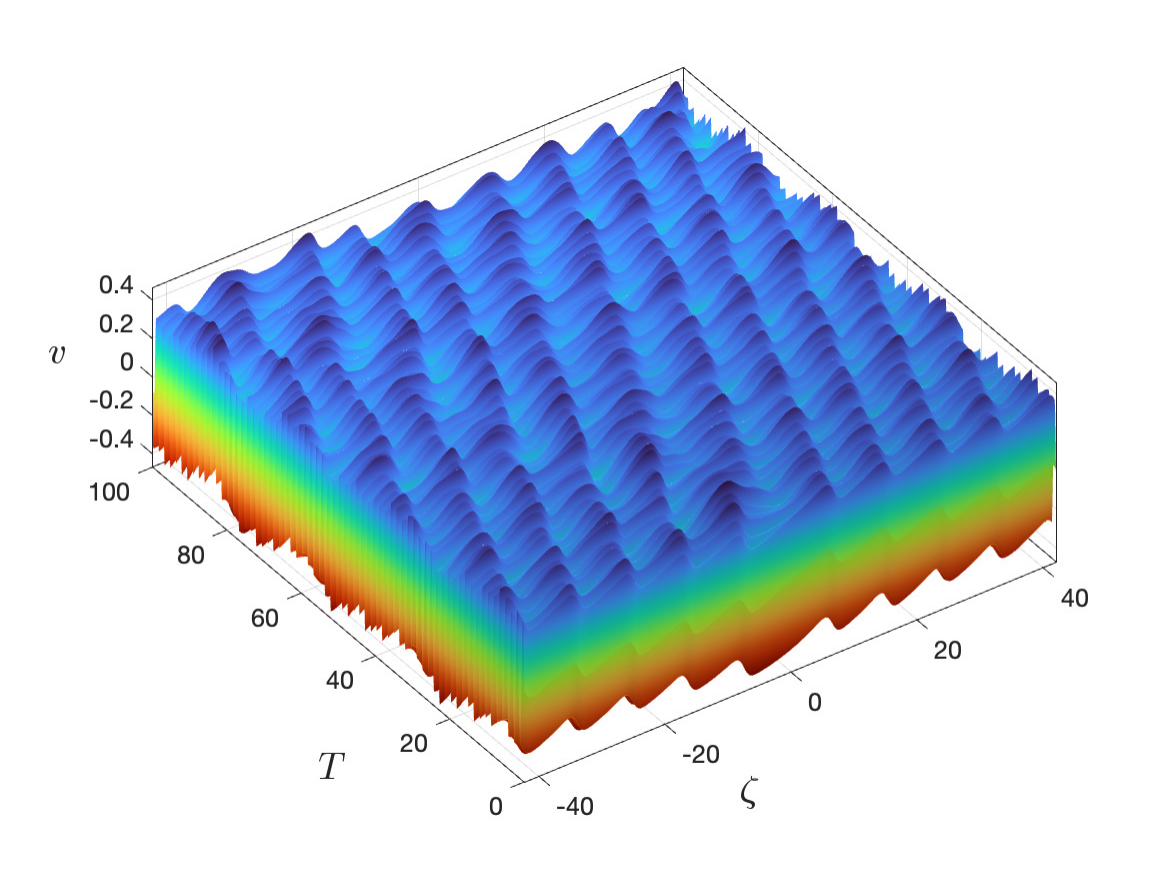}  
        \caption{Numerical solution for a cnoidal wave with a contraction (left) and  expansion (right) defect initial condition  (view from below). First row: interfacial displacement in the absence of rotation. Second row: interfacial displacement under the effect of rotation. Third / fourth row: shear  in the direction of wave propagation / orthogonal direction, under the effect of rotation.
      }
    \label{fig:A21.2}
\end{figure}

Motivated by Figure \ref{fig:1}, we consider cnoidal waves close to their solitonic limit and introduce  two types of periodicity defects (see Figure \ref{fig:WS} in Appendix A): contraction and expansion defects, depending on whether the distance between the two neighbouring peaks is shorter or longer than the period of the cnoidal wave. The first defect is introduced by symmetrically cutting away a small part close to the trough between the two neighbouring peaks, and gluing together the remaining parts of the solution. Naturally, the resulting  function  has discontinuous {\color{black} first} derivative at one point within the computational domain, but the jump in the derivative is small because we have cut close to extremum (smoothed in numerical simulations). The second defect is introduced by cutting the graph at the trough and symmetrically inserting a piece of a straight line.
% between two points close to the trough. 
The resulting function has continuous first derivative, and {\color{black} discontinuous second derivative at two points within the computational domain}. To our surprise, these functions evolved almost like travelling waves of the KdV equation. Both types of waves were long-lived, and {\color{black} cnoidal wave with an expansion defect, smoothed in a pseudospectral simulation, was stable, with no visible changes at the end of the long run (the difference between the numerical solution and the initial condition travelling at the speed of the cnoidal wave was of order $10^{-4}$).} The numerical results initiated with such initial conditions in the absence of rotation are  shown in the first row of Figure \ref{fig:A21.2} for the contraction defect (left column) and the expansion defect (right column).  Intrigued by this observation, we managed to prove that for the constructed functions (and their natural generalisations) all (infinitely many) conservation laws of the KdV equation, {\color{black} understood in the sense of a sum of integrals}, are satisfied exactly. This approach was inspired by the recent work by \citet{GNST2020,GS2022,GNS2024}, where {\color{black} interesting generalised } solutions with singularities satisfying the Weierstrass--Erdmann corner conditions (e.g. \citealt{CF1954}) and  requiring the continuity of the derivative at the junction, were constructed  in the context of  non-integrable Benjamin--Bona--Mahoney and conduit equations. To the best of our knowledge, the discontinuous {\color{black} generalised (shock-like) } travelling wave solutions of the KdV equation in the form of cnoidal waves with periodicity defects constructed in our paper have not been discussed before. {\color{black} A cnoidal wave with an expansion defect satisfies the corner condition exactly, it's smoothed counterpart was extremely stable in our numerical runs, but a smoothed counterpart of a cnoidal wave with a contraction defect also turned out to be long-lived. } Further discussion of solutions of the KdV equation with defects (both stable and unstable) can be found in Appendix \ref{sec:Appendix A}. Importantly, we conclude that, for the duration of our long simulations, such natural periodicity defects do not lead to the focusing of energy in the absence of rotation.

%%% short-periodicity Defect %%%

\begin{figure}
    \centering
    \includegraphics[width=0.47\linewidth]{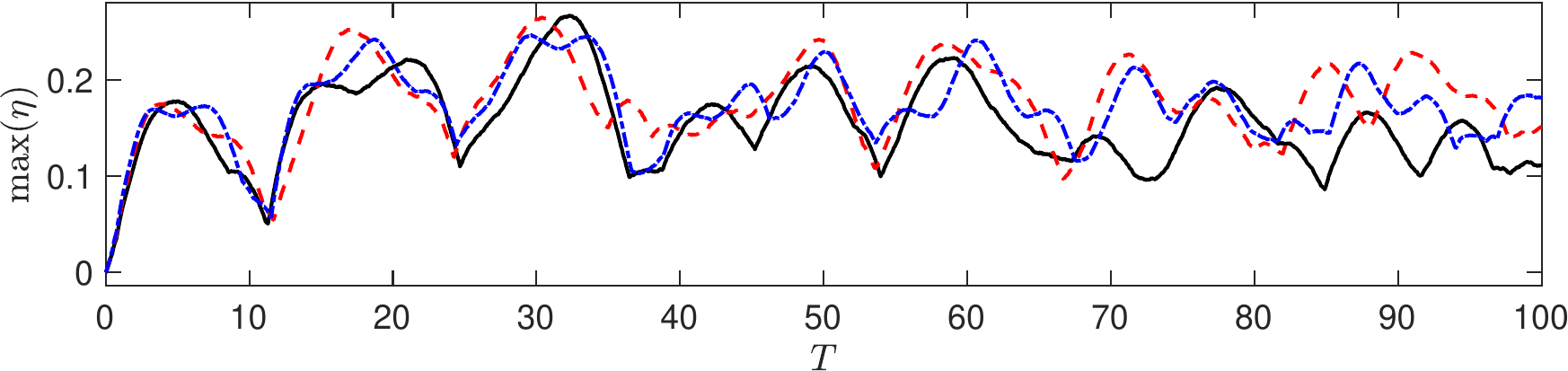}
         \includegraphics[width=0.47\linewidth]{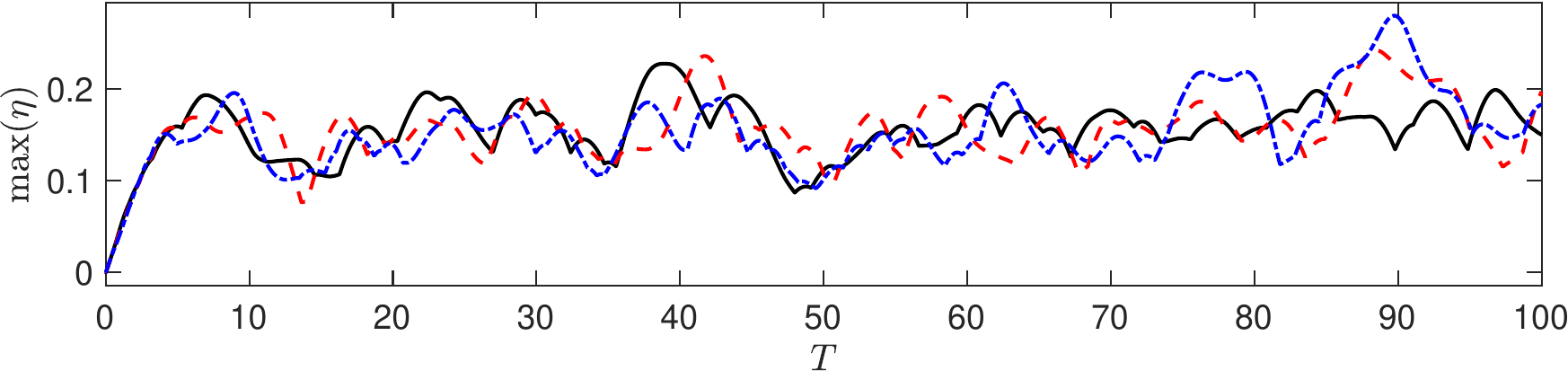}\\[0.5cm]
      
    \includegraphics[width=0.47\linewidth]{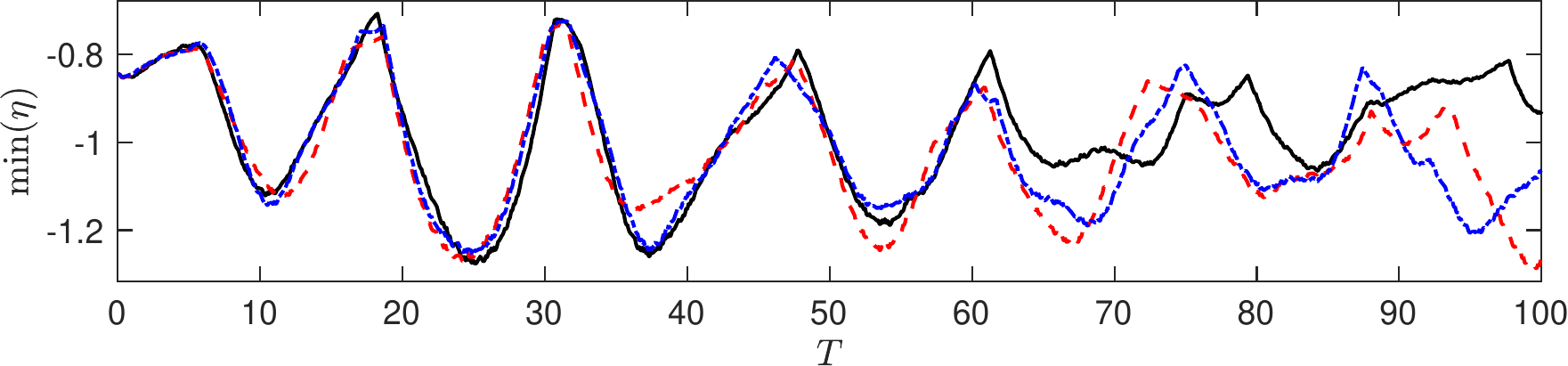}
       \includegraphics[width=0.47\linewidth]{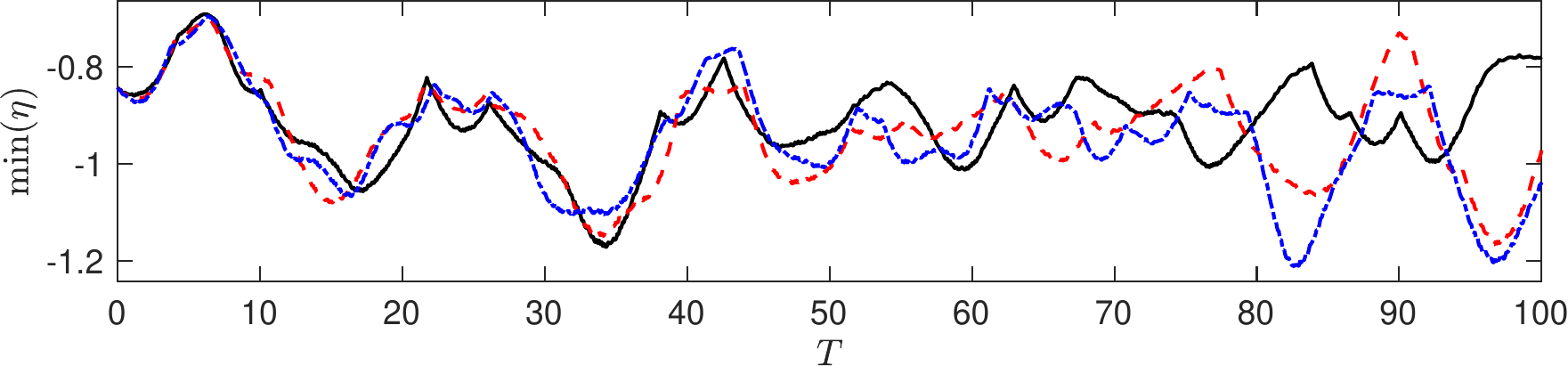}\\[0.5cm]

    \includegraphics[width=0.47\linewidth]{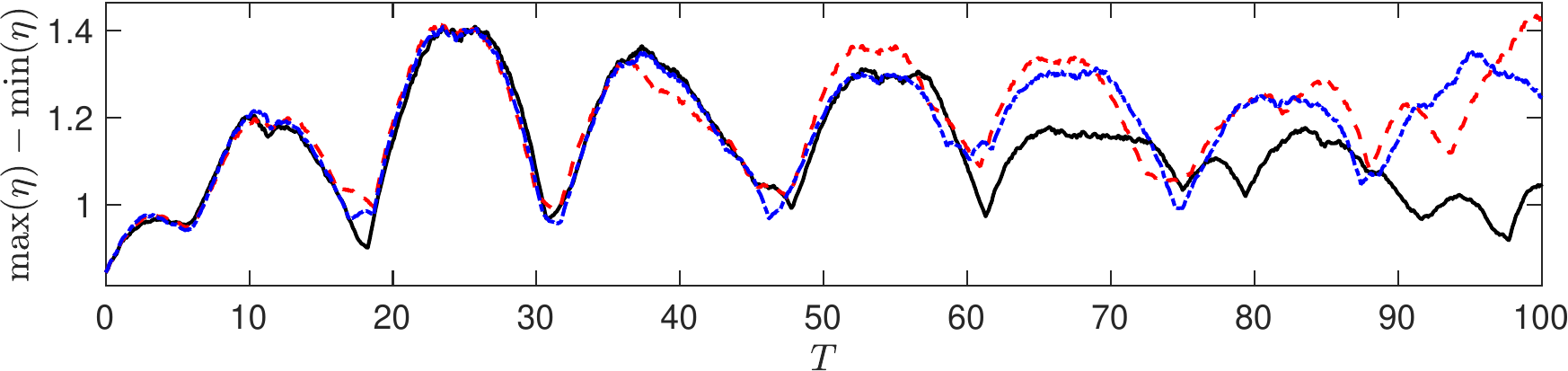}
 \includegraphics[width=0.47\linewidth]{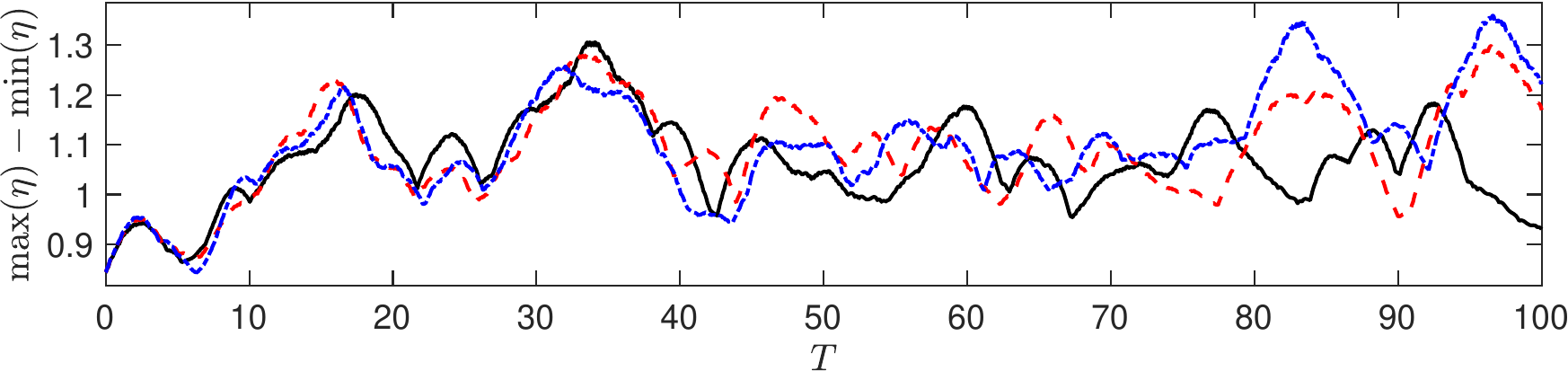}
    \caption{The effect of rotation on the evolution of the maximum, minimum and amplitude of the interfacial displacement for a cnoidal wave with a contraction (left) and expansion (right) defect initial condition. Black solid, red dashed and blue dot-dashed lines correspond to 5, 7 and 9 peaks in the domain, respectively.
    }
  \label{fig:A22}
\end{figure}

\begin{figure}
    \centering
    \includegraphics[width=0.5\linewidth]{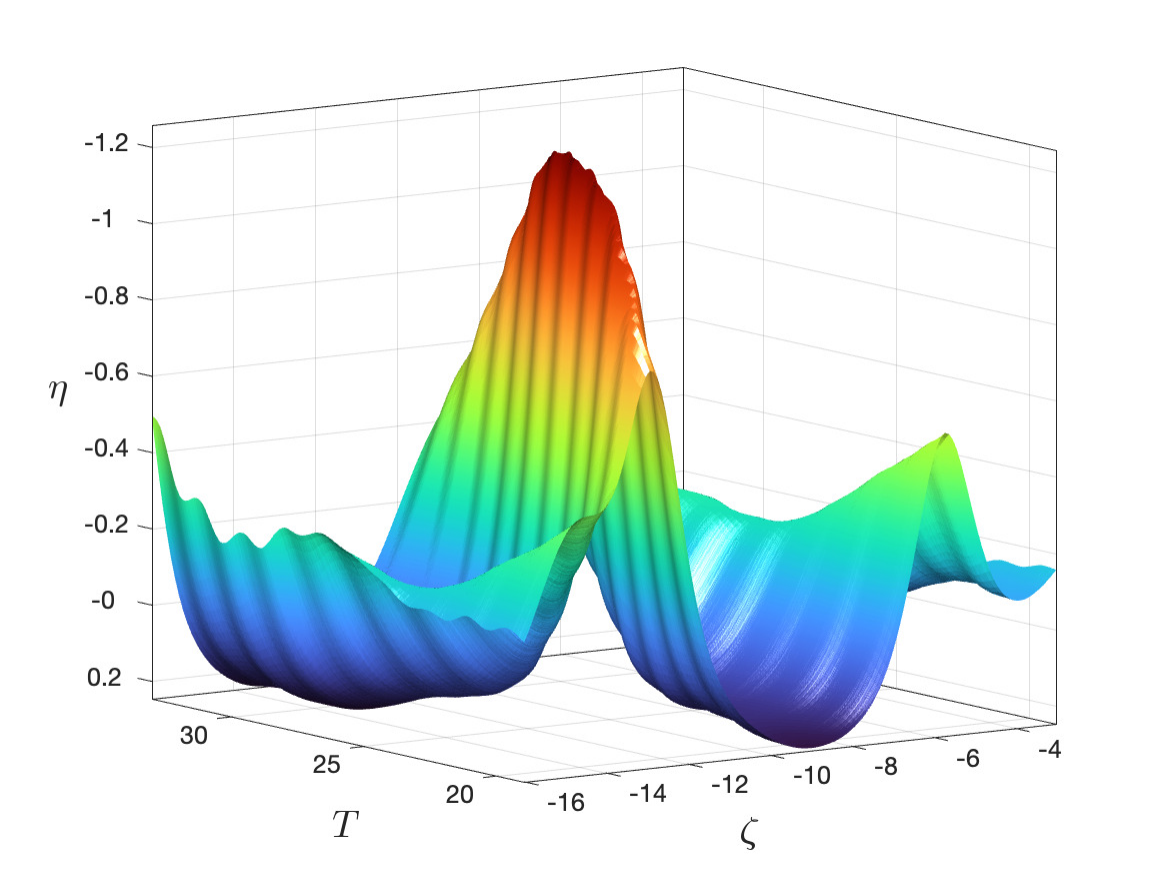}\hspace{-0.5cm}
     \includegraphics[width=0.5\linewidth]{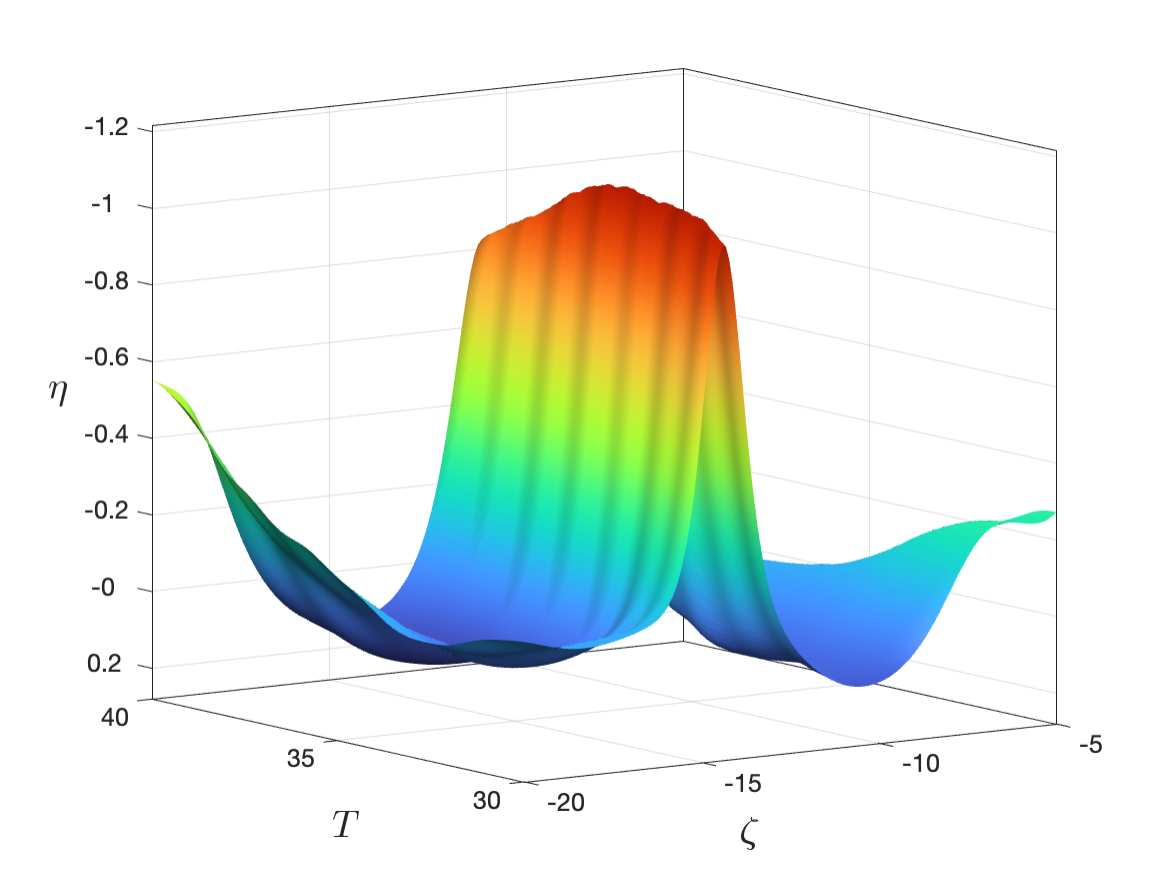}
      \caption{Close-up view from below of the large burst in the interfacial displacement for a cnoidal wave with a contraction (left) and expansion (right) defect around $T = 25$ (left) and $T = 32$ (right), respectively. }
        \label{fig:A25}
\end{figure}

{\color{black} Next, we use the cnoidal waves with expansion and contraction defects as the initial conditions in the presence of rotation.}
% for the Ostrovsky equation (\ref{AS33}). 
The effect of rotation on the cnoidal wave with a contraction and expansion defects is shown in the subsequent rows of  the same Figure \ref{fig:A21.2}, in the left and right columns, respectively, for the interfacial displacement (first row), shear in the direction of wave ropagation (third row) and shear in the orthogonal direction (fourth row).  Here, we see the formation of a strong burst of the interfacial waves and shear in the direction of wave propagation. In the moving reference frame the burst  propagates to the left.  Qualitatively, this is similar to the behaviour observed in the previous section, where numerical runs were initiated using the breather on a cnoidal wave initial conditions, but the important difference is that this effect is solely due to rotation and no bursts of any kind are present in the absence of rotation.

Figure \ref{fig:A22} shows the time evolution of the maximum, minimum and amplitude of the interfacial displacement for both cases of a contraction (left) and expansion (right) periodicity defects. We show that the observed effects of rotation are structurally stable with regard to the size of the computational domain. We experimented with 5, 7 and 9 major peaks in the domain. In the contraction defect case, the computational domains are $2L= 37.42$ (5 peaks), $2L = 55.02$ (7 peaks), $2L = 72.86$ (9 peaks). In the expansion defect case, the computational domains are $2L= 45.44$ (5 peaks), $2L = 63.04$ (7 peaks), $2L = 80.64$ (9 peaks). The remaining numerical parameters are the same as before. The burst forms soon after the initiation of the numerical runs, and the wave continues to grow for a long time after that. The close-up view of the large wave visible in the free surface elevation is shown in Figure \ref{fig:A25} for the time around {\color{black} $T=25$ (contraction defect, left), with approximately 49\% increase in the amplitude compared to the initial condition and  around $T=32$ (expansion defect, right), with an approximately $31\%$ increase in the amplitude compared to the initial condition.} Hence, we conclude that under the effect of rotation both contraction and expansion defects present in the otherwise regular cnoidal wave can lead to the emergence of strong bursts of interfacial waves and shear currents in the direction of propagation of the cnoidal wave.

\subsection{The effect of rotation on cnoidal waves with generic localised defects}

\begin{figure}
    \centering
        \includegraphics[width=0.51\linewidth]{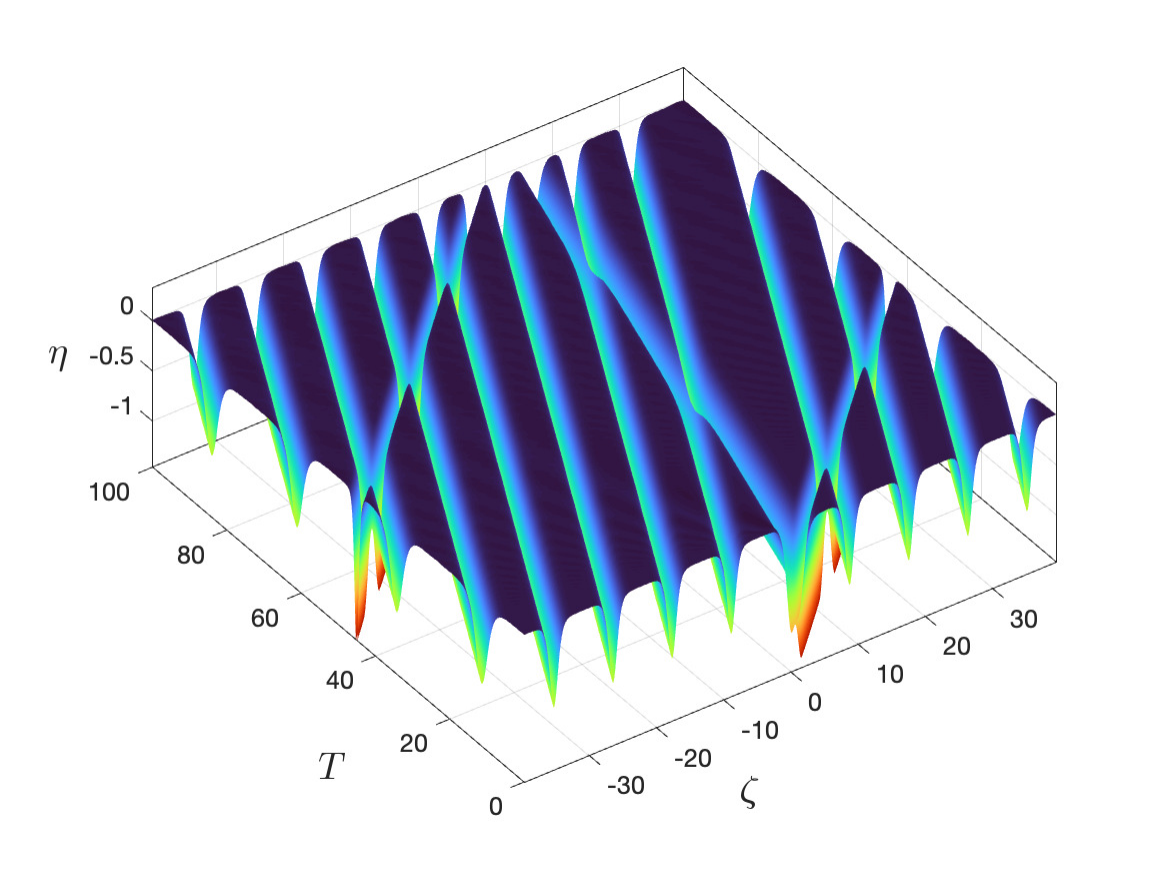}\hspace{-0.5cm}
    \includegraphics[width=0.51\linewidth]{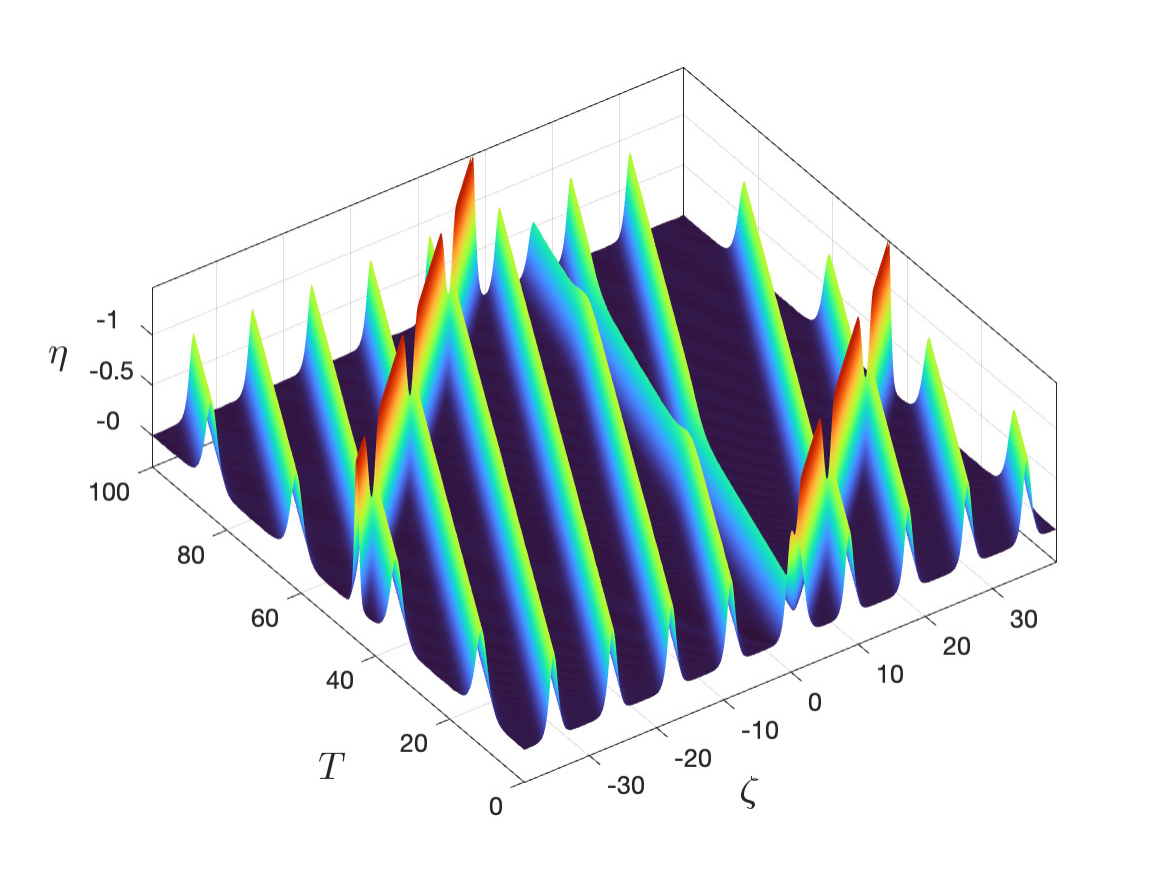}\\[-0.3cm]
    
    \includegraphics[width=0.51\linewidth]{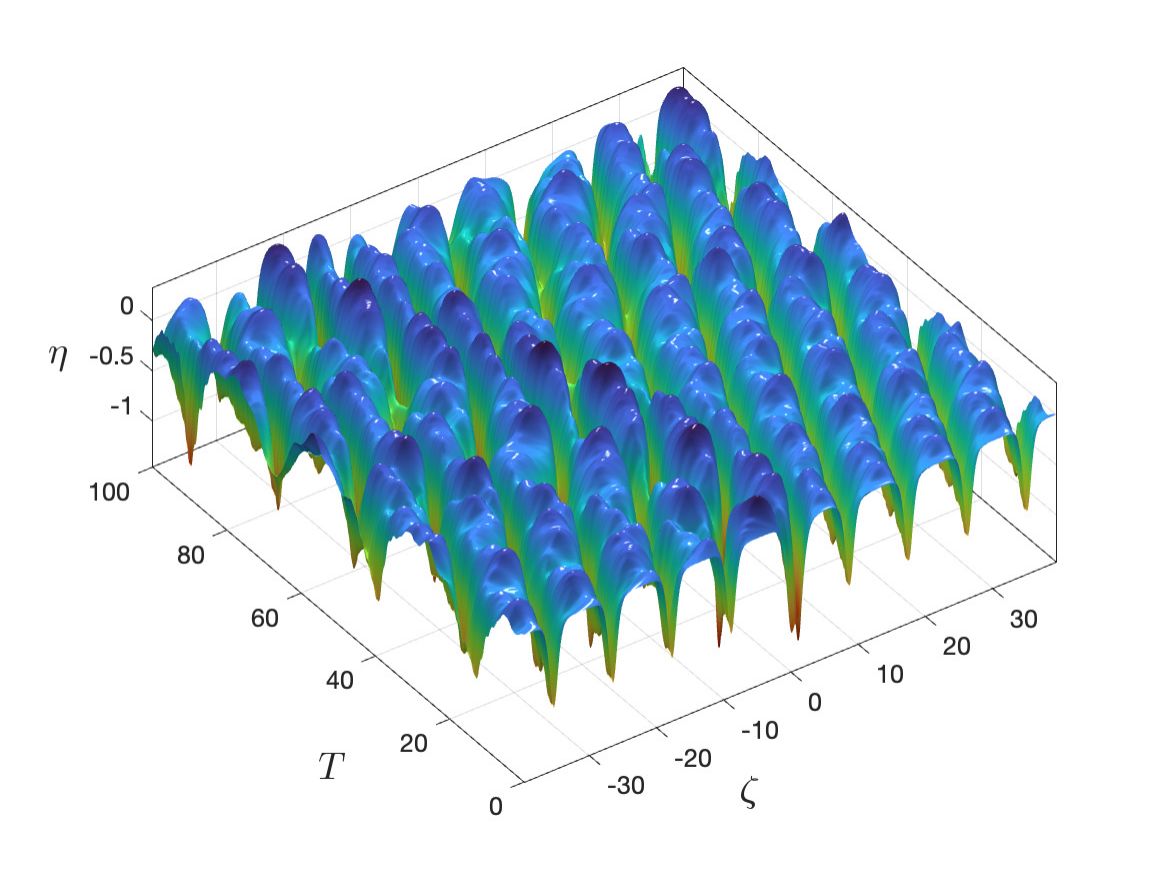}\hspace{-0.5cm}
    \includegraphics[width=0.51\linewidth]{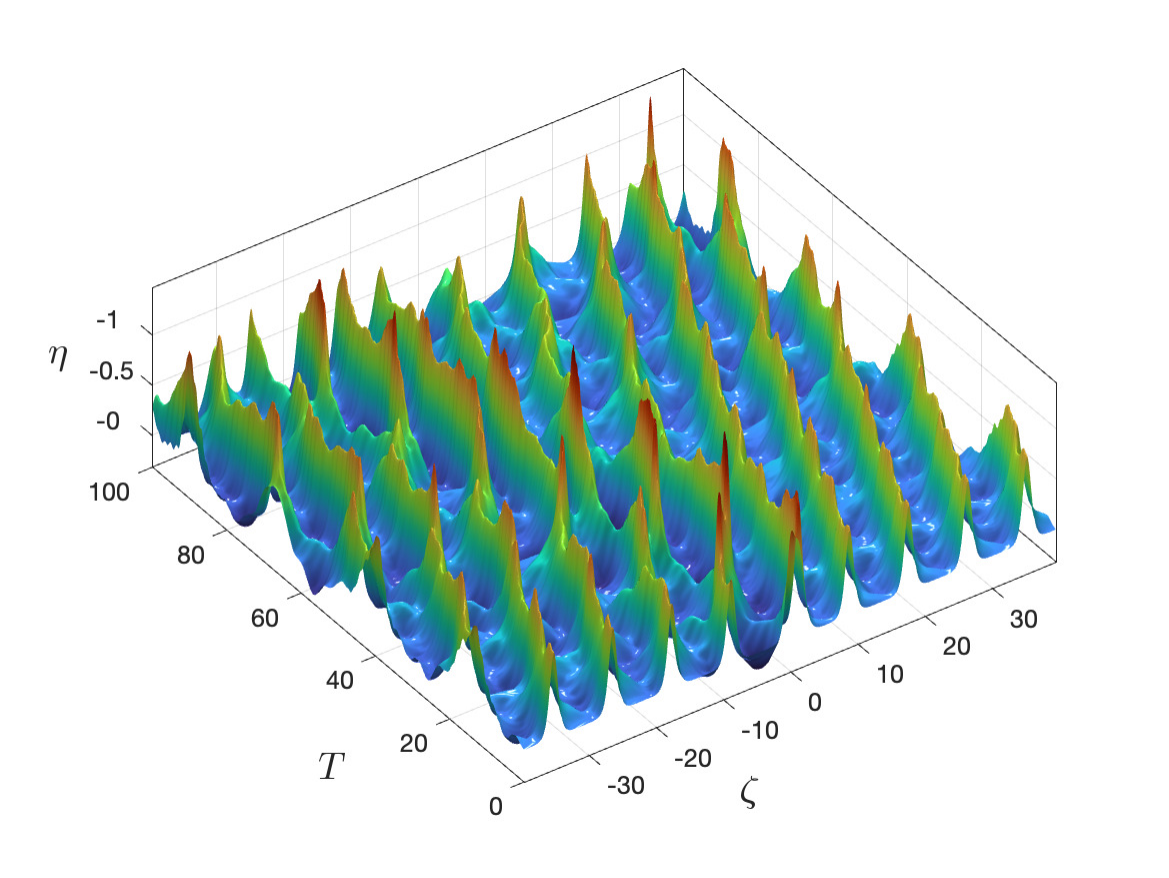}\\[-0.3cm]
   
    \includegraphics[width=0.51\linewidth]{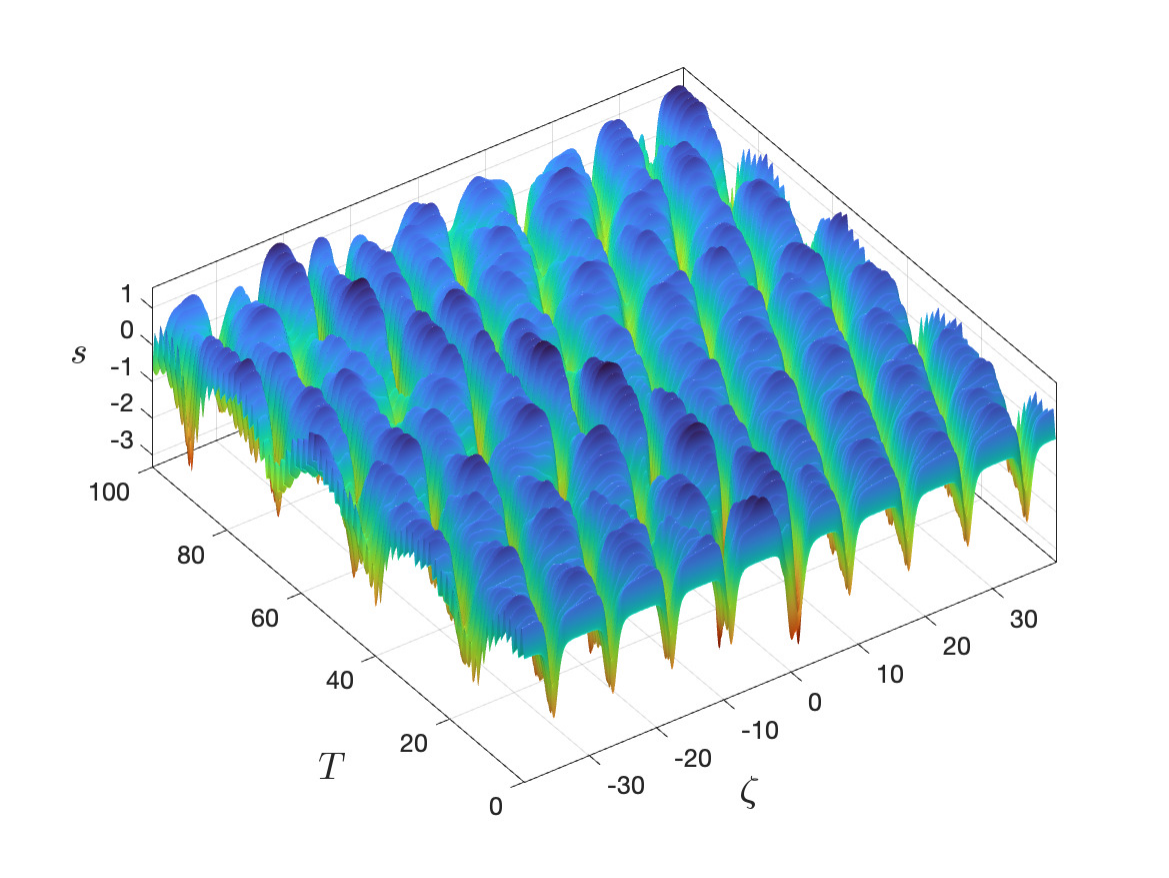}\hspace{-0.5cm}
    \includegraphics[width=0.51\linewidth]{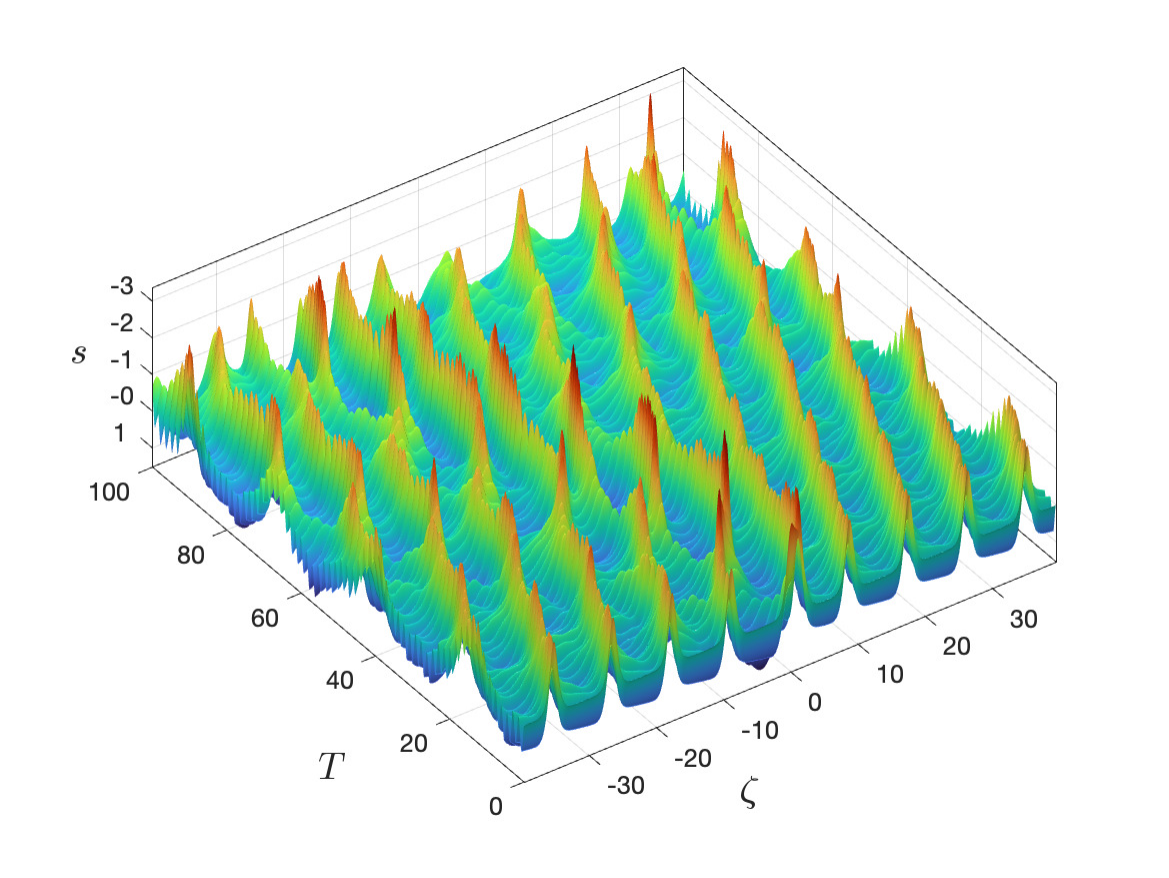}\\[-0.3cm]

    \includegraphics[width=0.51\linewidth]{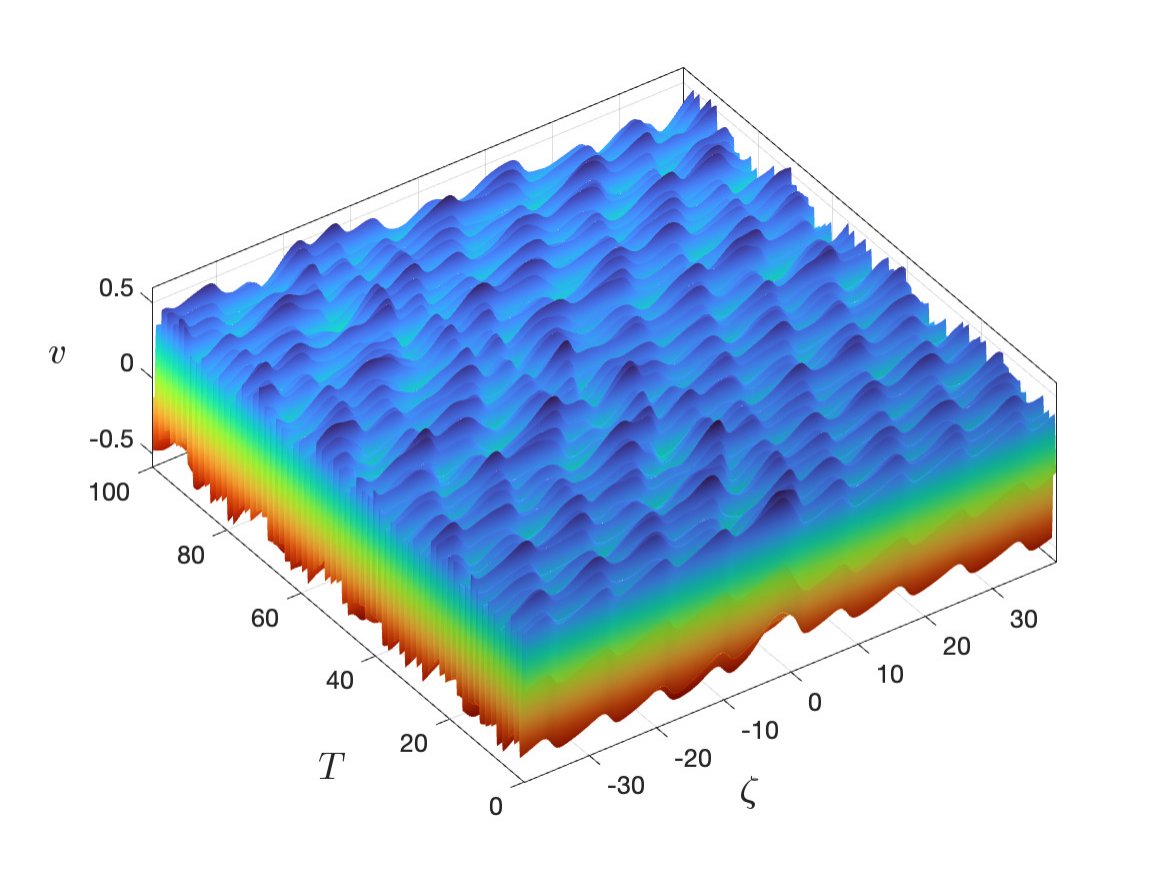}\hspace{-0.5cm}
    \includegraphics[width=0.51\linewidth]{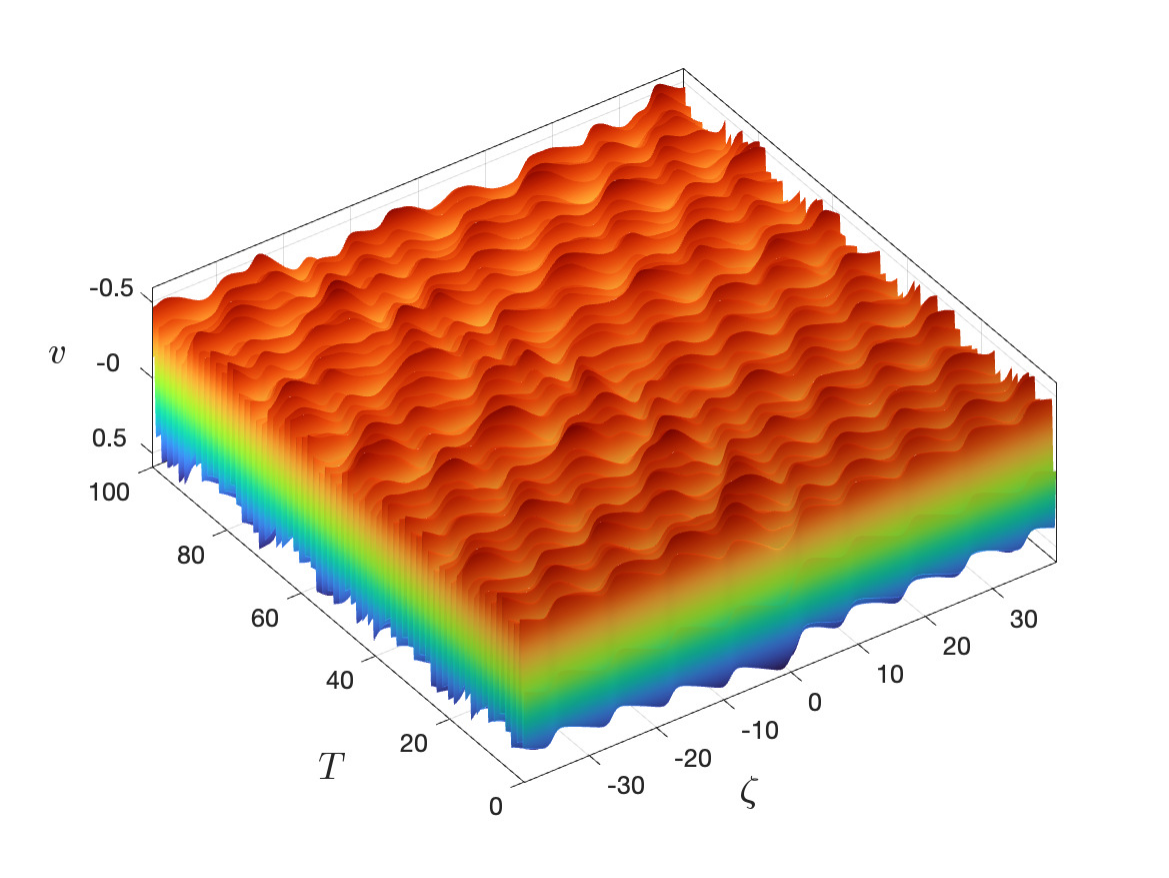}
            \caption{Numerical solution for a cnoidal wave with a generic localised defect initial condition (waves of depression). First row: view from above (left) and below (right) of the interfacial displacement in the absence of rotation. Second row: view from above (left) and below (right) of the interfacial displacement under the effect of rotation. Third  / fourth row: view from above (left) and below (right) of the shear in the direction of wave propagation / orthogonal direction, under the effect of rotation. }
    \label{fig:A21.1}
\end{figure}

\begin{figure}
    \centering
        \includegraphics[width=0.51\linewidth]{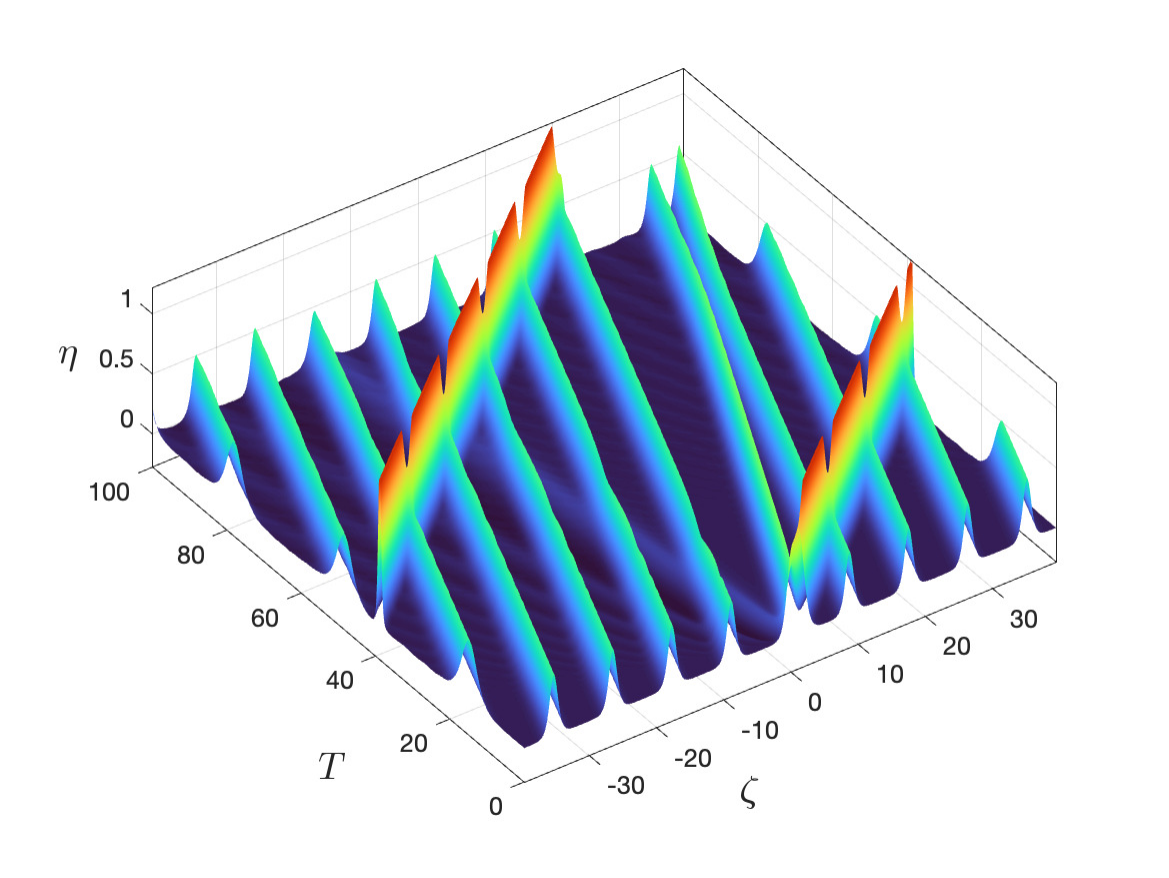}\hspace{-0.5cm}
    \includegraphics[width=0.51\linewidth]{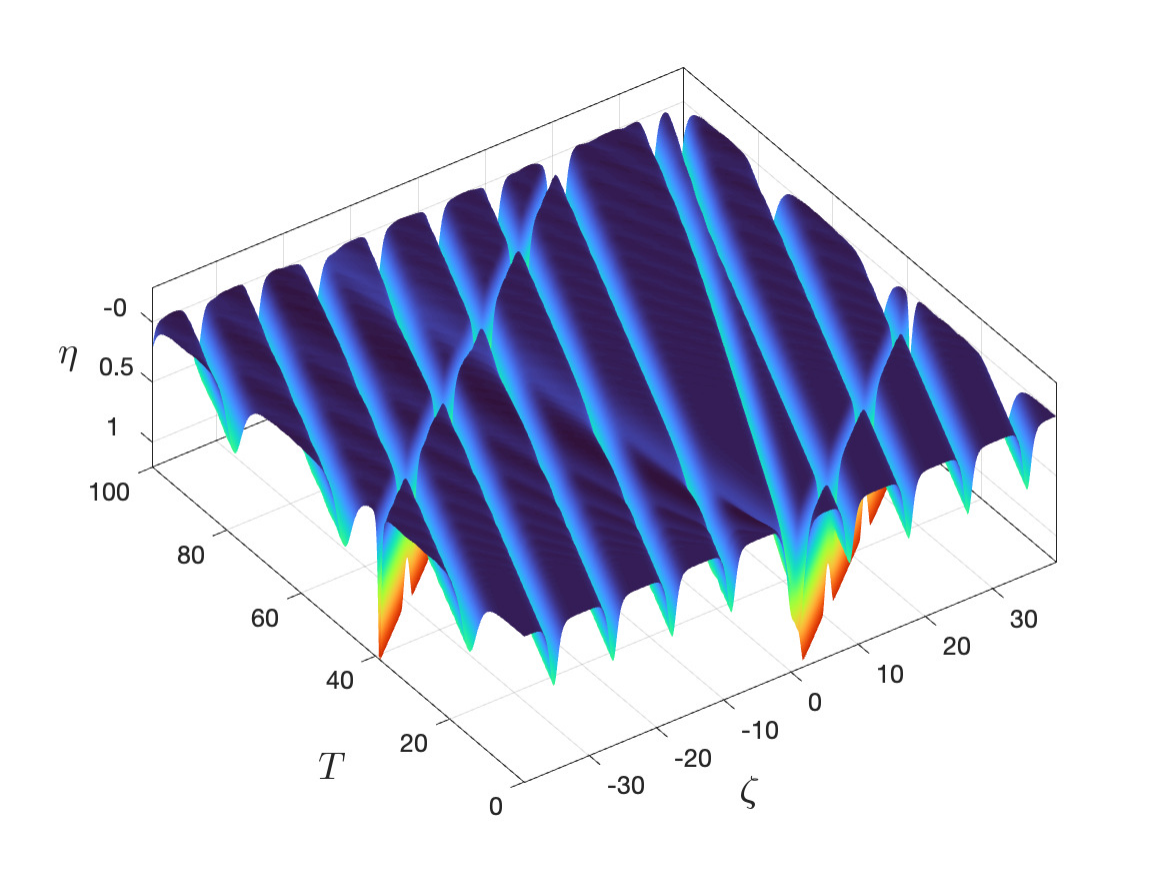}\\[-0.3cm]
    
    \includegraphics[width=0.51\linewidth]{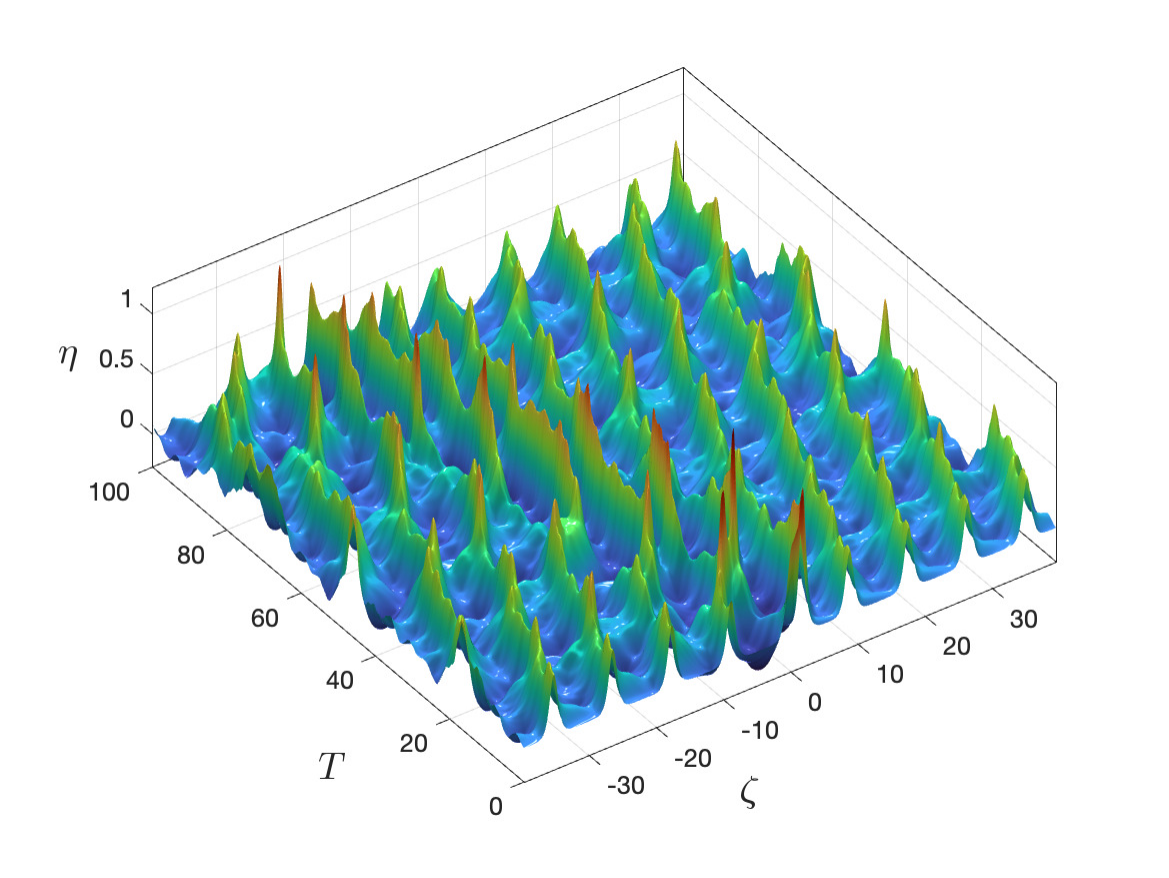}\hspace{-0.5cm}
    \includegraphics[width=0.51\linewidth]{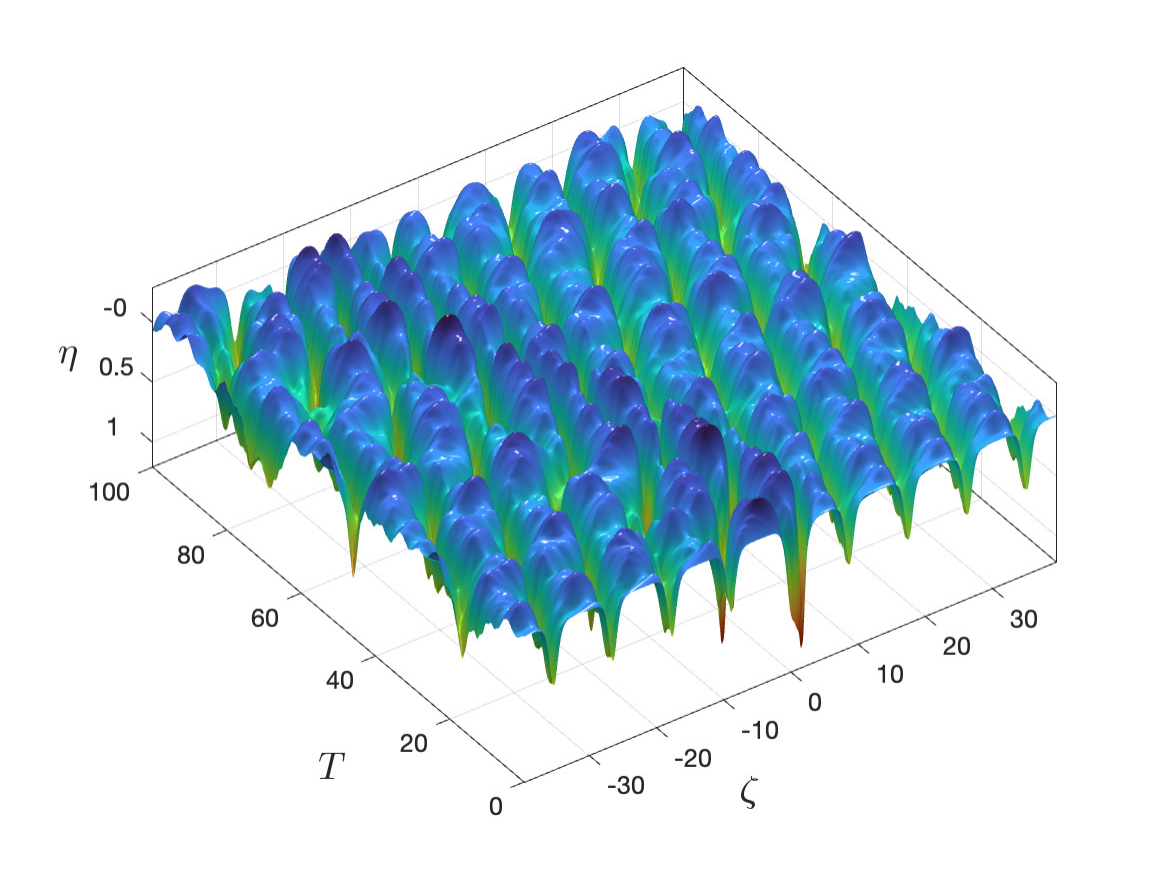}\\[-0.3cm]
   
    \includegraphics[width=0.51\linewidth]{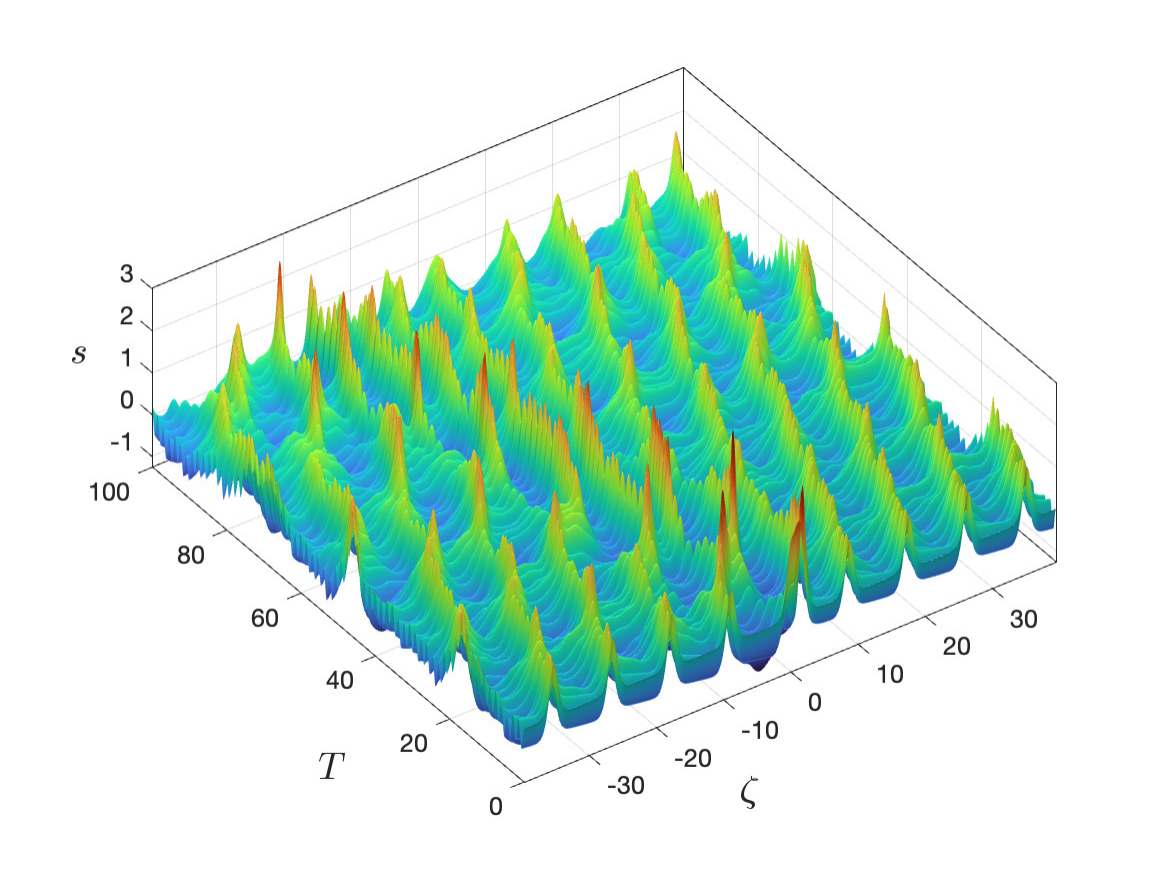}\hspace{-0.5cm}
    \includegraphics[width=0.51\linewidth]{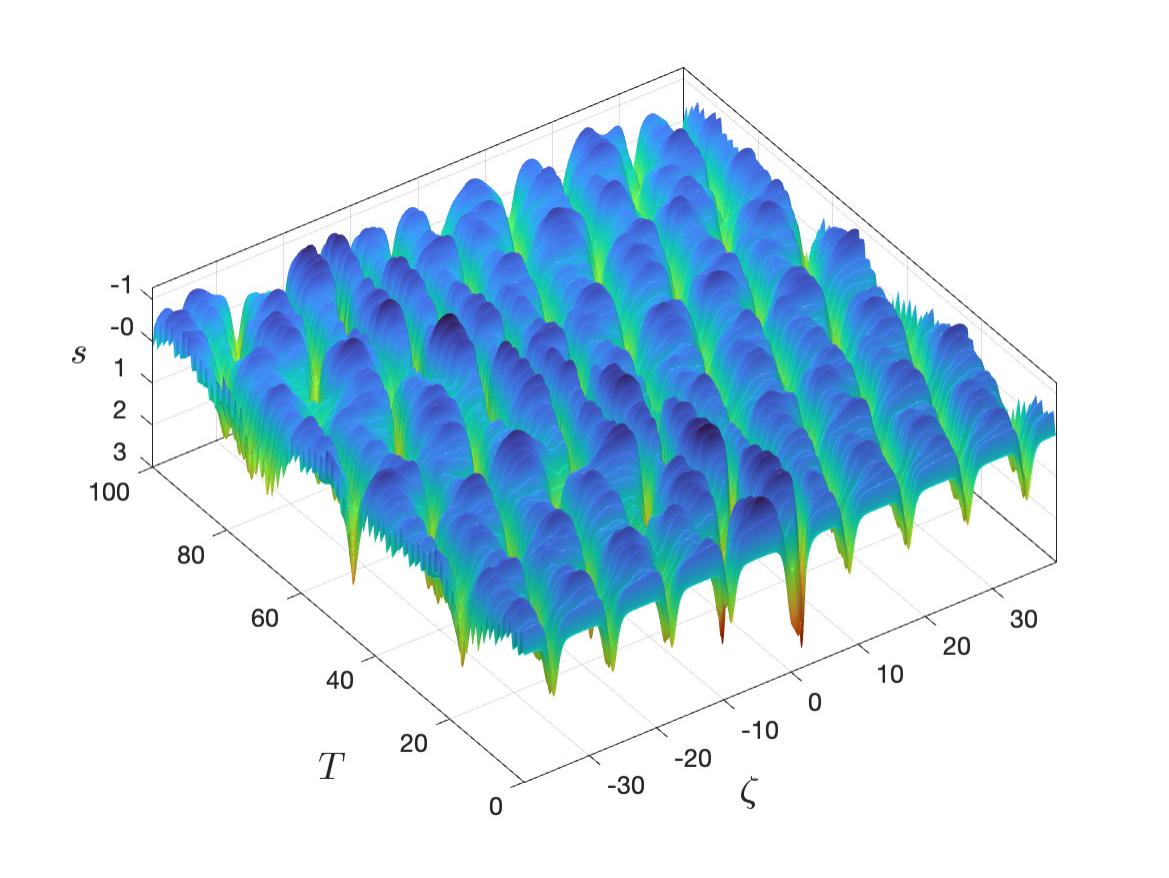}\\[-0.3cm]

    \includegraphics[width=0.51\linewidth]{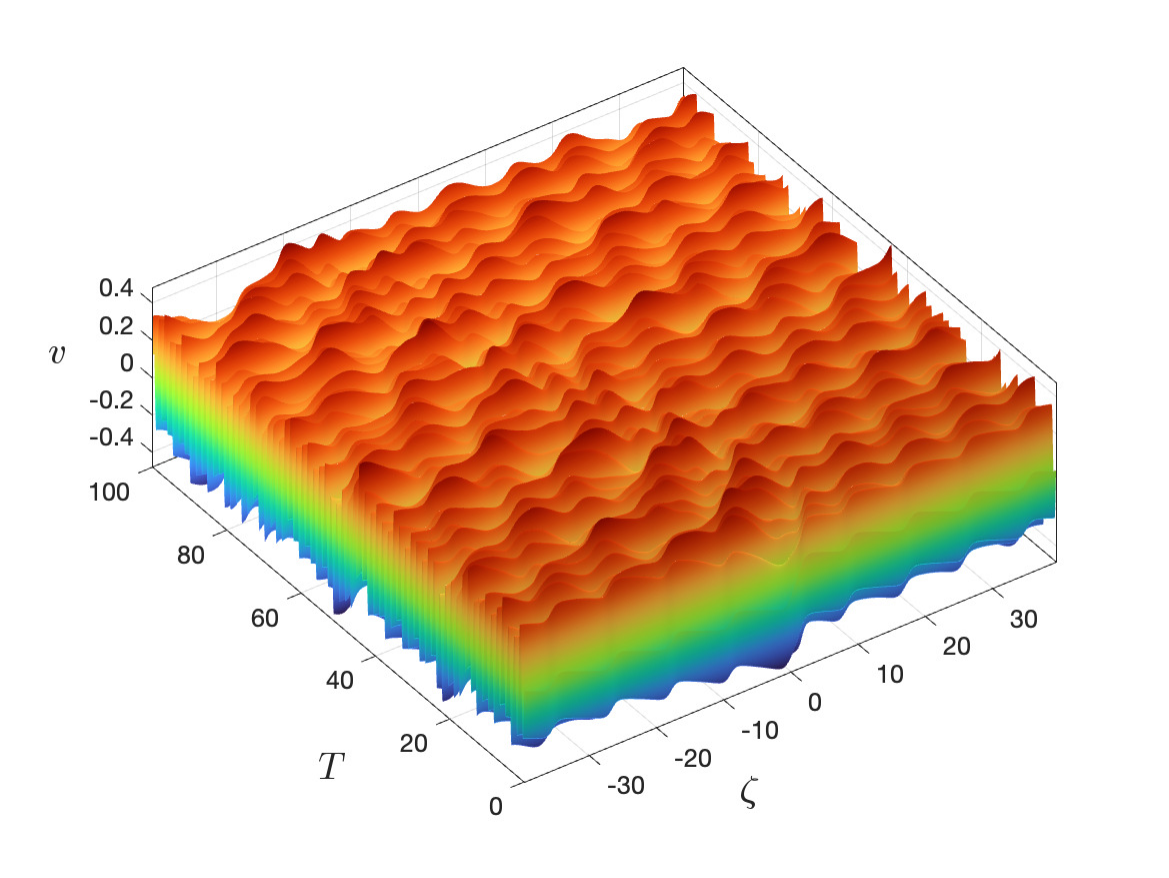}\hspace{-0.5cm}
    \includegraphics[width=0.51\linewidth]{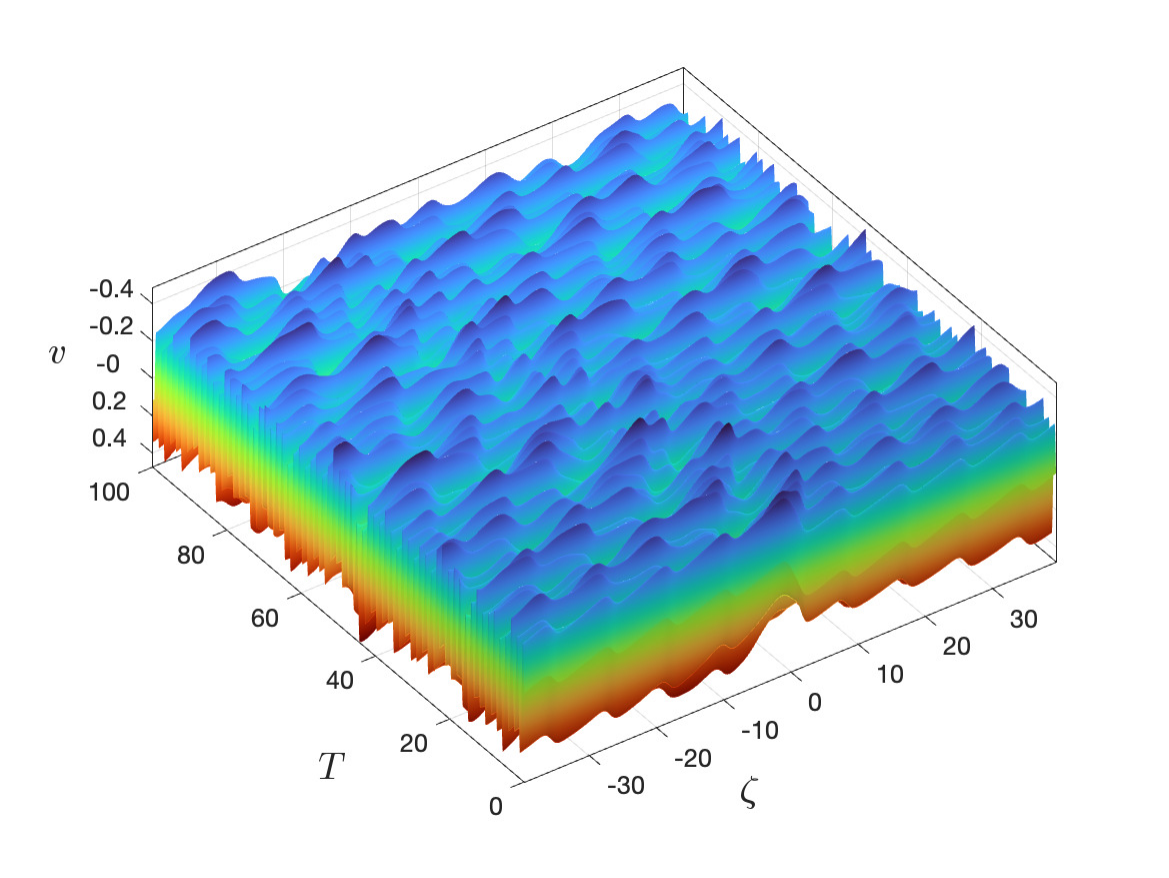}
        \caption{Numerical solution for a cnoidal wave with a generic localised defect initial condition (waves of elevation). First row: view from above (left) and below (right) of the interfacial displacement in the absence of rotation. Second row: view from above (left) and below (right) of the interfacial displacement under the effect of rotation. Third  / fourth row: view from above (left) and below (right) of the shear in the direction of wave propagation / orthogonal direction, under the effect of rotation. }
        \label{fig:A30}
\end{figure}

\begin{figure}
    \centering
    \includegraphics[width=0.47\linewidth]{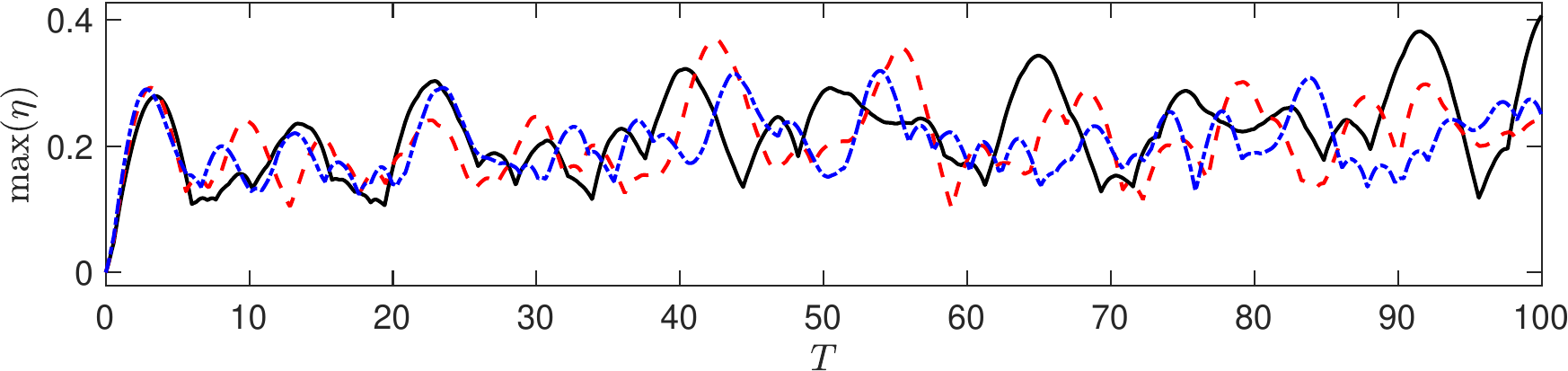}
         \includegraphics[width=0.47\linewidth]{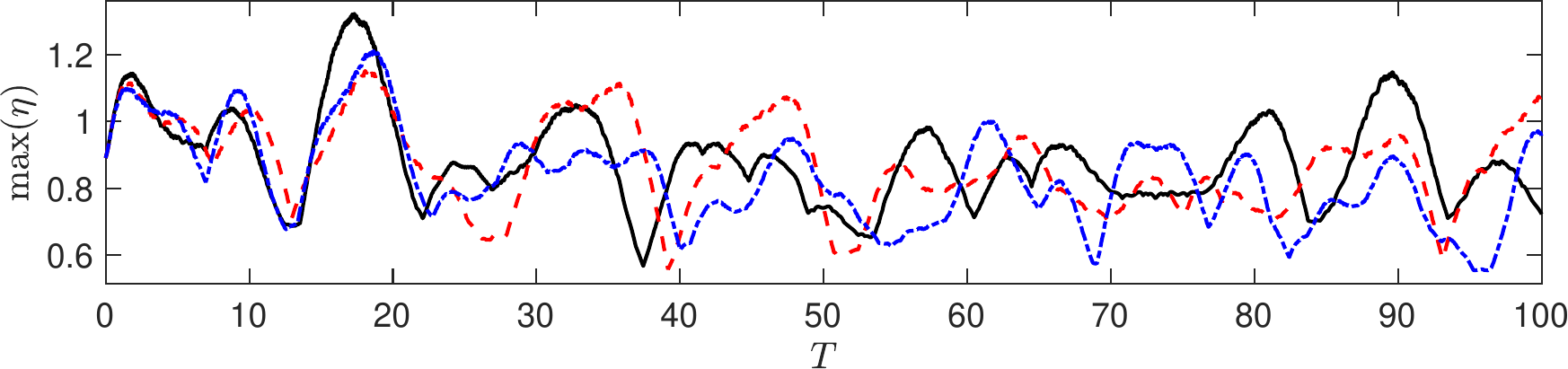}\\[0.5cm]
      
    \includegraphics[width=0.47\linewidth]{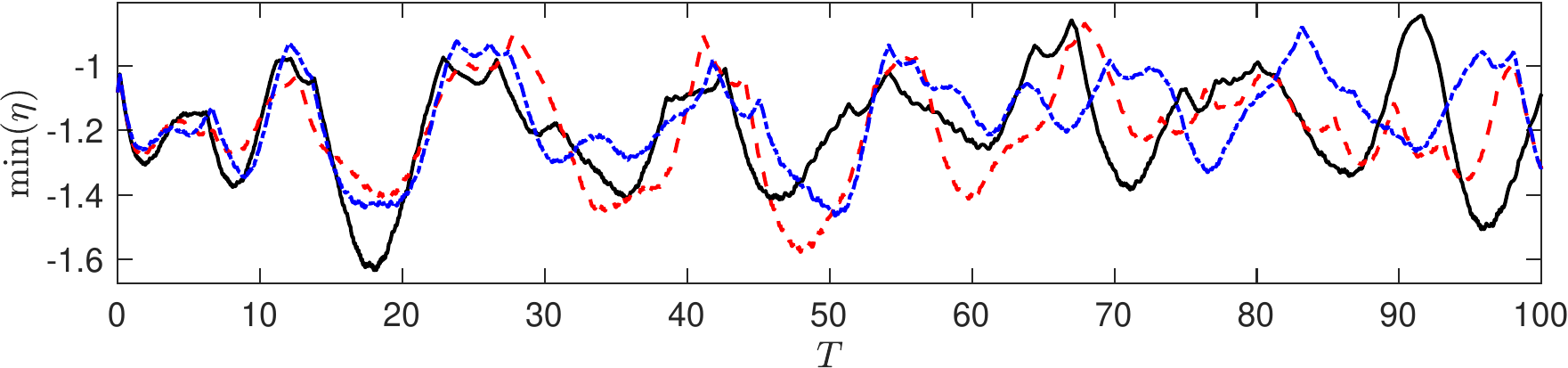}
       \includegraphics[width=0.47\linewidth]{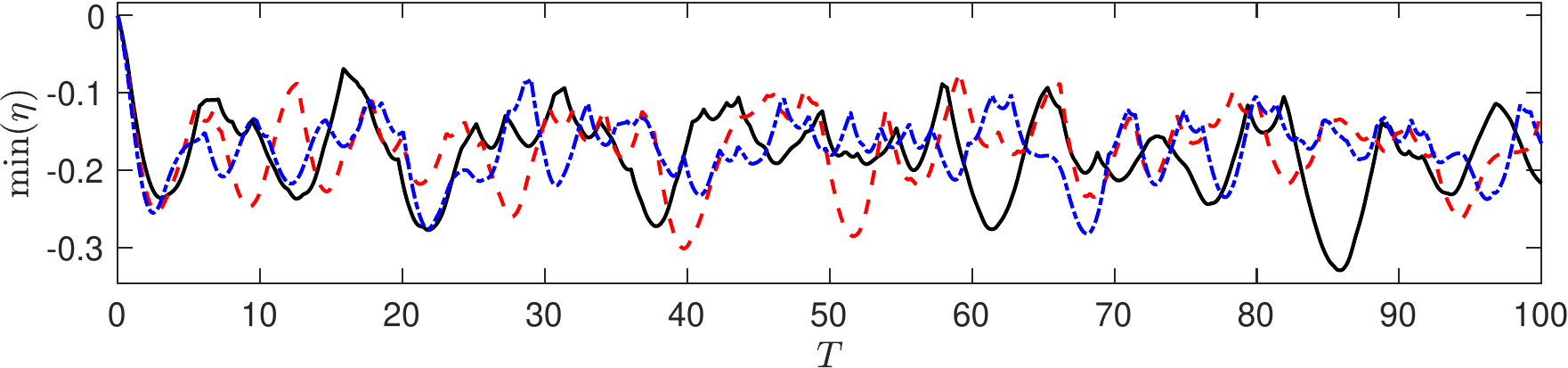}\\[0.5cm]

    \includegraphics[width=0.47\linewidth]{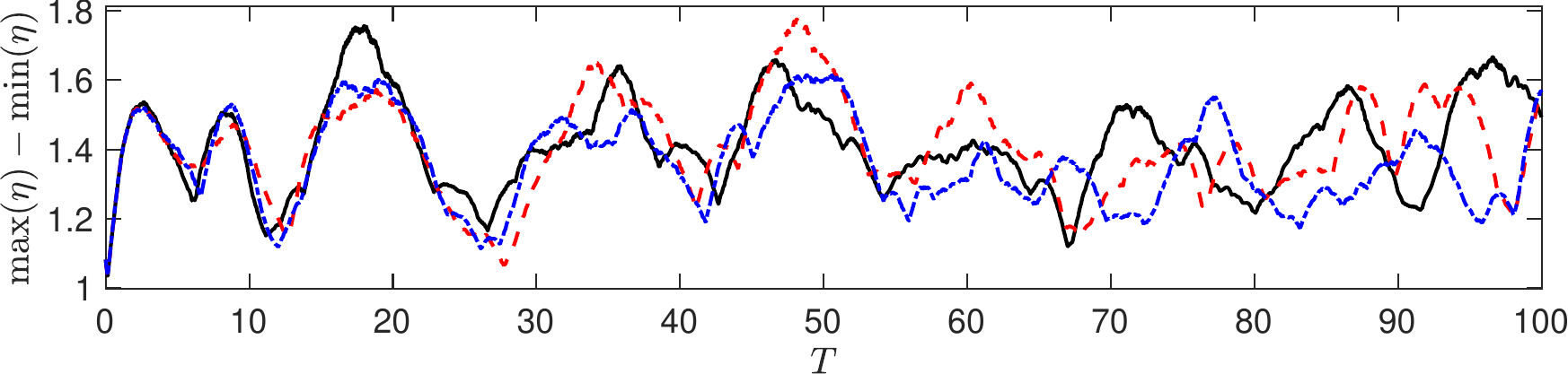}
        \includegraphics[width=0.47\linewidth]{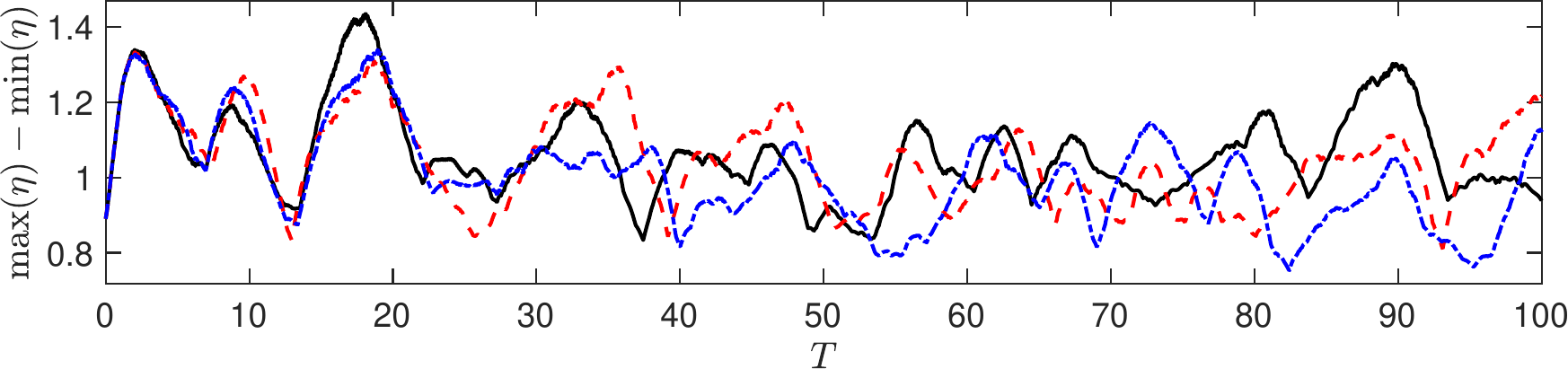}
    \caption{The effect of rotation on the evolution of the maximum, minimum and amplitude of the interfacial displacement for a cnoidal wave with a generic localised defect initial condition: waves of depression (left) and waves of elevation (right). Black solid, red dashed and blue dot-dashed lines correspond to 5, 7 and 9 peaks in the domain, respectively.     }
  \label{fig:A24}
\end{figure} 

\begin{figure}
    \centering
    \includegraphics[width=0.5\linewidth]{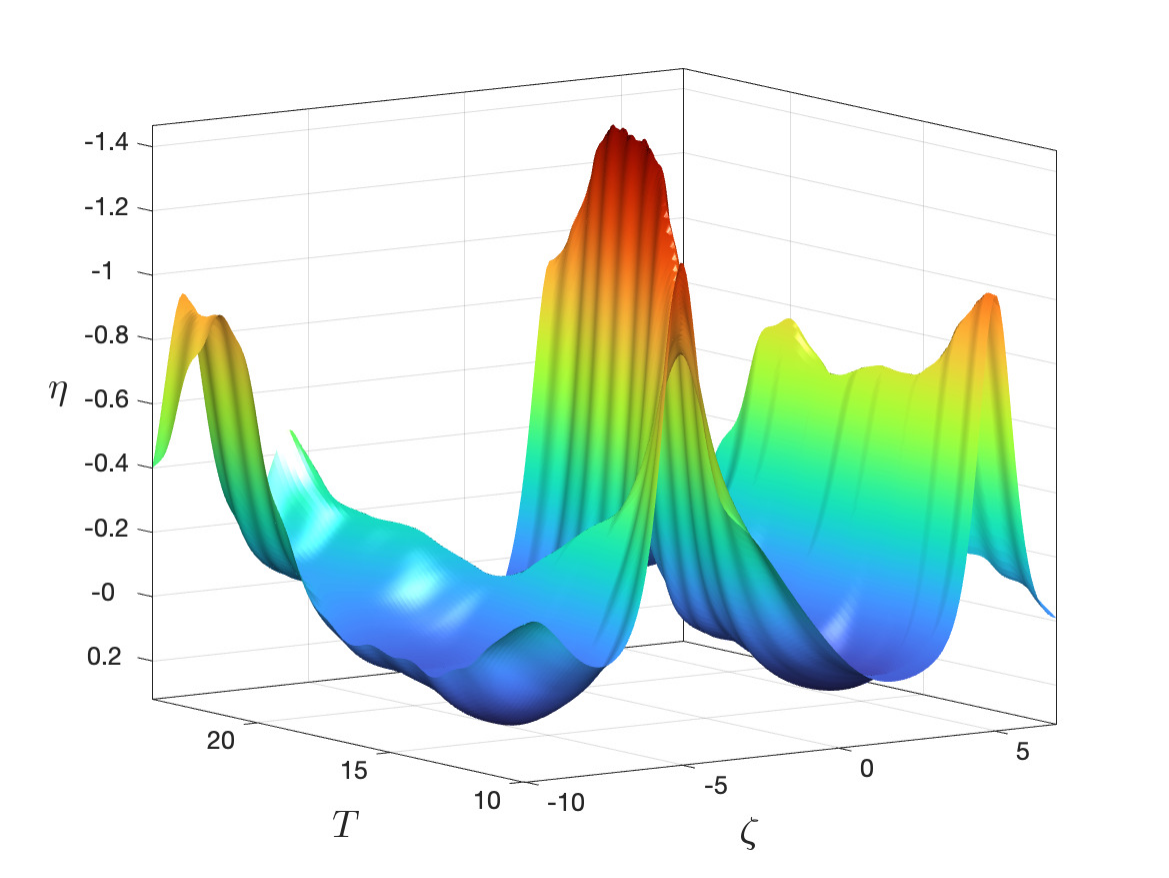}\hspace{-0.5cm}
    \includegraphics[width=0.5\linewidth]{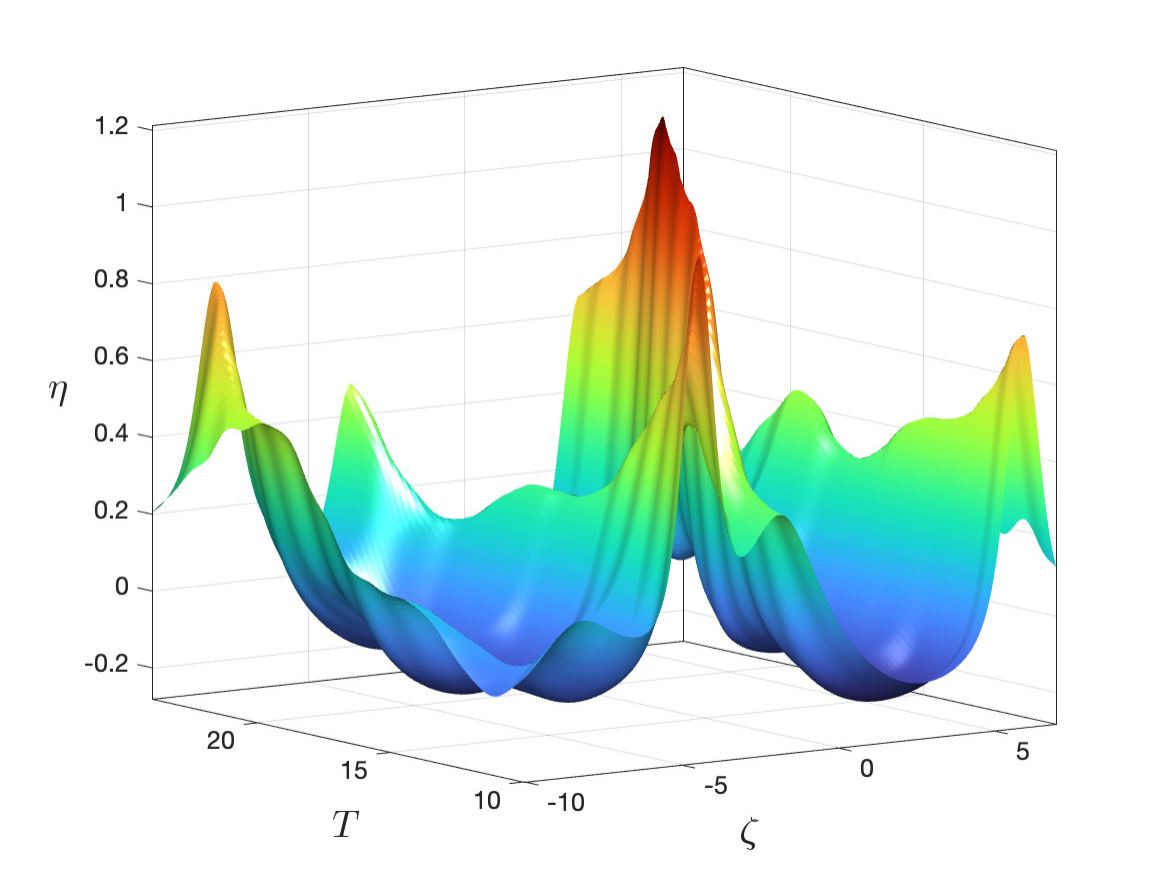}
      \caption{Close-up view from below (left) and above (right) of the large bursts in the interfacial displacement for a cnoidal wave with a generic localised defect initial condition  around $T = 19$ for the waves of depression (left) and waves of elevation (right). }
        \label{fig:A24.1}
\end{figure}

Finally in this section, motivated by the recent numerical and experimental generation of breathers in a fluid conduit during the interaction of a cnoidal wave with a soliton by \citet{MH2016, MCXH2023} and generalising this scenario, we initiate numerical runs for our problem with the initial condition which can be described as a cnoidal wave with a generic localised perturbation. Namely, the initial condition is given by the function 
\begin{align}
    \eta^{(0)}|_{T=0} = \dfrac{6\beta_1}{\alpha_1}\Bigg\{ u_2 + (u_3 - u_2) \text{ cn}^2\Big[ (\zeta {\color{black} + \frac{\alpha_1 \hat \eta^{(0}}{3} T} - v_c T) \sqrt{\dfrac{u_3 -u_1}{2}} ; m \Big] \Bigg\}_{T=0} \nonumber  \\
    + A_1 \text{sech}^2 \Big[ A_2 (\zeta +\zeta_0)  \Big], 
    \label{cnoidal_ad}
\end{align}
and $u_1 < u_2 < u_3$ are real, $v_c = 2 \beta_1 (u_1 + u_2 + u_3)$, 
%$\alpha_1 = \frac{3\sigma}{2c_0}$, $\beta_1=\frac{ \beta c_0^3}{6}$ are the coefficients of the Ostrovsky equation \eqref{eq:ost}, 
$m ={(u_3-u_2)}/{(u_3-u_1)}$, where $A_1$, $A_2$ and $\zeta_0$ are arbitrary constants.  

We experiment with both sets of physical parameters discussed at the beginning of Section \ref{sec: 4}. For the first set, numerical solutions initiated with the initial condition with a generic localised defect are shown in Figure \ref{fig:A21.1}. The computational domain is $2L = 79.20$, with the number of modes $M=790$, the spatial step $\Delta x \approx 10^{-1}$, the total simulation time $T_{max} = 100$, and the temporal step $\Delta T = 10^{-2}$. The parameters used in the initial condition are $u_1 = -10^{-3},\, u_2=0,\, u_3=3,\, A_1 = -0.80,\, A_2 = 1.00$, and  $\zeta_0 = -1.30$. 

The initial condition introduces a localised perturbation to the otherwise regular cnoidal wave. In our runs, in the absence of rotation, the introduction of such a localised defect leads to the formation of a pair of bright and dark breathers, as well as the clearly noticeable expansion defect, which is shown in the first row of the Figure \ref{fig:A21.1}, showing both the view from above (left) and the view from below (right).  Qualitatively, this is similar to the waves generated in the interaction of a cnoidal wave with a solitary wave in a conduit in \cite{MCXH2023}.  In the moving reference frame, the bright breather propagates to the right, while the dark breather propagates to the left, and the expansion defect moves with the speed of the cnoidal wave.  The effect of rotation on the evolution of the same initial condition is shown in the subsequent rows of the same figure, for the top (left)  and bottom (right) views of the interfacial displacement (second row), shear in the direction of wave propagation (third row) and shear in the orthogonal direction (fourth row). There are strong bursts both in the interfacial displacement and shear in the direction of wave propagation, which can be associated with the previously considered types of ``defects" in the otherwise regular cnoidal wave. The signal in the shear in the orthogonal direction is again weak. Hence, we conclude that the mechanism of formation of the bursts of strong interfacial displacements and shear in the direction of wave propagation  in these simulations can be interpreted as formation of KdV-type breathers and the expansion defect on a cnoidal wave background, with the subsequent effect of rotation.

In order to test our theoretical framework  further, we also modelled the effect of rotation on internal waves of elevation, using our second set of parameters discussed at the beginning of Section \ref{sec: 4}. 
In Figure \ref{fig:A30}, the computational domain is $2L = 79.20$, with the number of modes $M=790$, the spatial step $\Delta x \approx 10^{-1}$, the total simulation time $T_{max} = 100$, and the temporal step $\Delta T = 10^{-2}$. The parameters used in the initial condition are $u_1 = -10^{-3},\, u_2=0,\, u_3=3,\, A_1 = 0.80,\, A_2 = 1.00$, and  $\zeta_0 = -1.30$. 
Here, the pycnocline is closer to the bottom, the upper layer depth takes $80\%$ of the total depth, whereas in previous cases it was $25\%$. %The results are shown in Figures \ref{fig:A30}. 
The results are similar to our previous simulations for internal waves of depression, but here we see the formation of a pair of bright and dark breathers, as before, but also the noticeable formation of both expansion and contraction defects. The comparison of the evolution of maxima, minima and amplitude of the waves shown in Figures \ref{fig:A21.1} and \ref{fig:A30} is given in Figure \ref{fig:A24},  illustrating the growth of the amplitude of the waves and further supporting our interpretation of the observed (in all cases) formation of bursts of strong interfacial displacements and shear in the direction of wave propagation. For both sets, the computational domains are $2L= 44.00$ (5 peaks), $2L = 61.60$ (7 peaks), and $2L = 79.20$ (9 peaks).  The remaining numerical parameters are the same as before. The results are structurally stable. The large waves formed at the early stage of our computations are shown in Figure \ref{fig:A24.1}. {\color{black} At around $T=19$, the first and second sets have approximately $33\%$ and $36\%$  increase, respectively, in the amplitudes relative to the initial conditions.}

%\clearpage
\section{Discussion}
\label{sec: 5}

In this paper, we addressed several issues related to the modelling of internal waves in the Ostrovsky regime, i.e.\ in the KdV regime with the account of rotation. {\color{black} We chose Helfrich's f-plane extension (MMCC-f) of the two-layer MMCC model as our parent system.}
% (it would be better to call the system MMCC model, recognising the contribution of  \citealt{M1989}). 
The MMCC-f model has similar properties to the full Euler equations with rotation with respect to construction of the weakly-nonlinear solution, but the technical details related to the derivation are more manageable. In our derivation, we represented all field variables as the sums of time-dependent mean values and both spatially- and time-dependent deviations from these evolving means. We considered periodic solutions, and mean-field equations were obtained by averaging the equations over the period of the problem. In contrast to our previous research within the scope of the Boussinesq-type equations, the resulting equations for the mean fields turned out to be coupled to the equations for the deviations, which presented a significant new challenge. We managed to construct a large class of solutions for uni-directional waves by introducing two slow-time variables, and simultaneously constructing asymptotic expansions in powers of the square root of the amplitude parameter (as opposed to the traditional derivation of the Ostrovsky equation with just one slow time and in powers of the amplitude parameter). Since the resulting reduced model, the Ostrovsky equation, has been derived for the zero-mean deviations by construction, the so-called ``zero-mean contradiction" meaning the existence of the zero-mean constraints on all field variables of the parent system has been by-passed.

Next, we used the constructed weakly-nonlinear solution combined with extensive numerical modelling using the Ostrovsky equation in order to study the effect of rotation on the evolution of internal waves with initial conditions in the form of the KdV cnoidal waves, but with various local defects. We built the phenomenological framework by considering pure bright and dark breather on a cnoidal wave initial conditions and cnoidal waves with contraction and expansion periodicity defects. The latter defects were introduced `by hand', by modifying the cnoidal wave solution near the trough in  between the two peaks, and we showed that such functions satisfy all (infintely many) conservation laws of the KdV equation, {\color{black} where the integration is understood as the sum over the natural subintervals (see Appendix A).} Moreover, the cnoidal waves with an expansion defect also have continuous first derivative, satisfying the so-called `corner condition' necessary for it to be a non-smooth extremal of the relevant variational problem. It must be noted that in our numerical runs {\color{black} initial conditions with  both the contraction and expansion defects led to the long-lived states, and a cnoidal wave with an expansion defect behaved very closly to a travelling wave of the KdV equation. } The important difference between the defects represented by dislocation- (bright and dark breather)  and periodicity- (contraction and expansion)  perturbations of the cnoidal wave is that the second type does not lead to formation of bursts in the absence of rotation. Qualitatively, both types of defects lead to formation of bursts of large interfacial displacements and shear in the direction of wave propagation, under the effect of rotation. 

Finally, we considered initial conditions in the form of cnoidal waves with generic localised {\color{black} perturbations}. Our modelling has shown that, in the absence of rotation, such initial conditions typically produce a pair of a fast-moving bright and slow-moving dark breathers and expansion and contraction periodicity defects, moving with the speed of the cnoidal wave. Under the effect of rotation, the splitting of a generic localised {\color{black} perturbation} into these `basic' defects is followed by  formation of  several bursts of interfacial displacements and shear in the direction of wave propagation, which can be associated with the effect of rotation on breathers and the periodicity defects. In all simulations the large bursts propagated  to the left in our moving reference frame, i.e. slower than the {\color{black} speed of our moving reference frame}. {\color{black} We note that we compared the evolution of a cnoidal wave with a localised perturbation in the simulations with periodic boundary conditions with the simulations in a bigger domain and with zero boundary conditions and sponge layers near the boundaries. In the latter case the initial condition coincided with the former one in the central part of the domain, and then was gradually reducing to zero towards the boundaries. Our results have shown good agreement between these two simulations in the central part of the latter simulation. Hence,  for the problems of that type simulations with periodic boundary conditions in a smaller domain can be used instead of the simulations in the large domain.}

\begin{figure}
    \centering
    \includegraphics[width=0.52 \linewidth]{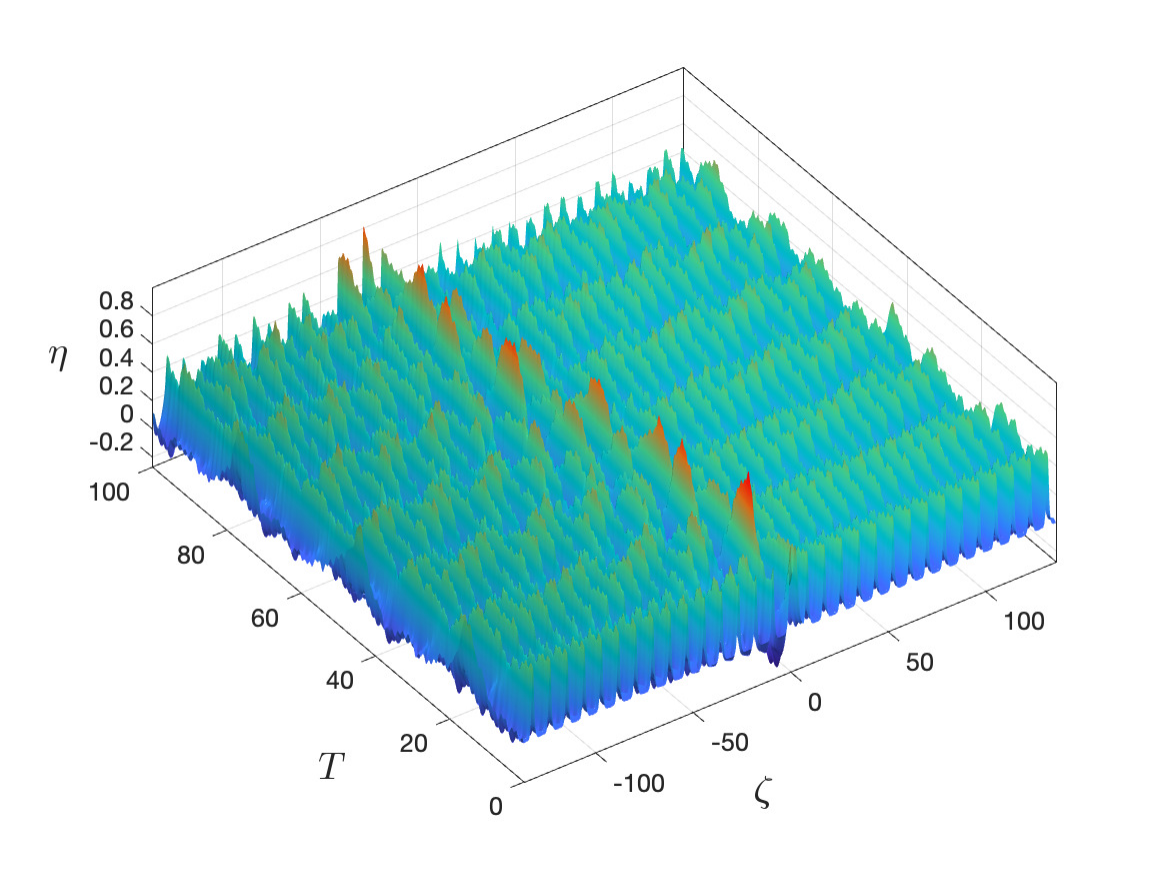}   
    \includegraphics[width=0.44  \linewidth]{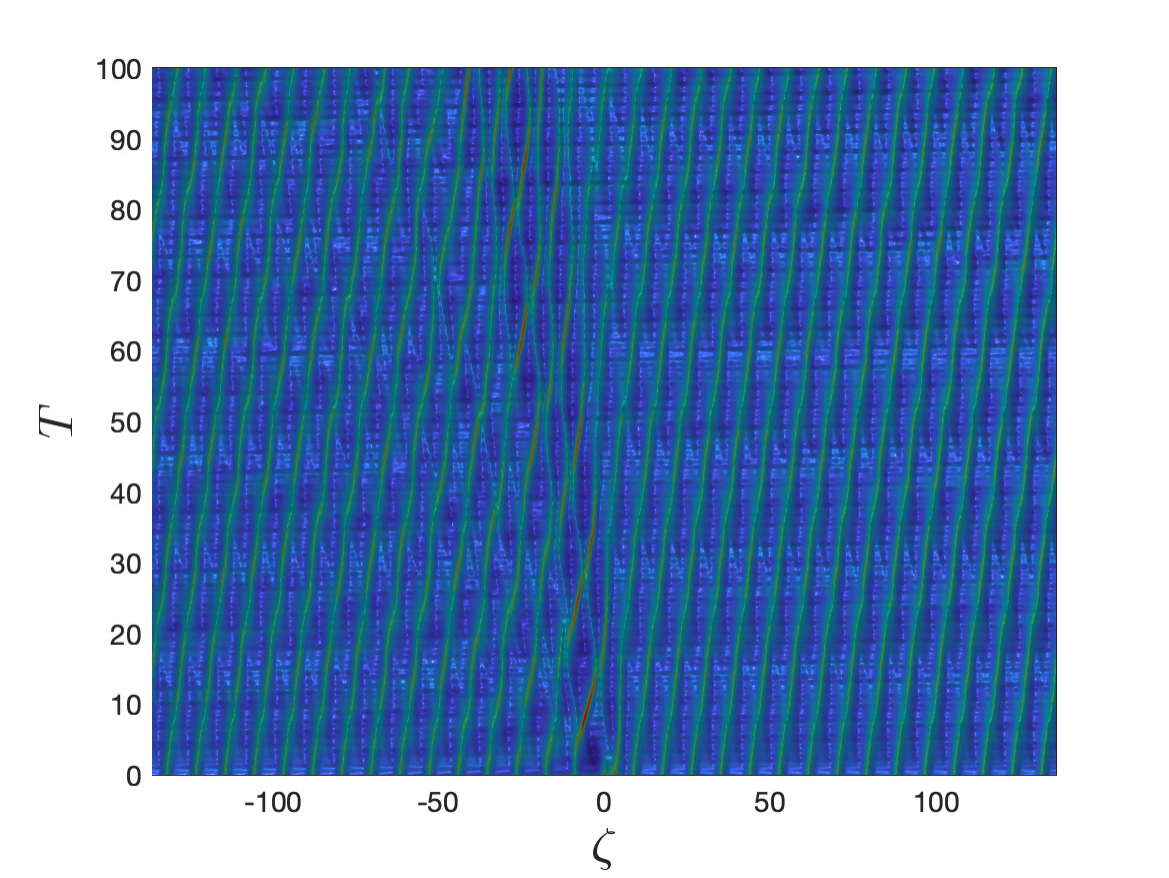}  \\[0.2cm]
    \includegraphics[width=0.47 \linewidth]{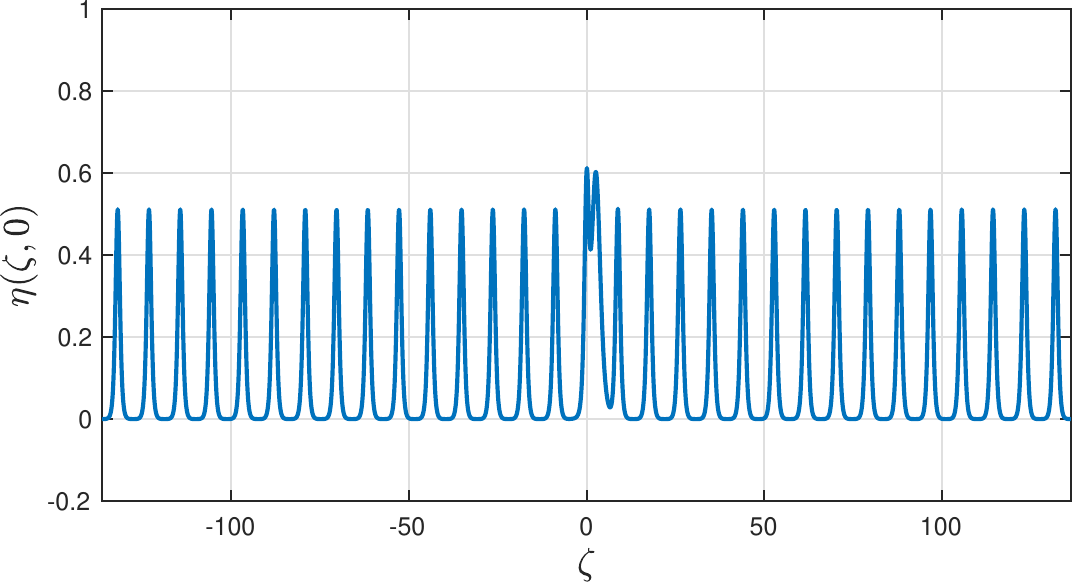}      \quad
    \includegraphics[width=0.47 \linewidth]{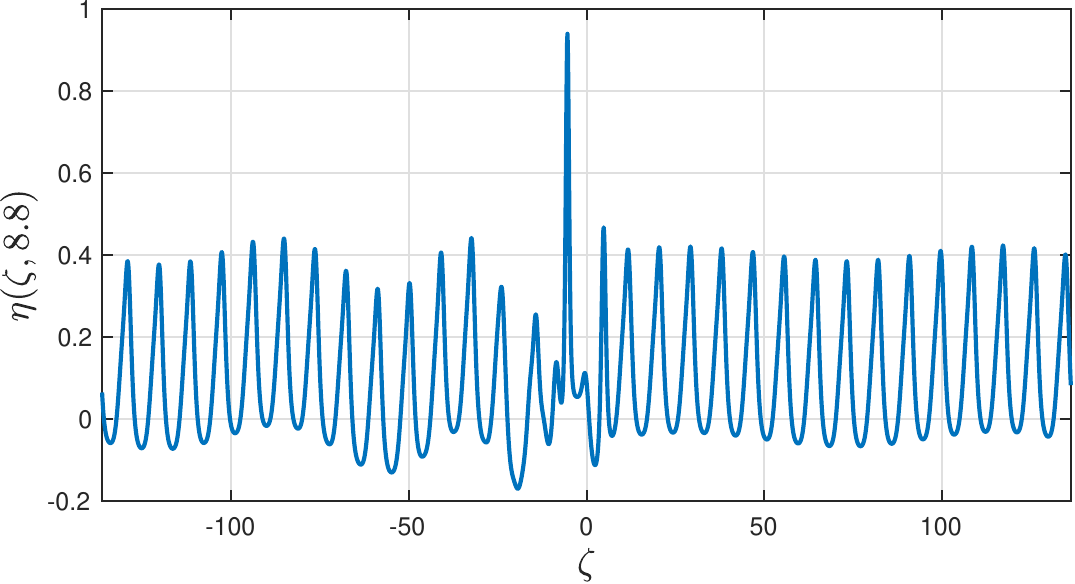}
      \caption{Rogue wave generation: 3D and 2D views of the interfacial displacement from above (top row) and initial condition at $T=0$ vs interfacial displacement at $T = 8.8$  (bottom row).  }
        \label{fig:RW}
\end{figure}

 In this study we did not aim to systematically investigate the probability of generation of rogue waves due to this scenario (in the sense of the classical definition of rogue waves as the waves with the amplitude being more than twice the significant wave height), but it is worth mentioning  the generation of a rogue wave shown in Figure \ref{fig:RW} using the same physical parameters as in Figure (\ref{fig:A30}) and initial condition (\ref{cnoidal_ad}) with $u_1 = -10^{-3},\, u_2=0,\, u_3=3,\, A_1 = 0.60,\, A_2 = 0.60$, and  $\zeta_0 = -2.60$. The computational domain is $2L=272.81$ with the number of modes $M=2726$, the spatial step $\Delta x \approx 10^{-1}$, the total simulation time $T_{max} = 100$, with the temporal step $\Delta T = 10^{-2}$. This could be an interesting direction of further research.

We hope that our study sheds light both on the typical structure of nearly cnoidal waves (with defects) observed in oceanic observations (see, for example, Figure 1 in the Introduction), and on the role of rotation in the formation of bursts of large interfacial displacements and shear in the direction of wave propagation. Further developments could concern bi-directional propagation and effects related to barotropic transport, which were discussed in the construction of the weakly-nonlinear solution, but were left out of scope of the subsequent modelling. Another issue which was left out of scope of the present study was the effect of the depth-dependent shear currents (see \citet{HKG2021, TABK2023} for the recent developments related to the presence of a long-wave instability,  in the absence of rotation). As a by-product of our study we constructed some curious {\color{black} generalised travelling waves} of the KdV equation, whose smoothed counterparts naturally emerged in the evolution of initial conditions in the form of cnoidal waves with local {\color{black} perturbations}.  It would be interesting to investigate whether similar solutions can be found in other integrable models, and what is their meaning and role in the relevant physical contexts. 

%\clearpage
\section{Acknowledgements}
Karima Khusnutdinova is grateful to Evgeny Ferapontov, Sergey Gavrilyuk, Boris Kruglikov and Victor Shrira for useful discussions, and  to Zakhar Makridin and Nikolay Makarenko for sending the paper \citet{M1989}. {\color{black} We also thank the referees for their helpful comments, and Jean-Claude Saut for the references to \citet{B1993} and \citet{K1983, KF1984}. } Korsarun Nirunwiroj would like to acknowledge the Royal Thai Government Scholarship supporting his postgraduate studies at Loughborough University. Karima Khusnutdinova would like to thank the Isaac Newton Institute for Mathematical Sciences, Cambridge, for support and hospitality during the programme ``Emergent phenomena in nonlinear dispersive waves", where partial work on this paper was undertaken. This work was supported by EPSRC grant EP/R014604/1.
\bigskip

\noindent
{\bf Declaration of Interests.} The authors report no conflict of interest.
\appendix

%%% Appendix B %%%
\section{}
\label{sec:Appendix A}

{\color{black}  The cnoidal waves with expansion / contraction defects  have a jump in the second / first derivative at some points within the domain.  Our analysis below follows the recent line of research developed in the papers by 
\citet{GNST2020,GS2022,GNS2024}.  The key idea is to treat the points of discontinuity similarly to shocks in the theory of {\it hyperbolic conservation laws}, i.e. to split the solution into several parts where the solution and all derivatives are smooth, and to match them at the points of discontinuity in such a way that the `generalised Rankine-Hugoniot (gRH)' conditions following from the {\it dispersive conservation laws} are satisfied. In the previous work developed within the scope of nonintegrable equations, e.g. the Benjamin-Bona-Mahoney equation considered by \citet{GS2022}, the conservation laws have led to some nontrivial gRH conditions, and numerical experiments have supported the importance of these conditions. Our observation reported in this Appendix is that the cnoidal waves of the KdV equation with expansion or contraction defects do not lead to any nontrivial jump conditions - all conservation laws are satisfied, in the same sense as in \citet{GS2022}. Moreover, the cnoidal wave with an expansion defect satisfies the Weirstrass-Erdmann corner condition and is a broken extremal of the associated variational problem. }

The KdV equation 
\begin{equation}
u_t - 6 u u_x + u_{xxx} = 0
\label{KdVA}
\end{equation}
can be represented in the form of a conservation law
\begin{equation}\label{1c}
\frac{\partial u}{\partial
t}+\frac{\partial}{\partial x}\left(-3u^2+u_{xx}\right)=0.
\end{equation}
Equation (\ref{1c}) implies conservation of 
`mass',
\begin{equation}\label{mass0}
 \frac{d}{d t}\int ^{L}_{-L} u dx=0,
\end{equation}
provided the function $u(x,t)$ either vanishes together with its
spatial derivatives as $L \to \infty$, or it is spatially
periodic on the interval $[-L, L]$.  The latter case is considered in our paper.

One also has  conservation laws for
 `momentum'
\begin{equation}\label{mom}
\frac{\partial}{\partial t}
\frac{u^2}{2}+\frac{\partial}{\partial
x}\left(uu_{xx}-\frac{1}{2}u_x^2 -2u^3 \right)=0
\end{equation}
and  `energy'
\begin{equation}\label{energy}
 \frac{\partial}{\partial t} \left(
u^3+\frac{1}{2}u^2_x \right)+\frac{\partial}{\partial x}\left(
-\frac{9}{2}u^4+3u^2u_{xx}-6uu_x^2 + u_x u_{xxx}-\frac{1}{2}u_{xx}^2\right)=0.
\end{equation}

It is well known that the KdV equation has infinitely many polynomial conservation laws, which can be proven using a generating function \citep{MGK1968}. {\color{black} We reproduce the main part of the proof below, since we need some formulae from it for our subsequent considerations. }The Gardner transformation
\begin{equation}
u=w+\epsilon w_x+ \epsilon^2 w^2,
\label{trans}
\end{equation}
 where 
$\epsilon$ is the grading parameter,  yields
\begin{equation}
u_t-6uu_x+u_{xxx}=(1+2\epsilon^2 w+\epsilon
\partial_x)[w_t+(w_{xx} - 3w^2-2\epsilon^2 w^3)_x]=0.
\end{equation}
Therefore, if $w(x,t)$ satisfies the conservation  law 
\begin{equation}\label{Gard}
w_t+(w_{xx} - 3w^2-2\epsilon^2 w^3)_x=0,
\end{equation}
then $u(x,t)$ satisfies the KdV equation.
Next, one represents $w$ in the form of an asymptotic expansion in
powers of $\epsilon$:
 \begin{equation}
 w = \sum_{n=0}^{\infty}\epsilon^n w_n.
 \label{w}
 \end{equation}
Collecting the coefficients of equal powers of  $\epsilon$ in (\ref{trans}) gives us
\begin{eqnarray}
&\displaystyle w_0=u \, , \qquad w_1=-w_{0x} =-u_x \, , \qquad
w_2=-w_{1x} -w_0^2 =u_{xx}-u^2,  &\\
&\displaystyle w_n = - w_{n-1, x} - \sum_{k=0}^{n-2} w_k w_{n- 2 - k} \quad \mbox{for} \quad n \ge 2. \label{w_n}&
\end{eqnarray}
Now, substituting expansion  (\ref{w})  into (\ref{Gard}) one obtains an infinite series of the KdV
conservation laws as the coefficients at even powers of $\epsilon$. The coefficients of odd powers are exact differentials \citep[referred to as `trivial conservation laws', see][for the details]{MGK1968}. 
The integrals
 $$I_n=\int ^{L}_{-L}w_{2n} dx\,,  \quad n=0,\,1,\,2,\, \dots $$ are called `Kruskal
integrals'. In particular, the first three integrals take the form
$$I_1=\int_{-L}^L u dx, \quad I_2 = \int_{-L}^L \frac 12 u^2 dx, \quad I_3= \int_{-L}^L \left (u^3 + \frac 12 u_x^2 \right ) dx.$$

Next, to prove that cnoidal waves with our contraction and expansion defects satisfy all conservation laws of the KdV equation, we also need a recurrence formula for the fluxes. Substituting  expansion (\ref{w}) into (\ref{Gard}) we obtain
\begin{eqnarray}
w_{xx} - 3 w^2 + 2 \epsilon^2 w^3 = \sum_{n=0}^\infty \varepsilon^n f_n,  
\end{eqnarray}
where
\begin{eqnarray}
&\displaystyle f_0 = w_{0, xx} - 2 w_0^2,  \quad f_1 = w_{1, xx} - 6 w_0 w_1,  \quad f_2 = w_{2, xx} - 3 w_1^2 - 6 {\color{black} w_0} w_2 + 2 w_0^3,\quad & \\
&\displaystyle f_n = w_{n, xx} - 3 \sum_{k=0}^n w_k w_{n-k} + 2 \sum_{k+l+m = n-2} w_k w_l w_m \quad \mbox{for} \quad n \ge 2 \label{f_n},&
\end{eqnarray}
with the last summation being over all triples of integers $(k, l, m)$ such that  $k+l+m = n-2$. {\color{black} The Kruskal integrals and the recurrence relations for the fluxes are invariant with respect to the Galilean transformations. }

{\color{black} Let us now consider a cnoidal wave with an expansion defect, as shown in the first row of Figure \ref{fig:WS}. This wave is constructed by cutting at a trough of the cnoidal wave between two peaks and symmetrically adding a segment of a straight line at the corresponding level. In the original $(x,t)$ frame, the entire wave moves with the constant speed of the cnoidal wave. We now consider it in the reference frame moving with this constant speed, then the wave is stationary. We place the origin in the middle of the defect. Then, the function describing the wave profile is even, see the top row of Figure \ref{fig:WS}.  Let us denote the period of the cnoidal wave by $\ell$. If we consider $m$ peaks in the domain, then without the defect $L=m \ell/2$. With the defect, given that the length of the inserted symmetric interval is $d$, then $L = (m \ell+d)/2$. The resulting function has a corner at two points in the periodic domain $[-L, L]$, and the graph is symmetric about the vertical axis passing through the middle of the interval which has been assumed to be $x=0$, with the function $w_0 = u$ being even in that frame. From the recurrence relation (\ref{w_n}), we observe that functions $w_{2n}$ are even 
%for the even values of $n$, and it is odd for the odd values of $n$. 
and $w_{2n+1}$ are odd.
Let the $x$-coordinates of the left and right endpoints of the line segment be denoted by $x_1$ and $x_2$, respectively. With our choice of the reference frame we  have $x_1 = - x_2 = - x_d$.  Consequently, treating the points of discontinuity similarly to shocks in the theory of hyperbolic equations (see  \citet{GNST2020,GS2022,GNS2024}), we represent the integral as the sum over three subintervals, and obtain
\begin{eqnarray}
\frac{d}{dt} I_n &=& - \int_{-L}^{-x_d} {f_{2n}}_ x dx -  \int_{-x_d}^{x_d} {f_{2n}}_xdx - \int_{x_d}^{L} {f_{2n}}_x dx\\
&=& - \left [f_{2n}\right]_{-L}^ {-x_d} - \left[f_{2n} \right]_{-x_d}^ {x_d} - \left[f_{2n}\right]_{x_d}^ {L}  = 0
\end{eqnarray}
by periodicity, the constant value of $f_{2n}$ on the interval $[-x_d, x_d]$, and since the flux $f_{2n}$ is an even function. 
We note that this can be generalised by first extracting a small part close to the trough, symmetrically inserting a segment of a straight line and gluing the remaining parts to it. The resulting functions will also satisfy all conservation laws of the KdV equation. Note that unlike the examples constructed in \citet{GS2022}  in the context of the Benjamin-Bona-Mahoney equation, there appear no nontrivial gRH conditions. 

Similarly we can  consider a cnoidal wave with a contraction defect, as shown in the second row of Figure \ref{fig:WS}, again in the reference frame where the wave is stationary. This wave is constructed by extracting a symmetric interval around a trough between  two neighbouring peaks and gluing the remaining parts together.  Let us again denote the period of the cnoidal wave by $\ell$ and consider $m$ peaks in the domain, then without the defect $L=m \ell/2$. With the defect, given that the length of the extracted symmetric interval is $d$, then $L = (m \ell-d)/2$. The resulting function has a corner at one point in the periodic domain $[-L, L]$, and the graph is symmetric about the vertical axis passing through that point. Let us denote this point by $x_0$.  We can then associate the origin of the moving reference frame with $x_0$, i.e.\ assume $x_0 = 0$, and the function $w_0 = u$ will be even in that frame. From the recurrence relation (\ref{w_n}), we observe that functions $w_{2n}$ are even 
%for the even values of $n$, and it is odd for the odd values of $n$. 
and $w_{2n+1}$ are odd.
Recalling that non-trivial conservation laws correspond to even values $2n$, we deduce from the recurrence relation (\ref{f_n}) that the corresponding fluxes $f_{2n}$ can only contain terms representing even functions. Therefore,
\begin{eqnarray}
\frac{d}{dt} I_n &=& - \int_{-L}^{0}{ f_{2n}}_x dx -  \int_{0}^{L}{ f_{2n}}_x dx \\
&=& - \left [f_{2n} \right]_{-L}^ {0} - \left [f_{2n} \right]_{0}^ {L} = 0
\end{eqnarray}
by periodicity and since the flux $f_{2n}$ is an even function. 

\begin{figure}
    \centering
    \includegraphics[width=0.28\linewidth]{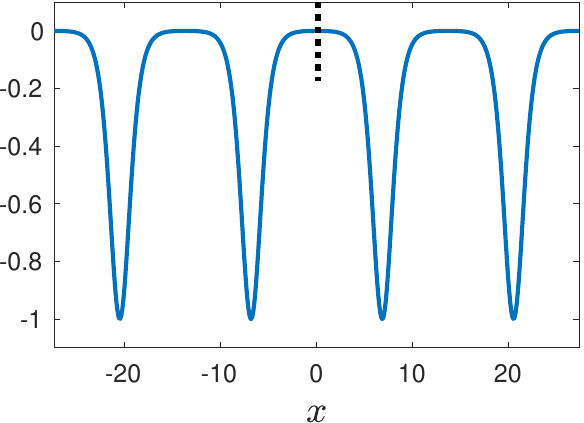} \quad 
    \includegraphics[width=0.28\linewidth]{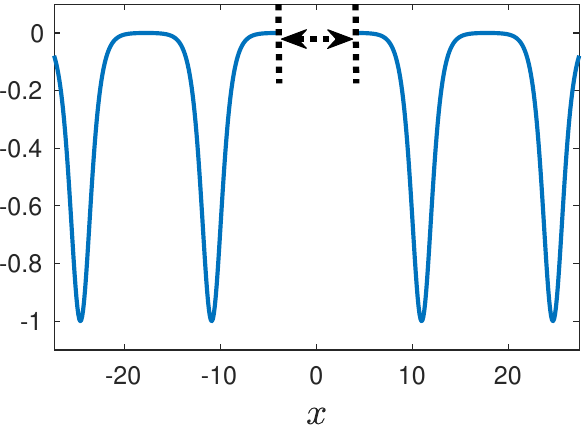} \quad
    \includegraphics[width=0.28\linewidth]{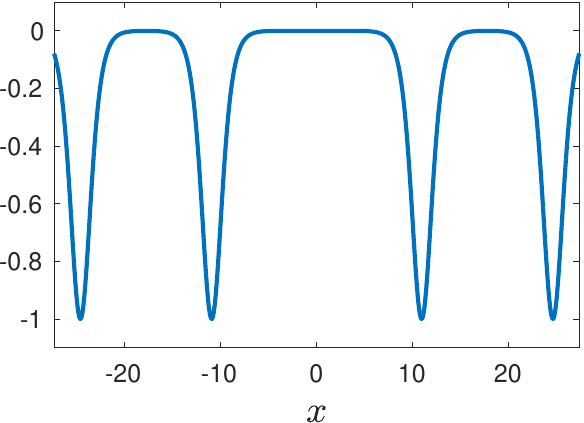}
\\[1ex]
 \includegraphics[width=0.28\linewidth]{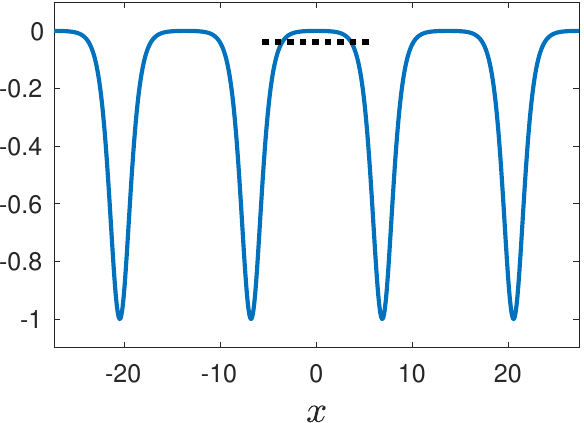} \quad 
    \includegraphics[width=0.28\linewidth]{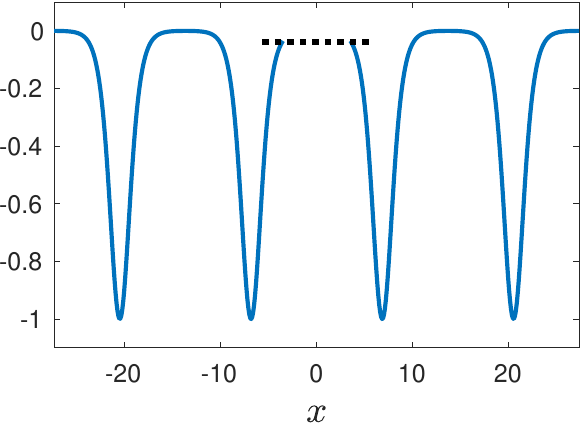} \quad
    \includegraphics[width=0.28\linewidth]{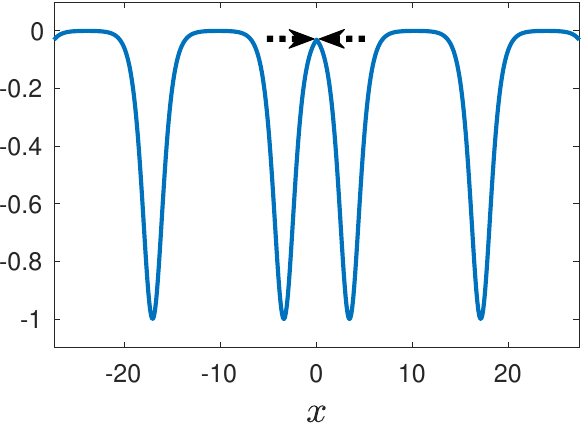}
\\[1ex]
    \includegraphics[width=0.28\linewidth]{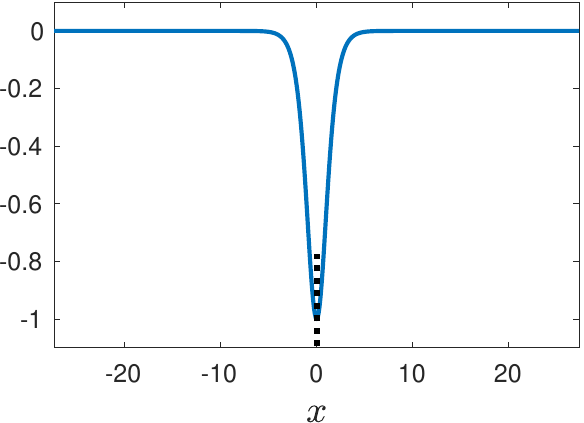} \quad 
    \includegraphics[width=0.28\linewidth]{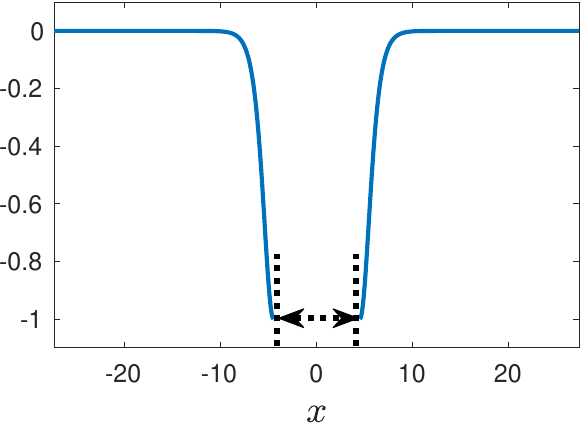} \quad
    \includegraphics[width=0.28\linewidth]{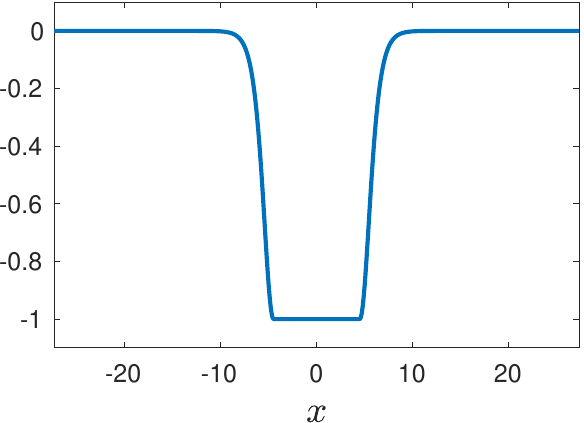}
\\[1ex]
     \includegraphics[width=0.28\linewidth]{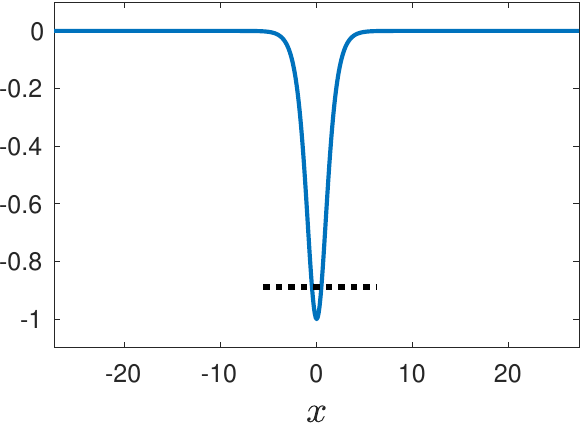} \quad 
    \includegraphics[width=0.28\linewidth]{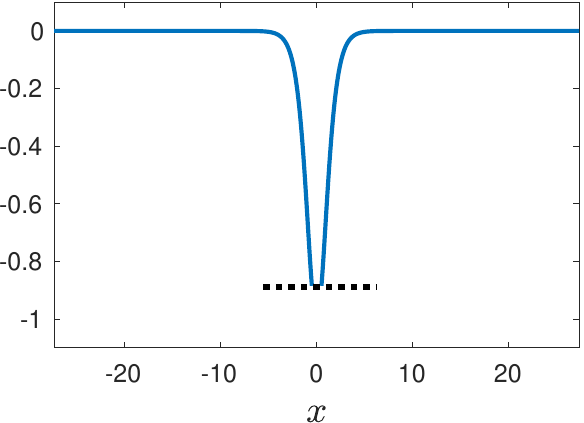} \quad
    \includegraphics[width=0.28\linewidth]{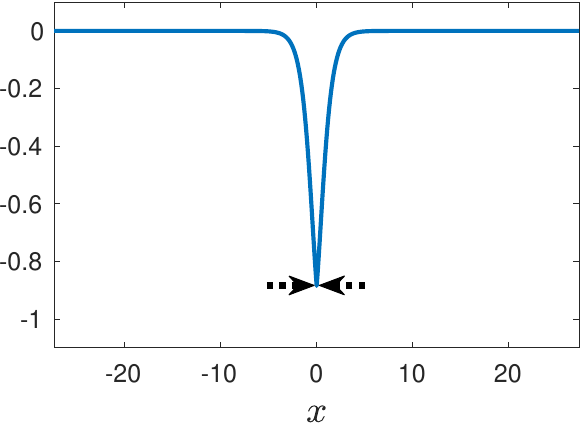}
    \caption{Schematic of construction of   long-lived approximate and exact weak solutions of the KdV equation in the form of a cnoidal wave with an expansion and contraction periodicity defects (first and second rows, respectively),  and a short-lived soliton with similar amplitude defects (third and fourth rows, respectively). {\color{black} The waves have negative polarity: the peaks have negative amplitude.}}
\label{fig:WS}
\end{figure}

Another way to proving the above relies on using that all non-trivial fluxes $f_{2n}$ are polynomial functions of even weight with respect to the scaling symmetries of the KdV equation. This also implies that the fluxes $f_{2n}$ are even functions, and the rest follows from that.

Next, the KdV equation (\ref{KdVA}) can be written in Lagrangian form
\begin{equation}
\delta \int {\cal L} dx dt = 0, \quad \mbox{where} \quad {\cal L} = \frac 12 \phi_t \phi_x - \phi_x^3 - \frac 12 \phi_{xx}^2, \quad u = \phi_x.
\label{LD}
\end{equation}
The cnoidal waves with expansion defects can be viewed as generalised (shock-like) travelling waves (see \citet{GS2022} for a relevant discussion in the context of the Benjamin-Bona-Mahoney equation). Then Lagrangian density (\ref{LD}) becomes a function of $u$ and $u_x$: ${\cal L} = {\cal L}(u, u_x)$.
The Weirstrass-Erdmann corner conditions for broken extremals require the continuity of
\begin{equation}
\frac{\partial \cal L}{\partial u_{x}} \quad \mbox{and} \quad {\cal L} - u_{x} \frac{\partial {\cal  L}}{\partial u_{x}}
\end{equation}
at each junction (e.g. \citealt{CF1954}) , which leads to the requirement of continuity of $u_x$, since $u$ is continuous.
This condition is satisfied exactly for a cnoidal wave with an expansion defect, and approximately for a cnoidal wave with a contraction defect, provided we cut close to extremum. The smoothed counterparts of cnoidal waves with both expansion and contraction defects were long-lived in our numerical simulations, with the former function generating states behaving very closely to a travelling wave.
%The cnoidal wave with an expansion defect is a generalised travelling wave in the terminology of \citet{GS2022}.
%Therefore, it makes sense to view the corresponding functions are the exact and approximate weak travelling wave solutions of the KdV equation, respectively. 

Having observed  that the constructed cnoidal waves with expansion defects generate long-lived states  close to travelling waves, it is also temping to try to construct similar solutions by cutting around the peak rather than trough. In contrast to the previous cases, this can be done for a single peak as well, starting from the exact soliton solution, as shown in the third and fourth rows of Figure \ref{fig:WS}.  However, in numerical runs, the smoothed counterparts of these solutions turned out to be unstable and short-lived, fissioning and giving rise to the usual solitons, in full agreement with the Inverse Scattering Transform (IST) predictions (\citealt{GGKM1967}, see also \citealt{DJ1989}) for the respective smoothed initial conditions. Similar behaviour is observed in the cnoidal wave with an insertion at a peak rather than though. Such smoothed initial conditions do not produce long-lived states close to travelling waves in pseudospectral simulations, they are unstable.

Finally, let us also make some remarks concerning a possible weak formulation. Consider a $2L$-periodic regular travelling wave described by a function $u = u(\xi), \xi = x - c t$ of the KdV equation (\ref{KdVA}):
\begin{equation}
(-c u - 3 u^2 + u_{\xi \xi})_{\xi} = 0.
\label{KdV1}
\end{equation}
%The cnoidal wave with an expansion defect is not a classical solution since it has discontinuous second derivative at two points in the periodic domain. 
Let us multiply equation (\ref{KdV1}) by a test function $\phi$, which is assumed to be smooth and $2L$-periodic, and integrate by parts. Then,
\begin{equation}
\int_{-L}^{L} (-c u - 3 u^2 + u_{\xi \xi})_{\xi} \phi d \xi = - \int_{-L}^L (- c u + u^2 + u_{\xi \xi}) \phi_{\xi} d \xi = 0,
\label{wf}
\end{equation}
where we used periodicity of the functions.
% and since $u$ is an even function. 
Consider a cnoidal wave with an expansion defect. It has a jump in the second derivative at two points in the periodic domain and does not satisfy the strong formulation (\ref{KdV1}). However, it does satisfy equation (\ref{wf}).
Indeed, let us consider the wave in the reference frame moving with the speed $c$. The wave is stationary and even in this reference frame. Then, it is natural to require the test function $\phi$ to be even too. The equation (\ref{KdV1}) is invariant with respect to the Galilean transformation. 
 The integral in the right-hand side of (\ref{wf}) is well-defined. Represent it as the sum of three integrals, over the intervals $[-L, -x_d], [-x_d, x_d], [x_d, L]$. In the first and third integrals, $ (- c u + u^2 + u_{\xi \xi})  = {\rm const}$ (cnoidal wave), and then the sum of these integrals is equal to zero using that $\phi$ is even and periodic. In the middle integral, $u = {\rm const}$ (expansion defect), and then again the integral vanishes using that $\phi$ is an even function. Hence, a cnoidal wave with an expansion defect satisfies a weak formulation, in the above sense.

If we relax the conditions on the test function, and require only the periodicity but do not require $\phi$ to be an even function, then we can still make the above argument work by requiring that the length of the defect is commensurate with the period $\ell$ of the cnoidal wave, i.e. $ x_d = n \ell$, where $n$ is a natural number. We note that we experimented with both commensurate and non-commensurate lengths of the defect, and there was no significant difference in the states produced at the end of long pseudospectral runs.

The global well-posedness of the Cauchy problem with periodic boundary conditions  in $L^2$, including uniqueness and continuous dependence with respect to the initial data, was proven by \citet{B1993} (the global existence of the weak $L^2$ solutions on the infinite line was proven by \citet{K1983,KF1984}).  We believe that our considerations provide strong evidence that cnoidal waves with expansions defects considered in this paper are weak solutions of the KdV equation. The states close to them naturally emerged in numerical modelling of the evolution of initial conditions given by cnoidal waves with localised perturbations (see, for example, the top rows of Figures \ref{fig:A21.1} and  \ref{fig:A30}, showing the evolution in the absence of rotation).
}

%%% Appendix A %%%
\section{}
\label{sec:Appendix B}
The Ostrovsky equation
\begin{align}
%    \big( \tilde{\eta}^{(0)}_{T} +\dfrac{3\sigma}{2c_0} \tilde{\eta}^{(0)}\tilde{\eta}^{(0)}_{\zeta} + \dfrac{ \beta  c_0^3}{6} \tilde{\eta}^{(0)}_{\zeta\zeta\zeta} \big)_\zeta = \dfrac{ \gamma^2}{2c_0} \tilde{\eta}^{(0)}
%    \label{eq:ost}
        \big( u_{T} + \alpha_1 u u_{\zeta} + \beta_1 u_{\zeta\zeta\zeta} \big)_\zeta = \gamma_1 u
    \label{eq:ost}
\end{align}
%where the characteristic variable $\zeta = \xi - \frac{ \sigma \hat{\eta}}{2c_0}T$, and slow time variable $T = \alpha t$, 
%derived from the MMCC-f equations, is obtained with the asymptotic solutions where each solution can be expressed in terms of $\tilde{\eta}^{(0)}$ and $\hat{V}$. The equation 
is numerically solved by pseudospectral methods (see \citealt{F1996, T2000}), which usually offer the highest accuracy and computational efficiency for smooth data on periodic domains. By implementing the Fast Fourier Transform (FFT) algorithm for spatial derivatives and the fourth-order Runge--Kutta scheme for time stepping, the method yields 
%coefficients of this combination can be computed, resulting in the 
accurate approximation of the solution to the differential equation.

The spatial domain is discretised by $M$ equidistant points with spacing $\Delta x = 2\pi /M$. Then, the discrete Fourier transform of the equation with respect to $\zeta$ gives
\begin{align*}
    \hat{u}_T - i\big(k^3 \beta_1  -\dfrac{\gamma_1}{k}\big)\hat{u} = - \dfrac{ik\alpha_1}{2} \hat{(u^2)}, 
\end{align*}
where 
%$u$ is $\tilde{\eta}^{(0)}$ and 
$k$ is the scaled wavenumber.
We use the 4th-order Runge--Kutta scheme for temporal integration. By the integrating factor method of Kassam and Trefethen (e.g.\ \citealt{T2000}), we multiply  the equation by $K=\exp[-i(k^3\beta_1-\frac{\gamma_1}{k})T]$  to obtain
\begin{align*}
    \hat{U}_T = -\dfrac{ik\alpha_1}{2} K \mathcal{F}\{ ( \mathcal{F}^{-1}\Big[ \dfrac{\hat{U}}{K} \Big])^2\},
\end{align*}
where
$\hat{U} = \exp[-i(k^3\beta_1-\frac{\gamma_1}{k})T] \, \hat{u} = K \hat{u}\,$ and $\mathcal{F}$ is the Fourier transform. Discretising the time domain as $T_n=n \Delta T$, and introducing the function 
\begin{align*}
    E = \exp[\frac{i}{2}(k^3\beta_1-\frac{\gamma_1}{k})\Delta T],
\end{align*}
the optimised Runge--Kutta time stepping has the form
\begin{align*}
    \hat{u}^{(n+1)} = E^2 \hat{u}^{(n)} + \dfrac{1}{6}[E^2 k_1 + 2E (k_2 + k_3) + k_4],
\end{align*}
where
\begin{align*} 
    k_1 &= -\dfrac{ik\alpha_1}{2} \Delta T \; \mathcal{F}\{ (\mathcal{F}^{-1} [ \hat{u}^{(n)} ] )^2\},\\
    k_2 &= -\dfrac{ik\alpha_1}{2} \Delta T \; \mathcal{F}\{ (\mathcal{F}^{-1} [ E (\hat{u}^{(n)} +k_1/2) ] )^2\},\\
    k_3 &= -\dfrac{ik\alpha_1}{2} \Delta T \; \mathcal{F}\{ (\mathcal{F}^{-1} [  E \hat{u}^{(n)} +k_2/2])^2\},\\
    k_4 &= -\dfrac{ik\alpha_1}{2} \Delta T \; \mathcal{F}\{ (\mathcal{F}^{-1} [  E^2 \hat{u}^{(n)} + E k_3 ])^2\}.
\end{align*}

\bibliographystyle{jfm}
%\bibliography{jfm2esam}

\end{document}